\documentstyle[12pt]{article}
\def\9{\phantom 0}      %%% for lining up numbers in columns
\renewcommand\linebreak{\unskip\break} %% breaks line & still justifies
 \newcommand{\be}[1]{\begin{equation}\label{#1}}
 \newcommand{\ee}{\end{equation}}
 \newcommand{\beqn}[1]{\begin{eqnarray}\label{#1}}
 \newcommand{\eeqn}{\end{eqnarray}}
\newcommand\beq{\begin{equation}}
\newcommand\eeq{\end{equation}}
\newcommand\st{\sin\theta}
\newcommand\ct{\cos\theta}
\newcommand\esc{{e \over{\st \ct}}}

\newcommand{\hc}{ {\rm h.c.} }
\newcommand{\ME}{ M_{ETC} }
\newcommand{\gE}{ g_{ETC} }

\def\Re{{\cal R \mskip-4mu \lower.1ex \hbox{\it e}\,}}
\def\Im{{\cal I \mskip-5mu \lower.1ex \hbox{\it m}\,}}
\def\ie{{\it i.e.}}
\def\eg{{\it e.g.}}

\def\etal{{\it et al.}}
\def\ibid{{\it ibid}.}
\def\sub#1{_{\lower.25ex\hbox{$\scriptstyle#1$}}}
\def\sul#1{_{\kern-.1em#1}}
\def\sll#1{_{\kern-.2em#1}}  
\def\sbl#1{_{\kern-.1em\lower.25ex\hbox{$\scriptstyle#1$}}}
\def\ssb#1{_{\lower.25ex\hbox{$\scriptscriptstyle#1$}}}
\def\sbb#1{_{\lower.4ex\hbox{$\scriptstyle#1$}}}

\def\to{\rightarrow}

\def\mh{\ifmmode m\sbl H \else $m\sbl H$\fi}
\def\mch{\ifmmode m_{H^\pm} \else $m_{H^\pm}$\fi}
\def\mt{\ifmmode m_t\else $m_t$\fi}
\def\mc{\ifmmode m_c\else $m_c$\fi}
\def\mz{\ifmmode M_Z\else $M_Z$\fi}
\def\mw{\ifmmode M_W\else $M_W$\fi}
\def\mnew{M_{\rm new}}
\def\mws{\ifmmode M_W^2 \else $M_W^2$\fi}
\def\mhs{\ifmmode m_H^2 \else $m_H^2$\fi}   
\def\mzs{\ifmmode M_Z^2 \else $M_Z^2$\fi}
\def\mts{\ifmmode m_t^2 \else $m_t^2$\fi}
\def\mcs{\ifmmode m_c^2 \else $m_c^2$\fi}
\def\mchs{\ifmmode m_{H^\pm}^2 \else $m_{H^\pm}^2$\fi}
\def\ztwo{\ifmmode Z_2\else $Z_2$\fi}
\def\zone{\ifmmode Z_1\else $Z_1$\fi}
\def\mtwo{\ifmmode M_2\else $M_2$\fi}
\def\mone{\ifmmode M_1\else $M_1$\fi}
\def\tb{\ifmmode \tan\beta \else $\tan\beta$\fi}
\def\xw{\ifmmode x\sub w\else $x\sub w$\fi}
\def\ch{\ifmmode H^\pm \else $H^\pm$\fi}
\def\lum{\ifmmode {\cal L}\else ${\cal L}$\fi}
\def\inpb{\ifmmode {\rm pb}^{-1}\else ${\rm pb}^{-1}$\fi}
\def\infb{\ifmmode {\rm fb}^{-1}\else ${\rm fb}^{-1}$\fi}
\def\epem{\ifmmode e^+e^-\else $e^+e^-$\fi}
\def\ppb{\ifmmode \bar pp\else $\bar pp$\fi}
\def\dmix{\ifmmode D^0-\bar D^0 \else $D^0-\bar D^0$\fi}
\def\dm{\ifmmode \Delta m_D \else $\Delta m_D$\fi}
\def\dmd{\ifmmode \Delta m_D \else $\Delta m_D$\fi}
\def\dg{\ifmmode \Delta \Gamma_D \else $\Delta \Gamma_D$\fi}
\def\bsg{\ifmmode b\to s\gamma \else $b\to s\gamma$\fi}

\newskip\zatskip \zatskip=0pt plus0pt minus0pt
\def\matth{\mathsurround=0pt}
\def\lsim{\mathrel{\mathpalette\atversim<}}
\def\gsim{\mathrel{\mathpalette\atversim>}}
\def\atversim#1#2{\lower0.7ex\vbox{\baselineskip\zatskip\lineskip\zatskip
  \lineskiplimit 0pt\ialign{$\matth#1\hfil##\hfil$\crcr#2\crcr\sim\crcr}}}

\def\AQ#1{\Pi_{\rm #1}(q^2)}
\def\A0#1{\Pi_{\rm #1}(0)}

\def\AZ#1{\Pi_{\rm #1}(\mz^2)}

\def\AP0#1{\Pi'_{\rm #1}(0)}
\def\APW#1{\Pi'_{\rm #1}(\mw^2)}
\def\APZ#1{\Pi'_{\rm #1}(\mz^2)}

\def\ANEW0#1{\Pi_{\rm #1}^{\rm new}(0)}
\def\ANEWP0#1{{\Pi'_{\rm #1}}^{\rm new}(0)}

\def\B0#1{\Pi'_{\rm #1}(0)}
\def\BW#1{\frac{\Pi_{\rm #1}(\mw^2)-\Pi_{\rm #1}(0)}{\mw^2}}
\def\BZ#1{\frac{\Pi_{\rm #1}(\mz^2)-\Pi_{\rm #1}(0)}{\mz^2}} 
\def\APP0#1{\Pi''_{\rm #1}(0)}
\def\APPP0#1{\Pi'''_{\rm #1}(0)}
\def\numu{\nu_\mu}
\def\numubar{\bar\nu_\mu}
\def\gl2{g_L^2}
\def\gr2{g_R^2}
\def\ganue{g_A^{\nu e}}
\def\gvnue{g_V^{\nu e}}
\def\QWCS{Q_W({}^{133}_{\phantom{1}55}{\rm Cs})}
\def\QWTL{Q_W({}^{205}_{\phantom{1}81}{\rm Tl})}
\def\MSbar{\overline{\rm MS}}

\def\seff{\sin^2\theta_{\rm eff}^{\rm lept}}
\def\AFB{A_{\rm FB}}
\def\sig0{\sigma_{\rm h}^0}
\def\Gamz{\Gamma_Z}
\def\Gamh{\Gamma_{\rm had}}
\def\Gamb{\Gamma_{b\bar{b}}}
\def\Gamc{\Gamma_{c\bar{c}}}

\def\Gaml{\Gamma_{\ell^+\ell^-}}
\def\Zbb{Zb\bar{b}}
\def\glb{g_L^b}
\def\grb{g_R^b}
\def\dglb{\delta g_L^b}
\def\dgrb{\delta g_R^b}
\def\ds{\delta s^2}
\def\das{\delta\alpha_s}

\def\xib{\xi_b}
\def\zetab{\zeta_b}

\def\beq{\begin{equation}}
\def\eeq{\end{equation}}
\def\beqa{\begin{eqnarray}}
\def\eeqa{\end{eqnarray}}
\def\tev{\,{\rm TeV}}
\def\gev{\,{\rm GeV}}
\def\mev{\,{\rm MeV}}

\def\nn{\noindent}
\def\kmue{K_L\rightarrow \mu^\pm e^\mp}
\def\kpimue{K^+\rightarrow \pi^+ \mu^\pm e^\mp}
\def\kpinunu{K^+\rightarrow \pi^+\nu\overline{\nu}}
\def\kmumu{K_L\rightarrow\mu^+\mu^-}
\def\klpiee{K_L\rightarrow\pi^0 e^+ e^-}
\def\klpinu{K_L\rightarrow\pi^0 \nu\overline{\nu}}
\def\kmu3{K^+\rightarrow\pi^0\mu\nu}

\begin{document}
\input psfig.sty
\newlength{\captsize} \let\captsize=\small % use \let\normalsize=\captsize
\newlength{\captwidth}                     % just before \caption{ ...
%%%%%%%%%%%%%%%%%%%%%%%%%%%%%%%%%%%%%%%%%%%%%%%%%%%%%%%%%%%%%%
\rightline{\vbox{\halign{&#\hfil\cr
&SLAC-PUB-7088\cr
&CERN-TH/96-56\cr
&March 1996\cr}}}
\vspace*{0.3cm}

\begin{center}
{\Large\bf INDIRECT PROBES OF NEW PHYSICS}\footnote{Work supported by the 
Department of Energy, Contract DE-AC03-76SF00515}

\vskip1.6pc

{\bf J.L. Hewett$^a$, T. Takeuchi$^b$, S. Thomas$^a$ }

\vskip1pc

$^a$ {\it Stanford Linear Accelerator Center, Stanford University,
Stanford, CA   94309\\}
$^b$ {\it CERN, TH-Division, CH-1211 Geneva 23, Switzerland}

\vspace*{0.3cm}

\small
\end{center}

\noindent{\underline {Contributors}}: 
G. B\'elanger,
C. Burgess,
D. Caldwell, 
U. Chattopadhyay,
R.S. Chivukula, 
T. Goto,
Y. Grossman,
D. Kennedy,
R.N. Mohapatra, 
P. Nath,
T. Nihei,
Y. Nir,
Y. Okada,
T.G. Rizzo, 
D. Silverman,
E.H. Simmons, 
J. Terning,
J.D. Wells

\vspace*{2cm}

\centerline{\bf ABSTRACT}

\vspace*{0.5cm}

\nn We summarize the indirect effects of new physics in a variety of 
processes.  We consider precision electroweak measurements,
the $g-2$ of the muon, rare decays, meson mixing, CP violation, lepton 
number violating interactions, double beta decay, and the electric dipole 
moments of atoms, molecules, and the neutron.  We include discussions of 
both model independent and dependent analyses where applicable.

\vspace*{1cm}

\nn To appear as a chapter in {\it Electroweak Symmetry Breaking and
Beyond the Standard Model}, edited by T. Barklow, S. Dawson, H.E. Haber, and
S. Siegrist, World Scientific.

\newpage

\begin{center}
{\Large\bf Table of Contents}
\end{center}
\vspace{0.25in}
\begin{flushleft}
1\quad Overview \dotfill 3\\
\bigskip
2\quad Description of Models \dotfill 3\\
\bigskip
3\quad Precision Electroweak Measurements \dotfill 8\\
\medskip
~~~~3.1 Introduction \dotfill 8\\
\medskip
~~~~3.2 The Measurements \dotfill 9\\
\medskip
~~~~3.3 Oblique Corrections \dotfill 11\\
\medskip
~~~~3.4 More Oblique Corrections \dotfill 17\\
\medskip
~~~~3.5 Non-Oblique Corrections \dotfill 19\\
\medskip
~~~~3.6 Extra Gauge Bosons \dotfill 31\\
\bigskip
4\quad A Model Independent Global Analysis \dotfill 32\\
\medskip
~~~~4.1 The Lowest-Dimension Effective Interactions \dotfill 32\\
\medskip
~~~~4.2 Comparing to Experiment \dotfill 34\\
\bigskip
5\quad $g-2$ of the Muon \dotfill 38\\
\bigskip 
6\quad Rare Processes in the Quark Sector \dotfill 40\\
\medskip
~~~~6.1 Kaons \dotfill 40\\
\medskip
~~~~6.2 Charm-Quark Sector \dotfill 44\\
\medskip
~~~~6.3 Bottom-Quark Sector \dotfill 53\\
\medskip
~~~~6.4 Top-Quark \dotfill 68\\
\bigskip
7\quad Electric Dipole Moments \dotfill 71\\
\bigskip
8\quad Lepton Number Violation \dotfill 74\\
\bigskip
9\quad Double Beta Decay \dotfill 76\\
\medskip
~~~~9.1 Experimental Situation \dotfill 77\\
\medskip
~~~~9.2 Neutrinoless Double Beta Decay and Physics Beyond the Standard Model
\dotfill 78\\
\bigskip
10\quad Summary \dotfill 84\\

\end{flushleft}

\newpage

\section{Overview}

The investigation of virtual effects of new physics provides an important
opportunity to probe the presence of interactions 
beyond the Standard Model (SM).
Various types of experiments may expose the existence of new physics, 
including the search for direct production of new particles at high
energy accelerators.  Although this scenario has the advantage in that it
would yield the cleanest observation of new physics, it is limited by the
kinematic reach and accumulated luminosity of colliders.  A complementary 
approach is offered by examining the indirect effects of new interactions 
in higher order processes and testing for deviations from SM predictions.  
In this case, one probes (i) the radiative corrections to perturbatively
calculable processes, as well as (ii) transitions which are either suppressed 
or forbidden in the SM.  Both of these scenarios carry the advantage of
being able to explore the existence of new physics at very high energy scales.  
In fact, studies of new loop induced couplings can provide a means of probing 
the detailed structure of the SM at the level of radiative corrections where
Glashow-Iliopoulos-Maiani (GIM) cancellations are important.  As will be
demonstrated below, in some cases the constraints on new degrees of 
freedom via indirect effects surpass those obtainable from collider searches.
In other cases, entire classes of models are found to be incompatible with
the data.  Given the large amount of high luminosity `low-energy' data which 
is presently available and will continue to accumulate during the next decade, 
the loop effects of new interactions in rare processes and precision 
measurements will play a major role in the search for physics beyond the SM.

In this report we will simultaneously follow both model independent and
dependent approaches, wherever possible, in determining the effects of
new physics.  In the following section, we first describe the general
features of the various models which we consider throughout the chapter.  
We then examine the capacity of precision electroweak measurements to probe 
new interactions, paying special attention to the current discrepancy between 
measurements and the SM prediction for the $Zb\bar b$ (and $Zc\bar c$) vertex,
as well considering oblique corrections.  Next we study the classic example
of precision tests, the $g-2$ of the muon.  We then turn our attention to
SM suppressed or forbidden processes, such as rare decays, meson mixing, and CP
violation, in the Kaon, charm-, bottom-, and top-quark systems.  The effects
of new physics on the electric dipole moments of atoms, molecules, and the
neutron are then investigated.  Finally, we focus on probes of the leptonic 
sector by examining lepton number violating processes, including 
double beta decay.

\section{Description of Models}

In this section we briefly summarize the general classes and
main characteristics of models containing new physics that will be discussed in 
this report.

\vspace{3mm}
\noindent{\bf $\bullet$ Additional Fermions}
\vspace{2mm}

New fermions are predicted to exist in many extensions of the SM.  In most
models they generally carry the usual baryon and lepton number assignments,
but can have unconventional electroweak quantum numbers.  They can be
classified according to their SU(2)$_L\times$U(1)$_Y$ assignments
in the following manner\cite{djtgr}: $(i)$ Sequential fermions.
The possibility of a fourth family of fermions has long been a popular 
extension to the SM.  LEP/SLC data restricts the fourth neutrino to be heavy, 
\ie, $m_{\nu_4}\gsim M_Z/2$, and the mixing between $\nu_4$ and $\nu_{e,\mu}$
to be small\cite{neutr}.  The recent LEP 1.5 run in November 1995 places
the preliminary 
constraint\cite{leptwo} $m_{\nu_4}>48.1-60.2$ GeV, depending on whether
the neutrino species is $e$ or $\mu$ and is Dirac or Majorana in nature.
It is worth noting that such a heavy fourth
neutrino could mediate a see-saw type mechanism\cite{hillpas} thus generating 
a small mass for $\nu_{e,\mu,\tau}$.  Constraints on the masses of the charged 
fourth generation fermions are\cite{pdg}: $m_{L_4}>45.1-46.4\gev$ from LEP I
(again, the LEP 1.5 run places constraints\cite{leptwo} up to 60 GeV on $L^\pm$,
with the exact limit depending on the mass of the associated neutral heavy 
lepton), $m_{b'}>85\gev$ from CDF assuming that it decays via charged current 
interactions, and $m_{t'}\gsim M_Z/2$ from LEP/SLC.  In principle one can
search for a $t'$-quark in the same manner as the top-search analyses at
the Tevatron, however, the results of such a search are not yet reported.
$(ii)$ Vector Fermions.  Numerous extensions of the SM contain fermions
whose left- and right-handed components transform identically under
SU(2)$_L$.  For example, $E_6$ grand unified theories contain\cite{jlhtgr} a 
vector-like color singlet weak iso-doublet as 
well as a vector, weak iso-singlet, $Q+-1/3$ color triplet.  Global
analyses of the bounds placed on these exotic fermions from flavor
changing neutral currents (FCNC) are performed in Ref. \cite{exotic}.
$(iii)$ Mirror Fermions.  The chiral properties of mirror fermions are
opposite to those of the ordinary fermions.  They appear\cite{mirrors} in some 
theories which restore left-right symmetry at the electroweak symmetry 
breaking scale, as well as in some grand unified and lattice gauge theories.  
Global constraints on their properties from precision electroweak measurements 
can be found in Csaki and Csikor\cite{mirrors}.  

\vspace{3mm}
\noindent{\bf $\bullet$ Extended Higgs Sector}
\vspace{2mm}

The possibility of an enlarged Higgs sector beyond the minimal one-doublet
version of the SM is consistent with data and has received substantial
attention in the literature\cite{hhg}.  In this report, we consider three
such classes of models.

The most economical case is that of Two-Higgs-Doublet Models (2HDM), which
contain 5 physical Higgs bosons,
2 scalars $h^0,\, H^0$, a pseudoscalar $A^0$,\, and 2 charged scalars
$H^\pm$.  Two such models
naturally avoid tree-level FCNC, and are denoted
as Model I, where one doublet ($\phi_2$) generates masses for all 
fermions and the second ($\phi_1$) decouples from the fermion sector, and 
Model II, where $\phi_2$ gives mass to the up-type quarks, while the down-type
quarks and charged leptons receive their mass from $\phi_1$.  Each doublet
receives a vacuum expectation value (vev) $v_i$, subject to the constraint that
$v_1^2+v_2^2=v^2_{\rm SM}$.  Here, we will mostly be concerned with the
$H^\pm$ interactions with the fermion sector, which are governed by the 
Lagrangian
\begin{eqnarray}
{\cal L} & = & {g\over 2\sqrt 2 M_W}H^\pm[V_{ij}m_{u_i}A_u\bar 
u_i(1-\gamma_5)d_j+V_{ij}m_{d_j}A_d\bar u_i(1+\gamma_5)d_j \nonumber \\
& & \quad\quad\quad\quad m_\ell A_\ell\bar\nu_\ell(1+\gamma_5)\ell]+h.c. \,,
\end{eqnarray}
with $A_u=\cot\beta$ in both models and $A_d=A_\ell=-\cot\beta(\tan\beta)$ in 
Model I(II), where $\tan\beta\equiv v_2/v_1$.  A review of the constraints
placed on such models from a variety of rare processes can be found in
Ref. \cite{bhp}.

Models with Three (or more) Higgs Doublets (3HDM) contain new CP violating 
phases, which can appear in charged scalar exchange.  These models can also 
avoid tree-level FCNC by imposing discrete symmetries or by requiring that only
one doublet couples to each quark sector.  In the latter case, the
interaction Lagrangian between the quark sector and the two physical
charged Higgs bosons can be written as
\begin{equation}
{\cal L}={g\over 2\sqrt 2 M_W}\sum_{i=1,2}H_i^+\bar U[Y_iM_uV_{CKM}(1-\gamma_5)
+X_iM_dV_{CKM}(1+\gamma_5)]D=h.c.\,,
\end{equation}
where $X$ and $Y$ are complex coupling constants that arise from the
diagonalization of the charged scalar mixing matrix and obey the relation
$\sum_{i=1,2}X_iY_i^*=1$.  A general phenomenological analysis of this
model can be found in Ref. \cite{threehdm}

The Higgs sector may also be extended without natural flavor conservation.
In these models the above requirement of a global symmetry which restricts
each fermion type to receive mass from only one doublet is 
replaced\cite{fcnch} by approximate flavor symmetries which act on the
fermion sector.  The Yukawa couplings can then possess a structure which
reflects the observed fermion mass and mixing hierarchy.  This allows
the low-energy FCNC bounds to be evaded as the flavor changing couplings
to the light fermions are small.  Here, we will employ the Cheng-Sher 
ansatz\cite{fcnch}, where the flavor changing couplings of the neutral
Higgs to two fermions of different flavor are $\lambda_{h^0f_if_j}=
(\sqrt 2G_F)^{1/2}\sqrt{m_im_j}\Delta_{ij}$, with the $m_{i(j)}$ being
the relevant fermion masses and $\Delta_{ij}$ representing a combination
of mixing angles.  The exact form of $\Delta_{ij}$ is calculable within a 
specific model.
 
\vspace{3mm}
\noindent{\bf $\bullet$ Supersymmetry}
\vspace{2mm}

Supersymmetry (SUSY) relates the properties of bosons and fermions and a result
of this symmetry is that all particles have supersymmetric partners with
the same mass and gauge interactions, but with spin differing by $1/2$.
For the SM particle content this predicts the existence of 
squarks, sleptons, gauginos, gluinos, and higgsinos.  These sparticles have
not yet been experimentally detected, and hence supersymmetry must be
broken.  There are theoretical and experimental reasons\cite{susyreport}
(associated with, for example, the stability of the scale hierarchy in grand
unified theories (GUTS), and with the consistency of measurements of the gauge 
couplings with these GUTS) to believe that the SUSY is broken near the scale
of 1 TeV.  The minimal supersymmetric standard model (MSSM) is the simplest
version of SUSY; it contains the minimal number of new particles with
the Higgs spectrum of the 2HDM Model II discussed above, and leads to the
conservation of a multiplicative quantum number denoted as R-parity.
Ordinary particles have R-parity of $+1$, while sparticles possess negative
R-parity.  Hence in MSSM, only {\it pairs} of sparticles can be produced or
exchanged in loops.  In SUSY GUTS the sparticle mass and
mixing spectrum can be described by a smaller set of parameters
which relate the physical particles at the GUT scale.  Assuming unification
at a high-energy scale, we can take these parameters to be the common
soft-breaking gaugino mass $m_{1/2}$, the universal scalar mass $m_0$,
the supersymmetric higgsino mass parameter $\mu$, the universal
trilinear soft-breaking term in the superpotential $A$, as well as
$\tan\beta$ defined above.  A general analysis of SUSY can be found
in Ref. \cite{susyreport}.

In non-minimal SUSY models R-parity can be broken; this leads to a very
different SUSY phenomenology as sparticles can now be singly produced or
exchanged in loops.  These models still contain the minimal superfield
content, but break R-parity either spontaneously, by the sneutrino
acquiring a vev, or explicitly through terms contained in the superpotential.
In the latter case, these terms take the form
\begin{equation}
W=\lambda_{ijk}L_iL_jE^c_k+\lambda^{\prime}_{ijk}L_iQ_jD^c_k
+\lambda^{''}_{ijk}U^c_iD^c_jD^c_k \,,
\end{equation}
where $ijk$ are generation indices, the $\lambda$'s are {\it a priori} unknown 
Yukawa coupling constants, and $Q,L,U,D,E$ represent the chiral superfields.
In order to preserve proton stability, the lepton and baryon number 
violating terms cannot simultaneously exist.  Restrictions on the value of
the Yukawa constants (the $\lambda$'s) have been obtained\cite{rpar} from 
a large variety of low-energy processes.  The typical bounds are found to lie
in the range
\begin{equation}
\lambda^{(','')}_{ijk}\le (0.01 - 0.50) {m_{\tilde f}\over 100\gev} \,,
\end{equation}
where $m_{\tilde f}$ represents the appropriate sparticle mass.

\vspace{3mm}
\noindent{\bf $\bullet$ GUTS Models}
\vspace{2mm}

There are many classes of models with extended gauge sectors.  One of the
most popular cases is that of the Left-Right Symmetric Model 
(LRM)\cite{mohap} which is based on the enlarged gauge group
$SU(2)_L\times SU(2)_R\times U(1)$.  Such theories have been fashionable
for many years, as both a possible generalization of the SM and in the
context of grand unified theories such as SO(10) and $E_6$.  One prediction
of these models is the existence of a heavy, charged, right-handed
gauge boson $W_R^\pm$, which in principle mixes with the SM $W_L^\pm$
via a mixing angle $\phi$ to form mass eigenstates $W^\pm_{1,2}$.  This
mixing angle is constrained\cite{wrcon} by data in polarized $\mu$
decay (in the case of light right-handed neutrinos) and from
universality requirements to be $|\phi|\lsim 0.05$.  As we will see
below,  the virtual exchange of a $W^\pm_R$ can be felt in a
variety of processes.

The Alternate Left-Right Symmetric Model (ALRM)\cite{ema} originates from
$E_6$ GUTS and is also based on the low-energy gauge group
$SU(2)_L\times SU(2)_R\times U(1)$.  However, since a single generation in
$E_6$ theories contains 27 2-component fermions (in contrast to the
16 fermions per generation in SO(10)), quantum number ambiguities arise
which allow the $T_{3L(R)}$ assignments of the usual SM fermions to differ
from those of the LRM for $\nu_{L,R}, e_L,$ and $d_R$.  This allows, for
example, the right-handed $W$ boson to couple the $u^i_R$ to the exotic charged 
$-1/3$, vector singlet, color triplet fermion $h_R$, which is present in
the ${\bf 27}$ of $E_6$.  This possibility can lead to some striking
signatures\cite{ema}.

\vspace{3mm}
\noindent{\bf $\bullet$ Technicolor}
\vspace{2mm}

In technicolor theories the fundamental Higgs boson of the SM is replaced
by fermion condensates which break the electroweak symmetry via vevs of
the form $\langle 0|\Psi\bar\Psi|0\rangle~\alpha~\Lambda^3_{TC}$.  The new
fermions $\Psi$ are known as technifermions and interact via a new
technicolor force.  The confinement scale of this new force is $\Lambda\sim
250$ GeV.  In order to generate masses for the SM fermions, additional
extended technicolor (ETC) interactions, which couple the SM fermions to
the technifermions, are usually introduced.  This results in masses for the
ordinary fermions of order $g^2_{ETC}\Lambda^3_{TC}/M^2_{ETC}$.  For the
typical value of $\Lambda_{TC}$ given above, it is clear that rather light
ETC bosons, $M_{ETC}\sim\sqrt{\Lambda^3_{TC}/m_f}\lsim $ TeV, are required
to achieve adequate values for the fermion masses $m_f$.  However, this TeV
mass range for the ETC bosons leads to large contributions to FCNC and
electroweak radiative corrections and hence potentially conflicts with 
experiment.  These problems are not insurmountable, and more realistic
technicolor models which address these issues are discussed in the sections
below.

\vspace{3mm}
\noindent{\bf $\bullet$ Leptoquarks}
\vspace{2mm}

Leptoquarks are color triplet particles which couple to a lepton-quark
pair and are naturally present in many theories beyond the SM which
relate leptons and quarks at a more fundamental level.  They appear in
Technicolor theories, models with quark-lepton substructure, horizontal
symmetries, and grand unified theories based on the gauge groups SU(5),
SO(10), and $E_6$.  In all these scenarios leptoquarks carry both
baryon and lepton number, but their other quantum numbers, \ie, spin, weak
isospin, and electric charge, can vary\cite{djtgr}.  They couple to
fermions via a Yukawa interaction with {\it a priori} unknown strength.
This interaction is usually parameterized in terms of the fine structure
constant as $\lambda_{LQ}^2/4\pi=F_{LQ}\alpha$.  An investigation of the
global constraints from FCNC on leptoquarks may be found in Ref. \cite{sacha}.

\vspace{3mm}
\noindent{\bf $\bullet$ Anomalous Couplings}
\vspace{2mm}

Possible deviations from the SM form for the trilinear $WW\gamma$ and $WWZ$ 
vertex has received much attention\cite{anomvvv}.  These potentially anomalous
vertices can be probed by looking for deviations from the SM in tree-level
processes such as $e^+e^-\to W^+W^-$, or in loop order processes, for example 
the $g-2$ of the muon.  In the latter case, cutoffs must generally be used
in order to regulate the divergent loop integrals and can introduce
errors by attributing a physical significance to the cutoff.  We will see
below that in some instances the GIM mechanism may be invoked to cancel
such divergences yielding cut-off independent results.   The 
CP-conserving interaction Lagrangian for $WWV$ interactions can be written as
\begin{eqnarray}
{\cal L}_{WWV}& = & ig_{WWV}\left[ \left( W^\dagger_{\mu\nu}W^\mu 
V^\nu-W^\dagger_\mu V_\nu W^{\mu\nu}\right) +\kappa_V W^\dagger_\mu W_\nu 
V^{\mu\nu}+{\lambda_V\over\mws}W^\dagger_{\lambda\mu}W^\mu_\nu V^{\nu\lambda}
\right. \nonumber \\
& & \quad\quad\quad\quad \left. -ig_5^V\epsilon^{\mu\nu\lambda\rho}\left(
W^\dagger_\mu\partial_\lambda W_\nu-W_\nu\partial_\lambda W^\dagger_\mu\right)
V_\rho \right] \,,
\label{anomw}
\end{eqnarray}
where $V_{\mu\nu}=\partial_\mu V_\nu-\partial_\nu V_\mu$, $g_{WWV}=gc_w(e)$ for
$V_\mu=Z_\mu(A_\mu)$, and the parameters ($\Delta\kappa_V\equiv\kappa_V-1$)
take on the values $\Delta\kappa_V,\lambda_V,g^V_5=0$ in the SM.

Anomalous couplings between the fermions and the gauge boson
sector may also be probed in loop processes.  In the case of the fermionic
coupling to a neutral gauge boson, a general Lagrangian 
(assuming operators of dimension-five or less, only) can be written as
\begin{eqnarray}
{\cal L}& = & e\bar f_i\left[ Q_fv_\gamma\gamma_\mu +
{i\sigma_{\mu\nu}q^\nu\over m_{f_i}+m_{f_j}}(\kappa_\gamma-
i\tilde\kappa_\gamma\gamma_5)\right] f_jA^\mu \\
& & +{g\over 2c_w}\bar f_i\left[ \gamma_\mu(v_Z-a_Z\gamma_5) +
{i\sigma_{\mu\nu}q^\nu\over m_{f_i}+m_{f_j}}(\kappa_Z-i\tilde\kappa_Z\gamma_5)
\right] f_jZ^\mu \,, \nonumber
\label{anomf}
\end{eqnarray}
where $Q_f$ represents the fermion's electric charge, $v_{\gamma,Z}, a_Z$
represent the fermion's vector and axial-vector coupling (where gauge
invariance dictates that the photon be off-shell 
in the case of $i\neq j$),
$\kappa(\tilde\kappa)_{\gamma,Z}$ represent the anomalous magnetic (electric) 
dipole moment, and $q^\nu$ corresponds to the momentum of the
gauge boson.  We will discuss the bounds placed on these anomalous couplings
below.

\section{Precision Electroweak Measurements}

\subsection{Introduction}

Virtual effects from new physics beyond the SM
can manifest themselves in a number of ways:
They can contribute to rare processes that are forbidden or
highly suppressed within the SM 
(as discussed extensively in a separate section), 
or they can affect well measured and perturbatively
calculable electroweak observables
through radiative corrections and lead to detectable deviations
between the SM predictions and the experimentally measured values. 
The presence of such deviations, or the absence
of an expected one, can give us important, albeit indirect, information
on the nature and particle content of the yet to be discovered
sectors beyond the SM.

This section is organized as follows:
In the next subsection, we first review the current status of
precision electroweak measurements.
We then apply the methods developed in Refs.~\cite{PT,MBL,TGR}
to the data and express the 
possible size of oblique and non--oblique 
corrections from new physics in terms of limits to a 
few relatively model independent parameters.

\subsection{The Measurements}

The precision measurements of electroweak observables have heretofore
been instrumental in verifying the validity of the electroweak sector 
of the SM, namely the $SU(2)_L\times U(1)_Y$ gauge theory of 
electroweak interactions \cite{WEINSALAM}.
Today, with the SM firmly established and the experimental errors
improving incrementally each year, they are our best hopes of
seeing the SM {\it fail} at some level of precision and 
thereby establish the existence of new physics effects.
In fact, such an effect may have already been seen at LEP/SLC
where the ratios $R_b \equiv \Gamb/\Gamh$ and 
$R_c \equiv \Gamc/\Gamh$ have been measured to deviate from
their SM predictions by $3.8\sigma$ and $2.4\sigma$, respectively.

In Table~\ref{TABLE1}, we list the most recent precision
measurements from the $e^+e^-$ and
$p\bar{p}$ collider experiments \cite{lepslc,WMASS},
$\numu$ and $\numubar$ deep inelastic scattering experiments
\cite{NUNUCLEON,PDB},
$\numu e$ and $\numubar e$ elastic 
scattering experiments \cite{PDB,CHARM2}, 
and atomic parity violation (APV) experiments \cite{ROSNER},
together with the predictions of the observables in the SM
with $\mt = 180 \gev$ \cite{TMASS} and $\mh = 300 \gev$. 
The accuracy of each measurement is shown in percentages and the
disagreement between the theoretical and experimental
central values are shown in units of the experimental error.

The SM predictions for the $W$ mass and the LEP/SLC
observables were obtained using
the program ZFITTER 4.9 \cite{zfit}, 
and the predictions for the low energy $\numu N$, $\numu e$,
and APV observables were calculated from the formulae given in
Ref.~\cite{RHOKAPPA}.
The values of the effective QED coupling constant
and the $\MSbar$ QCD coupling constant at the $Z$ mass scale 
were chosen to be
$\alpha^{-1}(\mz) = 128.9 \pm 0.1$ \cite{UCLA},
and 
$\alpha_s(\mz) = 0.123\pm 0.006$ \cite{BETHKE} respectively.
The errors on the SM predictions of 
$\Gamz$, $\sig0$, and $R_\ell$ are from the
uncertainty on the value of $\alpha_s(\mz)$.
Additional errors due to the uncertainty in the values of
$\alpha^{-1}(\mz)$, quark masses, etc. are not shown.
 
Several comments are in order:

\begin{itemize}

\item{The value of $\Gaml$ is derived from the values of 
the four line--shape parameters
\beq
\mz,\qquad 
\Gamz,\qquad 
\sig0 = \frac{12\pi}{\mz^2}\frac{\Gaml\Gamh}{\Gamz^2},\qquad
R_\ell = \frac{\Gamh}{\Gaml},
\eeq
and their correlations.
It is not an independent measurement, but the SM prediction
of $\Gaml$ has the advantage of being free of the QCD uncertainties
which plague the predictions for $\Gamz$, $\sig0$, and $R_\ell$.}

\item{The forward backward asymmetries $\AFB^{0,f}$ are defined as
\beq
\AFB^{0,f} = \frac{3}{4}A_e A_f\,,
\eeq
with
\beq  
A_f = \frac{ 2g_{Vf}g_{Af} }{g_{Vf}^2 - g_{AF}^2}
    = \frac{g_{Lf}^2 - g_{Rf}^2}{g_{Lf}^2 + g_{Rf}^2}\,,
\eeq
where $g_{Vf} = g_{Lf} + g_{Rf}$ and $g_{Af} = g_{Lf} - g_{Rf}$
are the effective vector
and axial--vector couplings of fermion $f$ to the $Z$.
QCD corrections have been removed from $\AFB^{0,b}$ and $\AFB^{0,c}$. }

\item{The effective weak angle $\seff$ is defined as
\beq
\seff \equiv \frac{1}{4}( 1 - g_{V\ell}/g_{A\ell} )\,.
\eeq
The LEP value of $\seff$\ is 
the average of values extracted from the leptonic asymmetries
$\AFB^{0,\ell}$, $A_e$, and $A_\tau$ only.
The SLC value of $\seff$\ is from $A_{LR}$.}

\item{The parameters $\gl2$ and $\gr2$ measured in
$\numu$-- and $\numubar$--nucleon scattering experiments
\cite{NUNUCLEON} are defined as
\beqa
R_\nu       & = & \gl2 + \gr2 r\,,   \cr
R_{\bar\nu} & = & \gl2 + \frac{\gr2}{r}\,,
\eeqa
where $R_\nu$, $R_{\bar\nu}$, and $r$ denote the following
cross section ratios:
\beqa
R_\nu     
& = & \frac{\sigma(\nu_\mu N \rightarrow \nu_\mu X)}
           {\sigma(\nu_\mu N \rightarrow \mu^-   X)}\,, \nonumber \\
R_{\bar\nu} 
& = & \frac{\sigma(\bar\nu_\mu N \rightarrow \bar\nu_\mu X)}
           {\sigma(\bar\nu_\mu N \rightarrow \mu^+   X)}\,, \\ 
r 
& = & \frac{\sigma(\bar\nu_\mu N \rightarrow \mu^+ X)}
           {\sigma(    \nu_\mu N \rightarrow \mu^- X)}\,. \nonumber
\eeqa
It is customary for $\numu N$ scattering 
experiments to report a value of $s_W^2 \equiv 1 - \mw^2/\mz^2$
which is obtained from $g_L^2$ and $g_R^2$ through the
relations \cite{LSMITH}
\beqa
\gl2 & = & \rho_{\nu N}^2
            \left( \frac{1}{2} - s_{\nu N}^2
                 + \frac{5}{9} s_{\nu N}^4
            \right)\,, \nonumber \\
\gr2 & = & \rho_{\nu N}^2
            \left( \frac{5}{9} s_{\nu N}^4
            \right)\,,
\eeqa
with
\beq
s_{\nu N}^2 = \kappa_{\nu N}s_W^2\,,
\eeq
where $\rho_{\nu N}$ and $\kappa_{\nu N}$ represent
radiative corrections \cite{RHOKAPPA}. 
($\rho_{\nu N}=\kappa_{\nu N}=1$ at tree level.)
While this mode of presentation is  
convenient for comparing the result with the $p\bar{p}$
measurement of $\mw$ and checking the consistency of the SM,
the possible presence of new physics contributions in 
$\rho_{\nu N}$ and $\kappa_{\nu N}$ can potentially
affect the extracted value of $s_W^2$.
We have therefore converted the reported values of $s_W^2$ 
back into values of $\gl2$ and $\gr2$ since these are 
directly measured quantities and will be unaffected 
by the presence of new physics.
The values quoted in Table~\ref{TABLE1} are the
global averages from Ref.~\cite{PDB}.}

\item{
Atomic parity violation (APV) experiments have 
already measured the amplitude of parity violating 
transitions in atoms
to an accuracy of 1\% to 2\% for cesium \cite{CESIUM},
bismuth \cite{BISMUTH}, lead \cite{LEAD}, and thallium 
\cite{THALLIUM1,THALLIUM2}.
Comparison of these APV results with the SM requires 
the extraction of the weak charge
\beq
Q_W = \rho'_{eq}\left[ Z ( 1-4{s'_{eq}}^2 ) - N \right]\,,
\eeq
which quantifies the coupling
of the nucleus to the $Z$ boson \cite{BOUCHIAT}.
However, this requires accurate atomic physics calculations
which is currently available for only cesium and thallium
\cite{APVTHEORY}.
The quoted value of $\QWCS$ is from Ref.~\cite{CESIUM}
using the result of Ref.~\cite{APVTHEORY},
and that of $\QWTL$ is the average of the value given in 
Ref.~\cite{THALLIUM1} and the value inferred by 
Rosner \cite{ROSNER} from the result of Ref.~\cite{THALLIUM2}.}

\end{itemize}

A quick glance through Table~\ref{TABLE1} will
give the reader a good idea on just how accurate modern precision
measurements have become and how successful the SM is
in predicting these results.   Though the experimental
error on many observables is a mere fraction of a percent,
the only observables that deviate from the SM by more than
$1\sigma$ are the LEP values of $R_b$ and $R_c$, 
the SLC values of $\seff$ and $A_b$, and $\QWCS$ from APV.

In the following, we will attempt to understand what
these agreements and deviations imply about the types and
sizes of possible radiative corrections from new physics.

\begin{table}
\begin{center}
\caption{\small Determination of electroweak parameters as of summer 
1995 [24].  The SM predictions are calculated for $m_t = 180$ GeV,
$m_H = 300$ GeV with $\alpha^{-1}(\mz) = 128.9$,
$\alpha_s(\mz) = 0.123$.}
\label{TABLE1}
\begin{tabular}{|l|c|c|c|r|}
\hline\hline
Observable  & Measurement & Error & 
SM prediction & Deviation \\
&& (\%) && ($\sigma$) \\
\hline
\underline{LEP}   &&&&\\
line--shape:      &&&&\\
$\mz$ [GeV]          & $91.1884 \pm 0.0022$
                     & $0.0024$ 
                     & used as input & \\
$\Gamz$[GeV]         & $2.4963 \pm 0.0032$ 
                     & $0.13$ 
                     & $2.4973 \pm 0.0032$     
                     & $-0.3$                          \\
$\sig0$ [nb]         & $41.488 \pm 0.078$
                     & $0.19$
                     & $41.454 \pm 0.032$
% \pm 0.001 \pm 0.032 $
                     & $0.4$                           \\
$R_\ell$             & $20.788 \pm 0.032$
                     & $0.15$
                     & $20.767 \pm 0.040$
                     & $0.7$                           \\
$\Gaml$[MeV]         & $83.93 \pm 0.14$        
                     & $0.17$
                     & $83.98$ 
                     & $-0.4$                           \\
&&&&\\
lepton &&&&\\ 
asymmetries: &&&&\\
$\seff$                 & $0.23160 \pm 0.00049$ 
                        & $0.21$
                        & $0.23179$
                        & $-0.4$                         \\ 
&&&&\\
$b$ and $c$ quark &&&&\\
results: &&&&\\
$R_b$                & $0.2219 \pm 0.0017$
                     & $0.77$ 
                     & $0.2155$
                     &  $3.8$            \\
$R_c$                & $0.1543 \pm 0.0074$
                     & $4.8$
                     & $0.1724$
                     & $-2.4$            \\
$\AFB^{0,b}$         & $0.0999 \pm 0.0031$
                     & $3.1$  
                     & $0.1016$
                     & $-0.5$            \\
$\AFB^{0,c}$         & $0.0725 \pm 0.0058$
                     & $8.0$
                     & $0.0725$
                     & $0.0$            \\
\hline
\underline{SLC}   &&&&\\
$\seff$              & $0.23049 \pm 0.00050$
                     & $0.22$
                     & $0.23179$
                     & $-2.6$            \\
$R_b$                & $0.2171 \pm 0.0054$
                     & $2.5$
                     & $0.2155$
                     & $0.3$            \\
$A_b$                & $0.841  \pm 0.053$
                     & $6.3$
                     & $0.9345$
                     & $-1.8$            \\
$A_c$                & $0.606  \pm 0.090$
                     & $15$   
                     & $0.6671$
                     & $-0.7$            \\
\hline
\underline{$p\bar{p}$ colliders} &&&&\\
$\mw$ [GeV]          & $80.26 \pm 0.16$
                     & $0.20$
                     & $80.35$
% \pm 0.01 \pm 0.00$
                     & $-0.6$            \\
\hline
\underline{$\nu N$ scattering} &&&&\\
$\gl2$               & $0.3017  \pm 0.0033$  
                     & $1.1$
                     & $0.304$              
                     & $-0.4$            \\
$\gr2$               & $0.0326  \pm 0.0033$
                     & $10$
                     & $0.030$  
                     & $0.8$             \\
\hline
\underline{$\nu e$ scattering} &&&&\\
$\ganue$             & $-0.506  \pm 0.015$
                     & $3.0$     
                     & $-0.507$               
                     & $0.07$            \\
$\gvnue$             & $-0.039  \pm 0.017$
                     & $44$
                     & $-0.037$
                     & $-0.1$            \\
\hline
\underline{APV}&&&&\\
$\QWCS$              & $-71.0 \pm 1.8$
                     & $2.5$
                     & $-73.2$
                     & $1.2$             \\
$\QWTL$              & $-116.3 \pm 3.1$
                     & $2.7$
                     & $-116.8$
                     & $0.2$             \\
\hline\hline
\end{tabular}
\end{center}
\end{table}

\subsection{Oblique Corrections}

Before confronting the data given in Table~\ref{TABLE1} 
and trying to extract limits on the  
radiative corrections from new physics,
it is worthwhile to discuss what type of corrections to expect.

We begin by noticing that all the precision electroweak
measurements conducted so far involve interactions which
at tree level are described by
light fermions exchanging a single electroweak gauge boson
($\gamma$, $W$, or $Z$). 
Radiative corrections
to such four--fermion processes
come in three classes: vacuum polarization corrections,
vertex corrections, and box corrections.
The vacuum polarization corrections are often called
`oblique' corrections, as opposed to the `direct' 
vertex and box corrections because they only 
affect the propagation and mixings of the gauge bosons
and do not change the form of the interaction itself.
They are independent of 
the external fermions and 
affect all processes that involve the exchange of 
the electroweak gauge boson universally.
On the other hand, direct corrections do depend on the
external fermions and are specific to each process.

For new physics to contribute to
oblique corrections, they need only to carry 
$SU(2)_L\times U(1)_Y$ quantum numbers and they will
contribute at the same order in $\alpha$ as the 
usual SM corrections.   Furthermore, as we will see below,
oblique corrections from new physics do not necessarily
decouple when the new physics scale is taken to infinity.

However,
for new physics to contribute to the direct 
vertex and box corrections
at the 1--loop level, they must couple directly to the light
external fermions.  
Such couplings can be expected to be highly suppressed:
if the `light' fermions coupled strongly to new and {\it heavy}
physics, they would not be so light.
(Recall that this is true for the Higgs sector of the SM.
The Yukawa couplings of the Higgs boson to light fermions are
suppressed by a factor of $m_f^2/\mw^2$ compared to the
gauge couplings.) 
One exception is the $b$ quark: 
Because the $b$ quark is the isospin partner of the $t$ quark,    
the mechanism responsible for the generation of the large
$t$ quark mass can also be expected to lead to a large correction
to the $Zb\bar{b}$ vertex.
Even within the SM the $\Zbb$ vertex receives an important
correction from the $t$-$W$ loop.

Following these considerations, it seems reasonable to
make the following three assumptions about radiative
corrections from new physics.
\begin{enumerate}
\item{The electroweak gauge group is the standard 
      $SU(2)_L \times U(1)_Y$.  The only electroweak gauge
      bosons are the photon, the $W^\pm$, and the $Z$.}
\item{The couplings of new physics to light fermions
      are highly suppressed so that `direct' corrections from
      new physics can be neglected (with the possible exception
      of processes involving the $b$ quark).
      Only oblique corrections need to be considered.}
\item{The new physics scale is large compared to the
      $W$ and $Z$ masses.}
\end{enumerate}
The first and second assumptions taken together means that
we only need to consider new physics contributions to 
four vacuum polarization functions, namely, the self energies
of the photon, $W$, and $Z$, and the $Z$--photon mixing.
Using the notation
\beq
\int d^4x \;e^{iq\cdot x}\langle J_X^\mu(x) J_Y^\nu(0)
                          \rangle
= ig^{\mu\nu}\AQ{XY} + (q^\mu q^\nu {\rm term})\,,
\eeq
where $J_X$ is the current that
couples to gauge boson $X$ ($X = \gamma,W,Z$),
we can write these functions as
$\AQ{\gamma\gamma}$, $\AQ{WW}$,
$\AQ{ZZ}$, and $\AQ{Z\gamma}$.  
(Note that we do not need to consider the $q^\mu q^\nu$
parts of the vacuum polarization tensors because
they only correct the longitudinal parts of the
gauge boson propagators which are suppressed 
by a factor of $m_f^2/M_{W/Z}^2$ compared
to the transverse parts when contracted  
with the external fermion currents.)

In general, the new physics contributions to
$\AQ{\gamma\gamma}$, $\AQ{WW}$,
$\AQ{ZZ}$, and $\AQ{Z\gamma}$ are complicated functions
of $q^2$.
However, the third assumption allows us to expand 
the new physics contributions to these functions
around $q^2=0$ in powers of $q^2/\mnew^2$, where
$\mnew$ is the scale of new physics, and to keep only 
the first few terms.
We will only keep the constant and linear terms in
$q^2$ since higher order terms will decouple as
$\mnew\rightarrow\infty$:
\beqa
\AQ{\gamma\gamma} 
& = & q^2 \AP0{\gamma\gamma} + \cdots \cr
\AQ{Z\gamma} 
& = & q^2 \AP0{Z\gamma} + \cdots \cr
\AQ{ZZ} 
& = & \A0{ZZ} + q^2 \AP0{ZZ} + \cdots \cr
\AQ{WW} 
& = & \A0{WW} + q^2 \AP0{WW} + \cdots
\label{LINEAR}
\eeqa
It is to be understood that we are expanding only the
part of the $\AQ{XY}$'s that arise from new physics.
This linear approximation permits us to express the
new physics contributions in terms of just six parameters,
namely
$\AP0{\gamma\gamma}$, $\AP0{Z\gamma}$,
$\A0{ZZ}$, $\AP0{ZZ}$, $\A0{WW}$, and $\AP0{WW}$.
Of these six,
three will be absorbed into 
the renormalization of the three input parameters
$\alpha$, $G_\mu$, and $\mz$ rendering them
unobservable. 
This leaves three parameters which are observable
and finite and can be expressed as linear combinations
of the original six.
One popular choice for these linear combinations is
given by \cite{PT}:
\beqa
\alpha S & = & 4s^2 c^2
               \left[ \AP0{ZZ}
                          -\frac{c^2-s^2}{sc}\AP0{Z\gamma}
                          -\AP0{\gamma\gamma}
               \right]\,,  \nonumber \\
\alpha T & = & \frac{\A0{WW}}{\mw^2} - \frac{\A0{ZZ}}{\mz^2}\,, \\
\alpha U & = & 4s^2
               \left[ \AP0{WW} - c^2\AP0{ZZ}
                         - 2sc\AP0{Z\gamma} - s^2\AP0{\gamma\gamma} 
               \right]\,. \nonumber
%               \phantom{\frac{e^2}{s^2}}   
\eeqa
This definition enjoys the property that
the parameters $T$ and $U$ will be zero when the new physics
does not break custodial isospin symmetry.
In fact, $\alpha T$ is just the shift of the $\rho$ parameter
due to new physics:
\beq
\rho = 1 + \delta\rho_{\rm SM} + \alpha T\,. 
\eeq
$S$, $T$, and $U$ can parameterize
the oblique corrections from various extensions 
of the SM as long as they satisfy the three assumptions
listed above.  While this excludes models which extend the
electroweak gauge group beyond the standard
$SU(2)_L \times U(1)_Y$ and which introduce new electroweak
gauge bosons, they still encompass a large class of models, including 
new generations of fermions or modifications of the SM Higgs sector.

The dependence of various observables on the values of $S$, $T$, and $U$ 
is most easily calculated using the formalism developed by
Kennedy and Lynn \cite{KL}.  
Since this is a straightforward procedure \cite{PT}, 
we will only list the results:
\beqa
M_W    & = & 80.35 - 0.29 S + 0.45 T + 0.34 U \quad[\gev] \,, \nonumber \\
\Gaml  & = & 83.98 - 0.18 S + 0.78 T \quad[\mev]\,, \nonumber \\
\seff  & = & 0.23179 + 0.00362 S - 0.00258 T\,, \nonumber \\
\gl2   & = & 0.304 - 0.0027 S + 0.0067 T\,,   \nonumber \\
\gr2   & = & 0.030 + 0.0009 S - 0.0002 T\,,   \\
\ganue & = & -0.507 - 0.0037 T\,, \nonumber  \\
\gvnue & = & -0.037 + 0.0068 S - 0.0051 T\,,   \nonumber \\
\QWCS  & = & -73.2 - 0.74 S  - 0.007 T \,,     \nonumber \\
\QWTL  & = & -116.8 - 1.1 S  - 0.08 T\,. \nonumber 
\label{STUDEP}      
\eeqa
By fitting these expressions to the data of Table~\ref{TABLE1},
we can obtain the experimentally preferred values of $S$, $T$, 
and $U$.   We will only use the purely leptonic observables
from LEP and SLC since their SM predictions are 
free from QCD uncertainties and are unaffected by possible
direct corrections from new physics to the $\Zbb$ vertex.
Using the ten data points for
$\mw$, $\Gaml$, $\seff$(LEP), $\seff$(SLC), $\gl2$, $\gr2$,
$\ganue$, $\gvnue$, $\QWCS$, and $\QWTL$, we obtain
the following:
\beqa
S & = & -0.33 \pm 0.19\,, \nonumber \\
T & = & -0.17 \pm 0.21\,, \\
U & = & -0.34 \pm 0.50\,. \nonumber
\label{STUFIT}
\eeqa
The correlation matrix between the variables in the fit is given in 
Table \ref{tatsu1}, and the quality of the fit is $\chi^2 = 4.5/(10-3)$.
\begin{table}
\centering
\begin{tabular}{|c||c|c|c|}\hline\hline
    &  $S$  &  $T$  &  $U$  \\
\hline
$S$ &  1    &  0.86 & -0.15 \\
$T$ &  0.86 &  1    & -0.27 \\
$U$ & -0.15 & -0.27 &  1    \\
\hline\hline
\end{tabular}
\caption{\small Correlation matrix for the fit to data for the variables
$S$, $T$, and $U$ as described in the text.}
\label{tatsu1}
\end{table}
The tightest constraints come from $\Gaml$ and $\seff$.
Indeed, fitting $S$ and $T$ to only the $\Gaml$ and $\seff$ data
yields:
\beqa
S & = & -0.29 \pm 0.19\,, \nonumber \\
T & = & -0.13 \pm 0.22   \,.
\label{LEPFIT}
\eeqa
The central values of $S$ and $T$ are negative because both
the $\Gaml$ and $\seff$ data are smaller than their SM predictions.

To give an example of how these bounds on $S$, $T$, and $U$ can
provide important constraints on possible new physics sectors,
consider the introduction of a new heavy fermion doublet $(N,E)$
with the usual left--handed couplings to $SU(2)_L$ , hypercharge $Y$,
and masses $m_N, m_E \gg \mz$ . The contribution of this doublet
to $S$, $T$, and $U$ is given by 
\beqa   
S & = & \frac{1}{6\pi}\left[ 1 - Y \ln\left(\frac{m_N^2}{m_E^2}
                                      \right)
                      \right]\,, \\
T & = & \frac{1}{16\pi s^2 c^2 \mz^2}
        \left[ m_N^2 + m_E^2 - \frac{2 m_N^2 m_E^2}{m_N^2 - m_E^2}
                               \ln\left(\frac{m_N^2}{m_E^2}
                                  \right)
        \right]\,, \nonumber \\
U & = & \frac{1}{6\pi}
        \left[ -\frac{5 m_N^2 - 22 m_N^2 m_E^2 + 5 m_E^2}
                     { 3( m_N^2 - m_E^2 )^2 }
        \right.  \nonumber\\
  &   & \phantom{\frac{1}{6\pi}}
        \left.
               +\frac{m_N^6 - 3 m_N^4 m_E^2 - 3 m_N^2 m_E^4 + m_E^6}
                     { (m_N^2 - m_E^2)^3 }
                \ln\left( \frac{m_N^2}{m_E^2}
                   \right)
        \right]\,. \nonumber
\eeqa
The above expression for $T$ is the usual contribution of a
fermion doublet to the $\rho$ parameter and is positive semidefinite.
Due to the constraint on $T$, this contribution must be constrained
by keeping the mass splitting within the doublet to be small:  
\ie, $\Delta m \equiv | m_N - m_E | \ll m_N, m_E$.    
In this case, the above expressions simplify to
\beqa
S & \approx & \frac{1}{6\pi} \approx 0.05\,, \nonumber \\
T & \approx & \frac{1}{12\pi s^2 c^2}
              \left[\frac{(\Delta m)^2}{\mz^2}\right]\,, \\
U & \approx & \frac{2}{15\pi}
              \left[\frac{(\Delta m)^2}{m_N^2}\right]\,, \nonumber
\eeqa
and we see that we are now in conflict with the fitted value for $S$.
Since the addition of each extra fermion doublet will contribute
additively to $S$, $T$, and $U$, circumventing these limits
quickly becomes a serious problem for theories such as technicolor
where a large number of extra fermion doublets must be introduced.
Furthermore, 
in the case of technicolor, strong interaction effects have
been estimated to enhance the value of $S$ by roughly a factor
of 2 \cite{PT}.

Several suggestions have been made as to how
one may introduce new fermions without conflicting with
Eq.~(21). Ref.~\cite{MAJORANA} shows that Majorana fermions can
simultaneously give negative contributions to both $S$ and $T$.
Ref.~\cite{REVENGE} argues that
if one introduces a complete generation of technifermions,
keeping the techniquarks degenerate while splitting the
masses of the technileptons will have the desired effect of
making $S$ negative while keeping $T$ in check.
Ref.~\cite{MAEKAWA} discusses the case where vectorlike, and
mirror fermions are introduced. 

Many other interesting suggestions 
and models have been presented as to how one may
extend the SM while conforming with the bounds on the values of $S$ and $T$
\cite{OTHERS}.
This has become increasingly difficult over the years due to
the ever improving determination of $S$ and $T$.  Calculating the values of
$S$ and $T$ has now become a standard viability test
for possible extensions of the SM.

\subsection{More Oblique Corrections}

Constraints on the values of $S$, $T$ and $U$, of course, do not apply to
models that do not satisfy the three conditions discussed above.
If we relax the third condition and allow the new physics scale
to be near the weak scale, then more parameters must be
introduced to express oblique corrections from new physics since
the linear approximation of Eq.~(\ref{LINEAR}) no longer applies.

In Ref.~\cite{MBL}, it was shown that it is sufficient to
introduce three more parameters and increase the total number to six.
Following the definition of Ref.~\cite{MBL},
we slightly modify the previous definitions of $S$ and $U$ as follows 
and introduce three additional parameters $V$, $W$, and $X$.
(The definition for $T$ remains unchanged.)
\beqa
\alpha S
& = & 4s^2 c^2\left[ \BZ{ZZ}
                    -\frac{c^2-s^2}{sc}\B0{Z\gamma}
                    -\B0{\gamma\gamma}
              \right]\,, \nonumber\\
\alpha T 
& = & \frac{\A0{WW}}{\mw^2} - \frac{\A0{ZZ}}{\mz^2}\,, \nonumber \\
\alpha U
& = & 4s^2\left[ \BW{WW} - c^2\BZ{ZZ} 
               - 2sc\B0{Z\gamma} - s^2\B0{\gamma\gamma}
          \right]      
          \phantom{\frac{e^2}{s^2}}\,, \nonumber\\
\alpha V & = & \APZ{ZZ} - \BZ{ZZ} \phantom{\frac{e^2}{s^2}} \,, \nonumber\\
\alpha W & = & \APW{WW} - \BW{WW} \phantom{\frac{e^2}{s^2}} \,, \\
\alpha X & = & -sc\left[ \frac{\AZ{Z\gamma}}{\mz^2} - \B0{Z\gamma}
                        \right]         \phantom{\frac{e^2}{s^2}}\,. \nonumber
\label{STUVWX}
\eeqa
In the limit $\mnew \rightarrow \infty$, $S$ and $U$
coincide with their original definitions, and $V$, $W$, and $X$
vanish.

To place limits on $S$ through $X$, we must see how they alter the 
expressions for various observables.
It turns out that the only relations in Eq.~(20)
that need to be modified are:
\beqa
\Gaml  & = & 83.98 - 0.18 S + 0.78 T + 0.65 V - 0.38 X \quad[\mev] \,, \nonumber
\\
\seff  & = & 0.23179 + 0.00362 S - 0.00258 T + 0.00776 X \,.
\label{VWXDEP}
\eeqa
The new parameter $W$ only appears in the $W$--width so we will not
be considering it any further \cite{MBL,WWIDTH}.
Fitting these expressions to the same set of data as before,
we obtain
\beqa
S  & = & -1.0 \pm 1.5   \,, \nonumber\\
T  & = & -0.68 \pm 0.80   \,, \nonumber\\
U  & = & -0.21 \pm 0.92   \,, \\
V  & = &  0.56 \pm 0.83   \,, \nonumber\\
X  & = &  0.13 \pm 0.51 \,, \nonumber
\label{NEWFIT}  
\eeqa
with the correlation matrix given in Table \ref{tatsu2}.
The quality of the fit is $\chi^2 = 4.0/(10-5)$.
\begin{table}
\centering
\begin{tabular}{|c||c|c|c|c|c|}\hline\hline 
    &  $S$  &  $T$  &  $U$  &  $V$  &  $X$  \\
\hline
$S$ &  1    &  0.79 &  0.48 & -0.76 & -0.94 \\
$T$ &  0.79 &  1    & -0.05 & -0.97 & -0.55 \\
$U$ &  0.48 & -0.05 &  1    &  0.05 & -0.68 \\
$V$ & -0.76 & -0.97 &  0.05 &  1    &  0.54 \\
$X$ & -0.94 & -0.55 & -0.68 &  0.54 &  1    \\ 
\hline\hline
\end{tabular}
\caption{\small Correlation matrix for the fit to data for the six
variables $S, T, U, V, W$, and $X$.}
\label{tatsu2}
\end{table}

At first sight, the above results seem to indicate
that the restrictions on the values of $S$, $T$ and $U$ are considerably
relaxed.  However, a careful look at the correlation
matrix shows a large correlation between T and V,
and S and X.  So in order for $S$ and $T$ to deviate 
from their central values, 
they must be accompanied by corresponding 
shifts in $V$ and $X$.  
An easier way to see this is to redefine the parameters as
\beqa
S' & = & S + 4s^2c^2 V + 4(c^2 - s^2) X\,, \nonumber\\
T' & = & T + V\,.
\eeqa
Then, Eq.~(\ref{VWXDEP}) will be reduced to
\beqa
\Gaml  & = & 83.98 - 0.18 S' + 0.78 T'  \quad[\mev]\,, \nonumber\\
\seff  & = & 0.23179 + 0.00362 S' - 0.00258 T' \,,
\eeqa
and the same limits as those in Eq.~(\ref{LEPFIT})
will apply to $S'$ and $T'$.  Thus $Z$-pole observables alone can only
probe the combination of $S'$ and $T'$.
Therefore, to accommodate large deviations of $S$ and $T$
from their central values,
the model must also predict large values for $V$ and $X$.
While it is possible to construct exceptional cases where
$V$ and $X$ are as large as or larger than $S$ or $T$ \cite{LARGEVWX},
in most cases of interest
they are still much smaller than $S$ or $T$
due to the natural suppression factor $\mz^2/\mnew^2$ \cite{TECHNIVWX}. 

It is interesting to note that new physics may be quite close to the
weak scale and yet makes little contribution to the oblique parameters.
As an example, we consider the low-energy sector of the string-inspired
SUSY-$E_6$ model wherein the ordinary particle spectrum of the MSSM is
augmented by three generations of exotic fermions and their SUSY partners
as described in Section 2.
If, for simplicity, we ignore mixing between these states and the SM fields
(and we also take them to be degenerate), we find that the new contribution to
$T$ vanishes automatically due to the vector-like nature of the exotic
fields.  The corresponding contributions to the other oblique parameters
are shown in Fig. \ref{exotics} and are seen to be small ($\lsim 0.1$)
for $m\gsim 150$ GeV.  This example shows that new physics may be lurking
nearby without it showing up in the oblique corrections.

\vspace*{-0.5cm}
\nn
\begin{figure}[htbp]
%\centerline{
%\psfig{figure=exotics_stuwxz.ps,height=9cm,width=12cm,angle=-90}}
\vspace*{8cm}
\caption{\small Contribution of 3 generations of degenerate $E_6$ exotic
fermions of mass $M$ and their SUSY partners to the oblique parameters.
From top to bottom the curves correspond to the parameter $-V, -W, -S,
X,$ and $-U$, respectively.}
\label{exotics}
\end{figure}
\vspace*{0.4mm}

\subsection{Non--Oblique Corrections}

Let us now turn to the problem of constraining
non--oblique direct corrections.
There has been considerable attention to non--oblique
corrections due to the $3.8\sigma$ and $2.4\sigma$
discrepancy between the SM and LEP/SLC values of $R_b$ and $R_c$.
The dependence of these observables on oblique corrections
(which is already severly constrained by $S$ and $T$) 
is weak, so this discrepancy must be explained by 
extra non--oblique corrections from new physics.
Many models of new physics do predict a large correction
to the $\Zbb$ vertex \cite{rb}
and hence to $R_b$ but often in the direction of
making the discrepancy even larger.
It is therefore useful to introduce 
parameters which describe these
non--oblique corrections in a model independent fashion
and constrain then with the
data to facilitate the comparison between theory and
experiment.

Here we will use the formalism of Ref.~\cite{TGR}; we note that
a similar formalism was developed earlier in Ref.~\cite{ABC}.
We will limit ourselves to the situation where
the only relevant non--oblique correction from new physics
is in the $\Zbb$ vertex.  We express corrections from new physics 
to the effective left-- and right--handed
couplings of the $b$ quark to the $Z$ as
\beqa
\glb  =  [\glb]_{\rm SM} + \frac{1}{3}\,\ds + \dglb\,, \nonumber\\
\grb  =  [\grb]_{\rm SM} + \frac{1}{3}\,\ds + \dgrb\,,
\eeqa
where $\ds$ expresses the shift of the effective weak angle
due to oblique corrections, and $\dglb$ and $\dgrb$ express
non--oblique vertex corrections from new physics.
In terms of $S$ and $T$, $\ds$ is given by:
\beq
\ds = \frac{\alpha}{c^2-s^2}\left[ \frac{1}{4}S - s^2c^2 T
                            \right]\,.
\eeq
It is convenient in practice 
to define the following linear combinations of $\dglb$ and $\dgrb$:
\beqa
\xib   & \equiv & (\cos\phi_b)\dglb - (\sin\phi_b)\dgrb\,, \nonumber\\
\zetab & \equiv & (\sin\phi_b)\dglb + (\cos\phi_b)\dgrb\,,
\label{xizetadef}
\eeqa
where 
\beq
\phi_b \equiv \tan^{-1}|\grb/\glb| = 0.182\,,
\eeq
since $\Gamb$ depends only on the linear combination $\xib$,
while $A_b$ depends only on the linear combination $\zetab$.

The dependence of the relevant observables on $\ds$, $\xib$, and
$\zetab$ are can be calculated just as in the $S$, $T$, $U$ case
and we find
\beqa
\sig0      & = & 41.454 + 4.4 \ds + 17  \xib - 5.1\das 
                  \quad [{\rm nb}] \,, \nonumber\\
R_\ell     & = & 20.767 - 18  \ds - 21  \xib + 6.4\das\,, \nonumber\\
R_b        & = & 0.2155 + 0.04\ds - 0.78\xib \,, \nonumber\\
R_c        & = & 0.1724 - 0.06\ds + 0.18\xib \,, \\
\AFB^{0,b} & = & 0.1016 - 5.6 \ds - 0.18\zetab \,, \nonumber\\
A_b        & = & 0.9345 - 0.64\ds - 1.6 \zetab \,, \nonumber\\
\seff      & = & 0.23179 + \ds \,. \nonumber
\eeqa
For $\sig0$ and $R_\ell$ we have also included a correction due to
the shift of the QCD coupling constant from its reference value:
$\alpha_s(\mz) = 0.123 + \das$.
We will not be considering $\Gamz$ in our analysis since it has
an extra dependence on oblique corrections coming from the $\rho$
parameter which is absent in the partial width ratios
$R_\ell$, $R_b$, and $R_c$, as well as $\sig0$, 

Fitting the above expressions to the LEP/SLC values of
$\sig0$, $R_\ell$, $R_b$(LEP), $R_b$(SLC), $R_c$
$\AFB^{0,b}$, $A_b$, $\seff$(LEP) and $\seff$(SLC) we find
\beqa
\ds    & = & -0.00084 \pm 0.00034 \,, \nonumber\\
\xib   & = & -0.006   \pm 0.002   \,, \nonumber\\
\zetab & = & \phantom{-}0.035   \pm 0.017   \,, \\
\das   & = & -0.023   \pm 0.008\,, \nonumber
\label{ZBBFIT}
\eeqa
with the correlation matrix in Table \ref{tatsu3}.
\begin{table}
\centering
\begin{tabular}{|c||c|c|c|c|}\hline\hline 
          & $\ds$ & $\xib$ & $\zetab$ &  $\das$  \\
\hline
$\ds$     &  1    &   0.02 &   -0.47 &  0.11  \\
$\xib$    &  0.02 &   1    &    0.06 &  0.87  \\
$\zetab$  & -0.47 &   0.06 &    1    &  0.02  \\
$\das$    &  0.11 &   0.87 &    0.02 &  1     \\
\hline\hline
\end{tabular}
\caption{\small Correlation matrix for the fit to the variables in 
Eq. (35).}
\label{tatsu3}
\end{table}
The quality of the fit is $\chi^2 = 9.8/(9-4)$
with $R_c$ alone contributing 6.0 to the overall $\chi^2$.
This means that $R_c$ still has a
$2.4\sigma$ discrepancy between theory and experiment
even with extra corrections to the $\Zbb$ vertex.
In order to explain this discrepancy, 
we must introduce extra non-oblique corrections to
the $Zc\bar{c}$ and perhaps other vertices as well.
(A most general fit including corrections to all
the vertices has been performed in 
Ref.~\cite{BURGESSETAL}.)

Assuming that the discrepancy in $R_c$ is just statistical
for the moment, we shall ignore it and 
convert the bounds on $\xib$ and $\zetab$ into constraints on 
the original $\dglb$ and $\dgrb$.   We find:
\beqa
\dglb & = & 0.0004 \pm 0.0037 \,, \nonumber\\
\dgrb & = & 0.036  \pm 0.017\,,
\eeqa
with a correlation of $0.84$.  Again, due to this large correlation,
care is needed when comparing these limits with theory.
For instance, many theories beyond the SM predict $\dgrb\approx 0$ (such
as Supersymmetry)
in which case the limit on $\dglb$ becomes
\beq
\dglb = -0.006 \pm 0.002 \qquad(\dgrb = 0)\,,
\eeq
and the disagreement between the SM and experiment is at the
$3\sigma$ level.

In order to explicitly display the discrepancy with the SM, Fig. 
\ref{delta_glgr} shows the result of fitting the values of $\delta g^b_{L,R}$
to the $Z\to b\bar b$ data set presented in Table \ref{TABLE1}.  For
simplicity, we now define 
\begin{eqnarray}
g^b_L & = & -{1\over 2} +{1\over 3}\sin^2\theta_w^b+\delta g^b_L \,,
\nonumber\\
g^b_R & = & {1\over 3}\sin^2\theta_w^b+\delta g^b_R \,,
\end{eqnarray}
where $\sin^2\theta^b_w$ is the b-quark effective weak mixing angle, and
use ZFITTER4.9 to calculate the SM predictions for different values of
$m_t$.  (Here, we take $m_H=300$ GeV, $\alpha_s(M_Z)=0.125$, and 
$\alpha^{-1}_{em}(M_Z)=128.896$, but our results are quite insensitive to
this particular choice of parameters.)  As can be seen from the figure, one
gains more in $\chi^2$ if we allow $\delta g^b_R\neq 0$.  We note in
passing that the SM point remains outside of the region selected by the data
even if we increase the C.L. to $99.9\%$. 
 
We now discuss the influence of some specific models on $R_b$.  A recent
summary of the effects of several classes of new physics on $R_b$ can be 
found in Ref.~\cite{rb}.

\vspace*{-0.5cm}
\nn
\begin{figure}[htbp]
%\centerline{
%\psfig{figure=delta_glgr.ps,height=9cm,width=12cm,angle=-90}}
\vspace*{8cm}
\caption{\small $95\%$ C.L. fit to the parameters $\delta g^b_{L,R}$
using the full LEP/SLC $Z\to b\bar b$ data set and ZFITTER4.9.  The dashed
(solid, dotted) curve corresponds to $m_t=170\, (180, 190)$ GeV.  The diamond
at the center is the SM prediction.  The three nearby diamonds are the 
$\chi^2$ minima for $m_t=170, 180, 190$ GeV (from top to bottom).  The
values of the other input parameters are as given in the text.}
\label{delta_glgr}
\end{figure}
\vspace*{0.4mm}

\subsubsection{$R_b$ and Supersymmetry}

Supersymmetric models with light super-partners could potentially 
account for the $R_b$ anomaly.  There are two well-separated 
regions~\cite{boulware91:2054}
of supersymmetric parameter space where large corrections to the
$Zb\bar b$ vertex are possible: low 
$\tan\beta\simeq 1$, and high $\tan\beta >50$.  This is illustrated
in Fig.~\ref{fig1:wells}.  The low $\tan\beta$
region makes use of the large top Yukawa entering in
the $\tilde t_R-\tilde H^+-b$ and $t-H^+-b$ interactions.  
The high $\tan\beta$ region makes
use of the large bottom Yukawa ($m_b/\cos\beta$) entering in the
$\tilde b-\tilde H^0-b$ and $b-b-A$ interactions.  In both cases, the
overall effect is to increase the theoretical prediction of $R_b$ over
the standard model value and bring theory more in line with experiment.
However, it has been argued in~\cite{wells95:372} that the high $\tan\beta$
region is not a viable solution to the $R_b$ anomaly because it
violates experimental constraints on 
$Z\to b\bar A$~\cite{djouadi91:175} 
and $b\to c\tau\nu_\tau$~\cite{btaunu}
decay data.  Fig.~\ref{fig2:wells} graphically demonstrates
the exclusion of high $\tan\beta$ models.
Only the low $\tan\beta$ region is allowed.

\vspace*{-0.5cm}
\nn
\begin{figure}[htbp]
%\centerline{
%\psfig{figure=fig1:wells.eps,height=8cm,width=12cm,angle=0}}
\vspace*{8cm}
\caption{\small The dependence of $R_b^{\hbox{\rm max}}$
on $\tan\beta$.  The maximum possible value for $R_b$ obtainable
as a function of $\tan\beta$ is plotted for $m_{\chi^\pm_1}=46\,\hbox{\rm GeV}$
(upper line) and $m_{\chi^\pm_1}=60\,\hbox{\rm GeV}$ (lower line). The upper
hatched region is the experimental $1\sigma$ range for $R_b$, while
the lower range represents the SM range consistent
with the $1\sigma$ bounds for $m_t$.  Only the very low and very
high values of $\tan\beta$ can significantly enhance the theoretical
prediction for $R_b$.}
\label{fig1:wells}
\end{figure}
\vspace*{0.4mm}

A successful low $\tan\beta$ supersymmetric model will have a light
Higgsino-like chargino, and a light top squark which is mainly the
super-partner of the right-handed top quark.  These requirements follow
directly from the need to maximize the coupling of the charginos
and stops to the bottom quarks.  The mass of the chargino and stop must
be below approximately 80 GeV (as shown in Fig.~\ref{fig3:wells})
to keep the loop integral from being suppressed.
Additional phenomenological implications 
arise from such models~\cite{wells95:372,ma95:338,sola95:339}.  For example,
the branching ratio of the top quark to top squark plus neutralinos
is at or above 0.4.  Although not currently excluded by Tevatron data,
this large branching fraction of top decays into particles other than
$b+W^+$ should be noticeable in the very near future.  

LEP2 should be able to find these light sparticles in pair production
roughly up to its kinematic limit.  In fact, preliminary results from
the recent LEP 1.5 run in November 1995 places\cite{leptwo} lower bounds on
the chargino mass of $\gsim 65$ GeV, if the chargino -- lightest neutralino 
mass splitting is greater than 10 GeV.  This null search clearly makes it 
more difficult for SUSY alone to accommodate the experimental value of 
$R_b$\cite{ellisrb}.  Perhaps larger SUSY GUT groups, such as SO(10) or
$E_6$, with a correspondingly larger particle content should be 
investigated.  

It is crucial to test that these models of supersymmetry which yield large
shifts in $R_b$ do not violate other constraints.  Most importantly,
one could imagine that light superpartners inducing large vertex
corrections could also substantially alter other precision data 
constraints, such as the $\rho$ parameter.  It has been shown by several
groups~\cite{kane95:350,chankowski95:304,garcia95:335,dabelstein95:251} 
that the global fit of electroweak observables in supersymmetry
not only allows models with higher $R_b$ but fits the compendium
of data better than that standard model as measured 
by total $\chi^2$.
	
As previously discussed, a discrepancy of $2.4\sigma$ between the experimental 
and SM predicted value of $R_c$ has also been reported.
Supersymmetry has no natural way to explain such a large deviation in 
$R_c$\cite{garcia95:335}, and so its ``true value'' is usually fixed to that 
of the SM value to maintain consistency in supersymmetry analyses. 
In addition to a possible statistical fluctuation in the measurement of $R_c$ 
and the large correlation of the measurement to $R_b$, another
compelling reason why the $R_c$ discrepancy might not be real 
is its correlation with $\alpha_s$.  Recall that additional physics which
brings the $R_b$ prediction closer to experiment also has the effect of
lowering the extracted $\alpha_s$ from the $Z$ line shape analysis --
a welcome development since the $Z$ line shape $\alpha_s$ in the standard
model is higher than what other methods seem to indicate.
If both $R_b$ and $R_c$ were to be accounted 
for by new physics, and their true values were fixed at the present measured 
central values, then the extracted $\alpha_s$ from the $Z$ line shape would 
actually go up, making it even more discordant with low energy 
$\alpha_s$ extractions. 

Another observable, the $b$-quark asymmetry 
$A_b=(g_L^2-g_R^2)/(g_L^2+g_R^2)$, is also demonstrating a $\sim 2\sigma$
deviation from the standard model prediction.  Given a supersymmetric
spectrum which produces a large shift in the $Zb\bar b$ couplings sufficient
to predict $R_b$ closer to the data, perhaps these light superpartners would
induce a more harmonious $A_b$ prediction as well.  A closer look shows this
not to be the case.  In fact, the supersymmetry prediction is indistinguishable
from the standard model case.  The reason is that shifts in $R_b$ are
most sensitive to corrections to the left-handed coupling of the $b$-quarks
to the $Z$, while the shifts in $A_b$ are most sensitive to corrections
to the right-handed couplings.  But supersymmetry can only significantly
change the left-handed coupling; therefore, consequential shifts of
$A_b$ in supersymmetry model are not expected.

A common framework employed in the study of supersymmetry phenomenology
imposes gauge coupling unification, common scalar masses at the high
scale (GUT scale or Planck scale), common gaugino masses at the high
scale, etc.~\cite{drees:dpf}.  This framework, sometimes called
a ``super-unified'' model, is motivated by
the apparent meeting of the gauge couplings at the high scale,
minimal supergravity boundary conditions from simple SUSY breaking paradigms, 
successful description of radiative electroweak symmetry breaking,
flavor changing neutral current constraints on the squark and slepton
masses, and more.  It turns out that the masses and mixings required
to yield a large result in $R_b$ are not exhibited in these super-unified
models, and a more general low energy Lagrangian framework must be
adopted~\cite{wells94:219,carena95:45}.  A natural connection between models 
yielding ``good $R_b$''
at the electroweak scale to more fundamental models at the high scale is
still unresolved, and perhaps will remain so until additional
observables weigh in.  

%\vspace*{-0.5cm}
\nn
\begin{figure}[htbp]
%\centerline{
%\psfig{figure=fig2:wells.eps,height=8cm,width=12cm,angle=0}}
\vspace*{7cm}
\caption{\small The high $\tan\beta$ exclusion plot as argued by 
Ref.~[55].  The $\delta R_b=0.003$ contour
is such that no supersymmetric solution below the contour
can provide $\delta R_b \geq 0.003$.  The region above the
$r=\tan\beta/m_{H^\pm}=0.52$ GeV$^{-1}$ contour is
excluded by $b\to c\tau\nu_\tau$ decay data.  The region to the
left of the vertical lines, which indicate contours of
$Z\to b\bar b A$ events, is perhaps already excluded by current data
as argued in the above reference.
Therefore, if one requires $\delta R_b >0.003$
then no region of the high $\tan\beta$
parameter space is 
simultaneously consistent with 
the $b\to c\tau\nu_\tau$ and $Z\to b\bar b A$
decay constraints.}
\label{fig2:wells}
\end{figure}
\vspace*{0.4mm}

\vspace*{-0.5cm}
\nn
\begin{figure}[htbp]
%\centerline{
%\psfig{figure=fig3:wells.eps,height=8cm,width=12cm,angle=0}}
\vspace*{6cm}
\caption{\small Contour of 
$\delta R_b=0.003$ in the $m_{\chi^\pm_1} - m_{\tilde t_1}$
plane with $m_t=170\,\hbox{\rm GeV}$ and $\tan\beta =1.1$.  Above
the contour no solution exists which yields $\delta R_b >0.003$.
Below the contour solutions do exist with $\delta R_b >0.003$ for
appropriate choices of parameters.  The numerical value of this
contour is enhanced (or diminished) by about
$(0.4/\sin\beta)^2(m_t/M_Z)^2$ for different choices of $m_t$ and
$\tan\beta$.}
\label{fig3:wells}
\end{figure}
\vspace*{0.4mm}

\subsubsection{$R_b$ and Extended Technicolor}

In extended\cite{ETC} technicolor\cite{tc} (ETC) models, the large mass of
the top quark generally arises from ETC dynamics at relatively low
energy scales.  Since the magnitude of the Cabibbo-Kobayashi-Maskawa (CKM)
matrix element $\vert V_{tb}\vert$ is nearly unity, $SU(2)_L$ gauge
invariance insures that ETC bosons coupling to the left-handed top
quark couple with equal strength to the left-handed bottom quark.   In
particular, the ETC dynamics which generate the top quark's mass also
couple to the left-handed bottom quark thereby affecting the $Zb\bar
b$ vertex.  This has been shown\cite{zbbone} to provide a powerful
experimental constraint on extended technicolor models -- particularly
on those models in which the ETC gauge group commutes with $SU(2)_L$.

Consider a model in which $m_t$ is generated by the exchange of  a
weak-singlet ETC gauge boson of mass $M_{ETC}$ coupling  with
strength $\gE$ to the current
\begin{equation}
{\xi} {\bar\psi^i}_L \gamma^\mu T_L^{ik}
+ {1\over\xi} {\bar t_R} \gamma^\mu U_R^k\ ,\ \ \ \ \ \ {\rm with}\ \ 
\psi_L\ \equiv\ \pmatrix{t \cr b \cr}_L\ \ 
T_L\ \equiv\ \pmatrix{U \cr D \cr}_L \,,
\label{tmasscur}
\end{equation}
where $U$ and $D$ are technifermions, $i$ and $k$ are weak and technicolor
indices, and $\xi$ is an ETC Clebsch expected
to be of order one.  At energies below $\ME$, ETC  gauge boson exchange
may be approximated by local four-fermion operators.   For example, $m_t$
arises from an operator coupling the  left- and right-handed currents
in Eq. (\ref{tmasscur}) 
\beq
   - {\gE^2 \over  \ME^2}  \left({\bar\psi}_L^i \gamma^\mu
T_L^{iw}\right) \left( {\bar U^w}_R \gamma_\mu t_R \right) + \hc\ \,.
\label{topff}
\eeq
Assuming, for simplicity, that there is only a doublet of
technifermions and that technicolor respects an $SU(2)_L \times
SU(2)_R$ chiral symmetry (so that the technipion decay constant, $F$,
is $v= 246$ GeV), the rules of naive dimensional analysis\cite{dimanal}
give an estimate of
\beq
   m_t\ = {\gE^2 \over \ME^2}
   \langle{\bar U}U\rangle\ \approx\ {\gE^2 \over \ME^2} (4\pi v^3)
\label{topmass}
\eeq
for the top quark mass when the technifermions' chiral
symmetries break.

The ETC boson responsible for producing $m_t$ also affects the $Zb\bar
b$ vertex\cite{zbbone} when exchanged between the two left-handed
fermion currents of Eq. (\ref{tmasscur}).
This diagram alters 
the $Z$-boson's tree-level coupling to left-handed bottom quarks by
\beq
\delta g_L^{ETC}\ = \ -{\xi^2 \over 2} {\gE^2 v^2\over\ME^2} \esc(I_3)
\ =\ {1\over 4} {\xi^2} {m_t\over{4\pi v}} \esc \,,
 \label{tb}
\eeq
where the right-most expression follows from applying Eq. (\ref{topmass}),
and $\theta$ is the weak mixing angle.
The ETC-induced shift in $R_b$ is then \cite{zbbone}
\beq
{\delta R_b \over R_b} \approx -5.1\% \xi^2 \left({m_t
\over 175{\rm GeV}} \right) \,,
\label{rbb}
\eeq
which is large enough to be detected by the LEP experiments.  
Since the experimental value of $R_b$
actually lies {\it above} the  SM prediction, any 
contribution from non-standard physics is positive: \ie, $\left[\delta R_b/
R_b\right]_{new} \approx + 3\%$, thereby excluding ETC models in
which the ETC and weak gauge groups commute.

%\vspace*{-0.5cm}
%\nn
%\begin{figure}[h]
%\centerline{
%\psfig{figure=fig_ehs.eps,height=5cm,width=5cm,angle=0}}
%%\vspace*{-1cm}
%\caption{\small Direct correction to the $Zb\bar b$ vertex
%from exchange of the ETC gauge boson that gives rise to the top quark
%mass.}
%\label{figehs:1}
%\end{figure}
%\vspace*{0.4mm}

\bigskip
It is also interesting to check how more realistic ETC models fare.
The following summary indicates how the $Zb\bar b$ vertex can
guide model builders.

$\bullet\ $ A slowly-running (`walking') technicolor beta-function is
included in ETC models to provide the light fermions with large enough
masses, while avoiding excessive flavor-changing neutral
currents\cite{walktc}.  Because a walking beta function enhances the
technifermion condensate $\langle \bar T T \rangle$, it leads to
larger fermion masses for a given ETC scale, $M_{ETC}$.  Enhancing
$m_t$ relative to $M_{ETC}$ does reduce the size of $\delta g_L$ --
but, generally, not enough to render the shift in $R_b$ invisible to
LEP \cite{walkrb}.

$\bullet\ $ In some ETC models, the ETC coupling itself becomes strong
before the scale $M_{ETC}$ and plays a significant role in electroweak
symmetry breaking \cite{strongETC}.  The spectra of strongly-coupled
ETC models include light composite scalars with Yukawa couplings to
ordinary fermions and technifermions \cite{comp} .  Exchange of the
composite scalars produces corrections to $R_b$ that are allowed by
experiment \cite{evans}.  The disadvantage of this approach is the
need to fine-tune the ETC coupling close to the critical value.

$\bullet\ $ ETC models also generally include `diagonal' techni-neutral ETC
bosons.  The effect of these gauge bosons on $R_b$ is discussed at
length in Ref. \cite{yoshi}.  Suffice it to say that while exchange of
the diagonal ETC bosons does tend to raise $R_b$, this effect is
significant only when the model includes large isospin violation --
leading to conflict with the measured value of the oblique parameter
$T$.

$\bullet\ $ Finally, we should recall that our analysis explicitly assumed that
the weak and ETC gauge groups commute.  More recent work
\cite{NCETCtwo} indicates that models in which ETC gauge bosons carry
weak charge {\it can} give experimentally allowed values of $R_b$ due
to cancellation between two contributions to the $Zb\bar b$ vertex
(from exchange of the ETC boson that generates $m_t$ and from $Z-Z'$
mixing).  Furthermore, these non-commuting ETC models can actually fit
the full set of precision electroweak data better than the standard
model!

\subsubsection{$R_b$ and Anomalous Couplings}

The possibility 
that the conventional SM $Zb\bar b$ vertex is modified due to the presence 
of weak electric ($\tilde\kappa_b$) and/or magnetic ($\kappa_b$) anomalous 
moment type couplings{\cite {anomc,frey}}  has been considered as a possible 
explanation{\cite {tom}} of the high measured value of $R_b$. 
Specifically, the $Zb\bar b$ interaction now takes the form  
\begin{equation}
{\cal L}={g\over {2c_w}}\bar b\left[\gamma_{\mu}
(v_b-a_b\gamma_5)+{i\over {2m_b}}
\sigma_{\mu\nu}q^{\nu}(\kappa_b-i\tilde\kappa_b\gamma_5)
\right]fZ^{\mu} \,,
\end{equation}
where $g$ is the standard weak coupling constant, $c_w=cos \theta_W$, 
$m_b$ is the $b$-quark  mass, $v_b(a_b)$ are the SM couplings and 
$q$ is the $Z$'s four-momentum. Note that the weak electric dipole moment 
coupling is $CP$-violating. 
If $\tilde\kappa_b$ and/or $\kappa_b$ were non-zero, 
$R_b$ and $A_{FB}^b$(the forward-backward asymmetry), as well as 
$A_{pol}^b$(the polarized forward-backward asymmetry), which can only 
be measured by SLD, would be found to differ from the expectations of the SM.
A description of all these observables and their 
dependencies on $\tilde\kappa$ and $\kappa$ is given in detail in 
Ref.{\cite {tom}}.  Using the
data presented at the EPS95 and Beijing summer conferences{\cite {lepslc}}, 
Fig. \ref{tgr1} displays the effects of these anomalous couplings. 
In particular, it shows the 
scaled ratios $R_b/R_b^{SM}$ and $A_b/A_b^{SM}$, where 
the latter quantity is the weighted combination of $A_{FB}^b/A_{FB}^b(SM)$ 
and $A_{pol}^b/A_{pol}^b(SM)$. Note that $R_b(A_b)$ is increased(decreased) 
by the existence of anomalous couplings. 
In this analysis we take $\alpha_s(M_Z)=0.125$, $\alpha_{em}^{-1}(M_Z)=
128.896$, and the SM Higgs boson mass ($m_H$) was set 
to 300 GeV. A modified version of ZFITTER4.9{\cite {zfit}} was used to 
obtain the predictions of the SM for these observables. Note that the presence 
of the anomalous couplings push the SM in the general direction of the data!

Allowing $\tilde\kappa_b$ and $\kappa_b$ to be non-zero, a 
$\chi^2$ fit was performed to determine the $95\%$ CL region 
for these anomalous couplings as shown in Fig. \ref{tgr2}. 
The SM lies outside the 
boundary of the allowed region due to the $>3\sigma$ discrepancy in 
the value of $R_b$ and the somewhat low ($\simeq 1.8\sigma$) value 
of $A_b$ from SLD. The data clearly prefers non-zero anomalous couplings. 
Performing the corresponding analyses for charm quarks and $\tau$ leptons 
yields the shown results in Fig. \ref{tgr3}. 
In the charm-quark case, $R_c$ is below the SM 
prediction while anomalous couplings can only produce an increase in $R_c$ 
thus leading to very strong constraints. The $\tau$ data on the otherhand 
is in complete agreement with the SM predictions which also produces tight 
bounds on the corresponding anomalous couplings.

\vspace*{-0.5cm}
\nn
\begin{figure}[htbp]
%\centerline{
%\psfig{figure=tgr1.ps,height=9cm,width=12cm,angle=-90}}
\vspace*{8cm}
\caption{\small $R_b$ vs. $A_b$ compared with the predictions of the SM 
for $m_t=170,~180,~190$ GeV, corresponding to the dotted, solid, dashed 
data point, respectively. The upper(lower) solid curve is the prediction 
for non-zero negative(positive) values of $\kappa_b$ with the points in 
steps of 0.01. The dashed line represents the corresponding case of non-zero 
$\tilde\kappa_b$.}
\label{tgr1}
\end{figure}
\vspace*{0.4mm}
\vspace*{-0.5cm}
\nn
\begin{figure}[htbp]
%\centerline{
%\psfig{figure=tgr2.ps,height=9cm,width=12cm,angle=-90}}
\vspace*{8cm}
\caption{\small Regions in the $\kappa_b$-$\tilde\kappa_b$ plane allowed 
at the $95\%$ CL by the Beijing and EPS95 data
for $m_t=170,~180,~190$ GeV, corresponding to the inside of the 
dotted, solid, and dashed curves, respectively. The diamonds mark the 
corresponding $\chi^2$ minima from left to right.}
\label{tgr2}
\end{figure}
\vspace*{0.4mm}
\vspace*{-0.5cm}
\nn
\begin{figure}[htbp]
%\centerline{
%\psfig{figure=tgrcharm.ps,height=9.1cm,width=9.1cm,angle=-90}
%\hspace*{-5mm}
%\psfig{figure=tgrtau.ps,height=9.1cm,width=9.1cm,angle=-90}}
\vspace*{8cm}
\caption{\small Allowed values of the anomalous couplings for charm and 
$\tau$ for the same top masses as in Fig.2. }
\label{tgr3}
\end{figure}
\vspace*{0.4mm}
\vspace*{-0.5cm}
\nn
\begin{figure}[htbp]
%\centerline{
%\psfig{figure=taupol_shift.ps,height=9.0cm,width=12.0cm,angle=-90}}
\vspace*{8cm}
\caption{\small Shift in the value of $x_w^{eff}=\sin^2\theta_{eff}^{leptons}$ 
extracted from the LEP determination of the $\tau$
polarization, $P_\tau$, due to non-zero values of $\kappa^Z_\tau$ (solid)
or $\tilde\kappa^Z_\tau$ (dotted).}
\label{tgr4}
\end{figure}
\vspace*{0.4mm}

Anomalous couplings could lead to interesting results elsewhere.
Although we obtain reasonably strong constraints on the anomalous couplings
 of the $\tau$ from precision measurements, it is interesting to contemplate
how such couplings might influence the values extracted for $x_w^{eff}=\sin^2
\theta_{eff}^{leptons}$ from various observables involving $\tau$'s.  As
an example, we consider the case of the $\tau$ polarization asymmetry, 
$P_\tau$.  Fig. \ref{tgr4} shows how non-zero values of either 
$\kappa^Z_\tau$ or $\tilde\kappa^Z_\tau$ can lead to an apparent shift in
the value of $x_w^{eff}$ extracted from $P_\tau$.  If $\kappa^Z_\tau=0.002$
then $\delta x_w^{eff}=0.001$, which means that the {\it true} value of
$x_w^{eff}$ is 0.001 lower than what would be extracted by naively using
the SM formulae.  It is interesting to note that the experimental value of
$x_w^{eff}$ extracted from $P_\tau$ is somewhat higher than that given by
either the leptonic forward-backward asymmetries or $A_{LR}$.

\subsection{Extra Gauge Bosons}

When the electroweak gauge group is extended beyond the
standard $SU(2)_L \times U(1)_Y$, and new electroweak
gauge bosons such as extra $Z'$s and/or right--handed $W$'s
are introduced, then one must consider corrections that are
not encompassed in the usual oblique and non--oblique 
correction framework.  These corrections are due to the direct exchange of
the new bosons between the external fermions,
and due to mixing among the new and 
ordinary gauge bosons,
which will affect their masses and couplings.
These corrections enter at both tree-level and at 1--loop.  
There are also additional loop corrections from the new gauge bosons 
as well as from extra fermions that must be introduced 
for anomaly cancellation purposes, 
and a more complicated Higgs sector necessary for
giving masses to all the gauge bosons except the photon, 
will all come into the picture.   This makes the analysis of
tree--level and 1--loop radiative corrections from
models with extra gauge bosons extremely complicated
and difficult to discuss in any simple model independent way\cite{jlhtgr}.   

However, a restricted $S - T$ type of analysis is 
possible\cite{altarelli,zprimestu} if we limit ourselves to only
leptonic observables at the $Z$ pole and $M_W$.  In this case
since there are only three observables under consideration, one is completely
free to parameterize their potential deviations from SM predictions in
terms of three variables which can be identified as $S, T$, and $U$.  
The following approximate relations for the shifted values of $S, T$, and $U$ 
due to $Z'$ exchange can be obtained (the exact relations are rather 
cumbersome and can be found in Ref.~\cite{altarelli}):
\begin{eqnarray}
\alpha\Delta S & \simeq & 2\phi[(1-2x_w)v'-(1+2x_w)a'] \,,\nonumber \\
\alpha\Delta T & \simeq & (\delta\rho-4a'\phi)\,,\\
\alpha\Delta U & \simeq & 4\phi(v'+3a')\,,\nonumber
\end{eqnarray}
where $v'$ and $a'$ are the charged lepton couplings to the $Z'$ 
(normalized as in the SM), $\phi$ is the $Z-Z'$ mixing angle, and $\delta\rho$
represents the shift in the effective $\rho$ parameter due to the $Z'$.  
$\delta\rho, \phi, v'$, and $a'$ are easily calculable within a specific 
extended gauge theory\cite{jlhtgr,zprimestu}.  It is possible 
to obtain consistency\cite{zprimestu}  with
the data in several models, including $E_6$ GUTS and the LRM, with $Z'$ masses
below 1 TeV.  It is important to remember here that all $Z$-pole and $M_W$ 
effects are due to $Z-Z'$ mixing.  Hence we also note here that general 
global electroweak analyses restrict\cite{zprimelep} the $Z-Z'$ mixing angle 
to be $|\phi|\lsim 0.01$, which can in turn provide model dependent bounds 
on the $Z'$ mass\cite{steve}.

The sensitivity of the weak charge $Q_W$, as determined in atomic parity
violation experiments, to the existence of a $Z'$ has been discussed by
several authors\cite{zprimeqw}.  It is found that (see also 
Ref.~\cite{zprimestu}) one of the best signatures of a $Z'$ arising from 
$E_6$ or the LRM is a small positive increase in the value for the weak
charge in Cesium, $\delta Q_W\simeq 0.2-0.3$, in comparison to the SM
prediction.  These effects can be important even if $Z-Z'$ mixing is absent, 
in contrast to the $Z$-pole data discussed above.  
Future experiments are expected to be sensitive to these effects.

\section{A Model-Independent Global Analysis}

A useful simplification occurs whenever all of the new particles which
arise in a model are heavy compared to the energies which are accessible in the
experiments of interest. In this case all of the model's predictions for
these experiments can be summarized by an effective Lagrangian, in which all of
the heavy particles are `integrated out'. The resulting effective interactions
amongst the light particles describe the virtual effects of all of the heavy
particles. Since the coefficients of higher-dimension interactions are 
suppressed by higher powers of the inverse of the heavy-particle masses in a 
computable way, this technique
of organizing calculations underlines the fact that only a comparatively few
combinations of the parameters of the model can contribute to low-energy
observables. 

These observations suggest a more model-independent way to explore the 
implications for new physics of current experiments. The approach is based on 
the observation that {\em all} models which share the same low-energy particle 
content are described by the same low-energy Lagrangian, differing only 
through the couplings they predict for each of the possible effective 
interactions. The predictions which are common to all such models may be 
obtained by working with the most general possible effective Lagrangian which 
is allowed by the low-energy particle content and symmetries, but using 
completely arbitrary couplings. The price to be paid for the model-independence 
of the resulting predictions is the loss of the predictive power which is 
possible when the effective Lagrangian is derived from a particular model.
The following discussion summarizes the constraints which may be obtained from
precision electroweak measurements by pursuing this type of model-independent
approach. 

\subsection{The Lowest-Dimension Effective Interactions}

We start with the most general possible effective Lagrangian which involves 
only the known light particles --- taken in what follows to be the 
SM complement, excluding the top quark and Higgs boson --- and 
which respects electromagnetic gauge invariance. For the present purposes, the 
effective couplings of the QCD gluons may be ignored.  Working up to and 
including mass dimension five, one finds that the most
important terms can be cast into the following form \cite{bigfit}:

\bigskip\noindent
{\it 1. Electromagnetic Couplings:}
The electromagnetic couplings of fermions are
\begin{equation}
{\cal L}_{\rm em} = -e  \Bigl[ \overline{f}_i \gamma^\mu Q_i \, f_i \; A_\mu +
  \overline{f}_i \sigma^{\mu\nu} ( d^{ij}_L \gamma_L + d^{ij}_R \gamma_R) 
  \, f_j \;  F_{\mu\nu} \Bigr],  
\end{equation}
where the indices $i$ and $j$ are to be summed over all possible flavors of
light fermions, $f_i$. Here $\gamma_L$ and 
$\gamma_R$ denote the usual projection matrices
onto left- and right-handed spinors. The effective couplings, $d^{ij}_L$
and $d^{ij}_R$, represent linear combinations of nonstandard magnetic- and
electric-dipole moment interactions.

\bigskip\noindent
{\it 2. Charged-Current Interactions:}
The fermion charged-current interactions become
\begin{eqnarray}
{\cal L}_{\rm cc} &=&  -{e\over \sqrt{2}s_w} \; \Bigl[ \overline{f}_i 
\gamma^\mu 
    (h_L^{ij} \gamma_L + h_R^{ij} \gamma_R )\, f_j \; W^\dagger_\mu \nonumber\\
    && \qquad \qquad + \overline{f}_i \sigma^{\mu\nu}
    ( c_L^{ij} \gamma_L + c_R^{ij}
    \gamma_R) \, f_j \; W^\dagger_{\mu\nu} \Bigr] + {\rm h.c.}, 
\end{eqnarray}
where $W_{\mu\nu} = D_\mu W_\nu - D_\nu W_\mu$ is the $W$ field strength
using electromagnetic covariant derivatives, $D_\mu$. The coupling coefficients,
$h_L$ and $h_R$ are given by
\begin{eqnarray}
h_L^{ij} &=&  \delta \tilde{h}_L^{ij}  
+ \tilde{V}^{ij} \left(1 - { \alpha S \over 4
( c_w^2 - s_w^2)} + { c_w^2\; \alpha T \over 2 (c_w^2 - s_w^2)}  + {\alpha
U \over 8 s_w^2} - { c_w^2\; (\Delta_e + \Delta_\mu) \over 2 (c_w^2 -
s_w^2)} \right), \nonumber\\
h_R^{ij} &=& \delta \tilde{h}_R^{ij}~, 
\end{eqnarray}
where $\tilde{V}^{ij}$ represents the unitary CKM matrices for the left-handed 
charged current interactions of the quarks ($\tilde{V}^{ij} = 
\tilde{V}_q^{ij}$) and leptons ($\tilde{V}^{ij} = \tilde{V}_\ell^{ij}$), 
assuming massive neutrinos. The coefficients $\delta 
\tilde{h}_{L(R)}^{u_i d_j}$ and $c^{ij}_{L(R)}$ represent a set of 
arbitrary nonstandard fermion-$W$ couplings.  The parameters $S$, $T$ and $U$ 
are the usual `oblique' corrections to the gauge boson vacuum polarizations as 
discussed above.  Finally, $\Delta_f$ (with $f=e$, $\mu$ or $\tau$) 
denotes the quantity $\Delta_f \equiv \sqrt{\sum_i
\left| h_L^{\nu_i f}  \right|^2} \; - 1 $.

\bigskip\noindent
{\it 3. Neutral-Current Couplings:}
The general interactions between the light fermions and the $Z$ boson can be
written as 
\begin{equation}
{\cal L}_{\rm nc} = -{e\over s_w c_w} \; \Bigl[ \overline{f}_i \gamma^\mu (
   g_L^{ij} \gamma_L + g_R^{ij} \gamma_R ) \, f_j 
   \; Z_\mu + \overline{f}_i \sigma^{\mu\nu} ( n_L^{ij} \gamma_L + n_R^{ij} 
   \gamma_R) \, f_j  \; Z_{\mu\nu} \Bigr], 
\end{equation}
where
\begin{eqnarray}
 g_{L(R)}^{ij} &=& \delta \tilde{g}_{L(R)}^{ij} + \Bigl( g_{L,R}^{ij}
   \Bigr)_{SM} \; 
   \left[ 1 + {1 \over 2} ( \alpha T - \Delta_e  - \Delta_\mu ) 
   \right] \\
   && - Q_i \, \delta^{ij}
   \; \left({ \alpha S \over 4 ( c_w^2 - s_w^2)} - { c_w^2s_w^2 \; \alpha T 
    \over c_w^2 - s_w^2} + {c_w^2 s_w^2 (\Delta_e + \Delta_\mu) \over
   c_w^2 - s_w^2} \right) . \nonumber
\end{eqnarray}
The SM couplings are $ \Bigl( g_L^{ij} \Bigr)_{SM} = \Bigl( T_{3i} - Q_i
\, s_w^2 \Bigr) \delta^{ij}$ and $\Bigl( g_R^{ij} \Bigr)_{SM} = \Bigl( - 
Q_i \, s_w^2 \Bigr) \delta^{ij}$. $Q_i$ is the electric charge of fermion 
$f_i$, and $T_{3i}$ is its eigenvalue for the third component of weak isospin. 
$Z_{\mu\nu}$ is the abelian curl: $\partial_\mu Z_\nu - \partial_\nu Z_\mu$. 
The effective coupling matrices $\delta \tilde{g}_{L(R)}^{ij}$ and 
$n_{L(R)}^{ij}$ are arbitrary sets
of nonstandard couplings between the fermions and the $Z$ boson.  

\bigskip\noindent
{\it 4. The $W$ Mass:}
Besides the direct changes to the couplings between fermions and electroweak
bosons, the low-energy Lagrangian also produces a deviation from the SM
prediction for $M_W$ in terms of the three inputs, $M_Z$, $G_F$ and $\alpha$
\cite{bigfit}
\begin{equation}
M_W^2 = (M_W^2)_{SM} \left[ 1 - {\alpha S \over 2 ( c_w^2 - s_w^2)}  + 
   {c_w^2 \; \alpha T \over  c_w^2 - s_w^2}  + { \alpha U \over 4 s_w^2}
   - {s_w^2 (\Delta_e + \Delta_\mu) \over c_w^2 - s_w^2} \right] .
\end{equation}

\bigskip
We briefly note that an 
alternative, and widely used, parameterization for four of the parameters 
which appear in the above effective interactions is given by 
\cite{altarelli,AB}
\begin{equation}
\delta \epsilon_1 = \alpha T, \qquad  \delta\epsilon_2 = -  \; { 
\alpha U \over 4
s_w^2},  \qquad  \delta \epsilon_3 = {\alpha S \over 4 s_w^2}, \qquad 
\delta \epsilon_b  = -2 \;  \delta \tilde{g}_L^{bb}. 
\end{equation}
The $\delta$ here indicates the deviation of these parameters from their SM 
values, computed using reference values, $\hat{m_t}$ and $\hat{m_H}$, for 
the top-quark and Higgs-boson masses. Of these two 
parameters, it is $m_t$ which is most important to follow because of the 
relatively strong dependence on it of low-energy observables. This dependence 
can be computed by determining the contributions of virtual top quarks to the 
various effective couplings. The dominant
part of the result is proportional to $G_F m_t^2$, and is explicitly given by
\cite{PT,AB,MRKLang}:
\begin{eqnarray}
\delta T_{SM} &\simeq& {3 \over 16 \pi s_w^2 c_w^2 } \; \left( {m_t^2 - 
\hat m_t^2 \over M_Z^2} \right), \nonumber \\   
\Bigl( \delta \tilde{g}^{bb}_L \Bigr)_{SM} &\simeq& {\alpha \over 16 \pi s_w^2 
c_w^2}\; \left( {m_t^2 - \hat m_t^2 \over M_Z^2} \right)  .
\end{eqnarray} 
The subdominant dependence on $m_t$ and $m_H$ has also been calculated, and
explicit formulae may be found in the references. 

\subsection{Comparing to Experiment}

Given these effective interactions it is straightforward to compute their
implications for experiments. The results to linear order in the small
nonstandard effective couplings which describe the deviations from the Standard
Model are given for a large number of low- and high-energy precision electroweak
observables in Ref.~\cite{bigfit}. These expressions may then be fit to the 
data. 

Not all of the possible effective interactions need be included in such a fit. 
For instance, many interactions cannot interfere with Standard Model 
contributions, and so cannot contribute to any observables to linear order in 
the new physics couplings. Among the interactions which do not appear for this 
reason are most of the flavor-changing interactions. Many of these can be 
independently constrained using
experimental limits on flavor-changing processes, and Ref.~\cite{bigfit}
summarizes the resulting bounds. 

The results of a fit of the general Lagrangian to the data, taking $\hat m_t=
150$ GeV and $\hat m_H=300$ GeV,  are summarized in 
Tables \ref{cliff1}, \ref{cliff2}, and \ref{cliff3}.  
Two types of fits are presented in these tables. For the 
`Individual Fit' the parameter in question has been considered in isolation,  
with all of the other parameters set to zero by hand. This kind of fit is not  
realistic, but has often been considered in the literature. The `Global Fit', 
on the other hand, allows all of the parameters to be floated while fitting 
the data.  Perhaps surprisingly, the resulting bounds on the various 
parameters are nevertheless reasonably good, expressing the general success of 
the SM description. Since this success
has been somewhat undermined in the most recent LEP results \cite{lepslc} 
--- most
notably in the branching ratio $R_b = \Gamma(Z \to \overline{b} b)/\Gamma(Z \to
\mbox{hadrons})$ --- these fits are currently being updated to include this more
recent data. 

One of the applications of the bounds given in Tables \ref{cliff1}-\ref{cliff3}
is to constrain
the parameters of a specific model for nonstandard physics. The logic of such a
constraint goes as follows. One first computes the effective Lagrangian which is
obtained when all of the undiscovered (and assumed heavy) particles are 
integrated out. This results in a series of expressions for the effective 
couplings as functions of the couplings and masses of the underlying model. 
Next, the bounds from Tables \ref{cliff1} through \ref{cliff3} are used to 
constrain the parameters of the underlying model.  The resulting bounds are 
generally weaker than those that would have been obtained if
the model were fit directly to the experiments, since the fit whose results are
described in these Tables permits all of the effective couplings to vary
independently. Direct comparison of the results obtained in these two ways
\cite{bigfit}, shows that the bounds obtained often do not differ by much. 
This is typically because only a few experiments are responsible for the 
strongest experimental limits.

\begin{table}
\centering
\begin{tabular}{|c|c|c|}\hline\hline
Parameter & Individual Fit & Global Fit \\ \hline
$S$ & $-0.10 \pm 0.16$ & $-0.2 \pm 1.0$ \\
$T$ & $+0.01 \pm 0.17$ & $-0.02 \pm 0.89$ \\
$U$ & $-0.14 \pm 0.63$ & $+ 0.3 \pm 1.2$ \\ \hline\hline
\end{tabular}
\caption{\small Oblique Parameters.
Results for the oblique parameters $S$, $T$ and $U$ obtained from the
fit of the new-physics parameters to the data. The second column gives
the result for the (unrealistic) case where all other parameters are 
constrained to vanish. Column three gives the result of a global fit in which 
all of the parameters of the effective Lagrangian are varied. }
\label{cliff1}
\end{table}

\begin{table}
\centering
\begin{tabular}{|c|c|c|}\hline\hline
Parameter & Individual Fit & Global Fit \\ \hline
$\Delta_e$ & $-0.0008\pm .0010$  & $-0.0011 \pm .0041$ \\
$\Delta_\mu$ & $+0.00047 \pm .00056$  & $+0.0005\pm .0039$ \\
$\Delta_\tau$ & $-0.018 \pm 0.008$ & $-0.018\pm .009$ \\
$\Re(\delta\tilde{h}^{ud}_L)$ & $-0.00041\pm .00072$ & $+0.0001\pm
.0060$\\ 
$\Re(\delta\tilde{h}^{ud}_R)$ & $-0.00055\pm .00066$ & $+0.0003 \pm
.0073$\\ 
$\Im(\delta\tilde{h}^{ud}_R)$ & $0 \pm 0.0036$ & $-0.0036 \pm .0080$\\
$\Re(\delta\tilde{h}^{us}_L)$ & $-0.0018\pm .0032$ & --- \\ 
$\Re(\delta\tilde{h}^{us}_R)$ & $-0.00088\pm .00079$ & $+0.0007\pm
.0016$\\ 
$\Im(\delta\tilde{h}^{us}_R)$ & $0 \pm 0.0008$ & $-0.0004\pm .0016$\\ 
$\Re(\delta\tilde{h}^{ub}_L)$,$\Im(\delta\tilde{h}^{ub}_L)$ 
& $-0.09\pm .16$ & --- \\
$\Sigma_1$ & --- &$+0.005\pm .027$ \\
$\Re(\delta\tilde{h}^{ub}_R)$ & --- & --- \\
$\Re(\delta\tilde{h}^{cd}_L)$ & $+0.11 \pm .98$ & --- \\
$\Re(\delta\tilde{h}^{cd}_R)$ & --- & --- \\
$\Re(\delta\tilde{h}^{cs}_L)$ & $+0.022\pm .20$ & --- \\
$\Re(\delta\tilde{h}^{cs}_R)$ & $+0.022\pm .20$ & --- \\
$\Re(\delta\tilde{h}^{cb}_L)$ & $+0.5\pm 4.6$ & --- \\
$\Sigma_2$ & --- & $+0.11\pm 0.98$ \\
$\Re(\delta\tilde{h}^{cb}_R)$ & --- & ---  \\ \hline\hline
\end{tabular}
\caption{\small Charged-Current Parameters.
More results of the fits of the
new-physics parameters to the data. The quantities $\Sigma_1$ and 
$\Sigma_2$ arise in tests for the unitarity of the CKM matrix, 
and are defined as: $\Sigma_1 \equiv \Re( \delta \tilde{h}_L^{us}) +
\left[\Re(V_{ub}) \Re(\delta \tilde{h}_L^{ub}) + \Im(V_{ub}) \Im(\delta
\tilde{h}_L^{ub}) \right]/ \vert V_{us}\vert $ and $\Sigma_2 \equiv
\Re(\delta \tilde{h}_L^{cd}) + \vert V_{cs} \vert \Re(\delta \tilde{h}_L^{cs} +
\delta \tilde{h}_R^{cs})/\vert V_{cd} \vert + |V_{cb}| \Re(\delta
\tilde{h}_L^{cb})/\vert V_{cd} \vert $. Blanks indicate where the
corresponding fit would be inappropriate, such as for when a parameter
always appears in a particular combination with others, and so 
cannot be individually fit. }\label{cliff2}
\end{table}

\begin{table}
\centering
\begin{tabular}{|c|c|c|}\hline\hline
Parameter & Individual Fit & Global Fit \\ \hline
$\delta\tilde{g}^{dd}_L$ & $+0.0016\pm .0015$ & $+0.003\pm .012$ \\
$\delta\tilde{g}^{dd}_R$ & $+0.0037\pm .0038$ & $+0.007\pm .015$ \\
$\delta\tilde{g}^{uu}_L$ & $-0.0003\pm .0018$ & $-0.002\pm 0.014$ \\
$\delta\tilde{g}^{uu}_R$ & $+0.0032\pm .0032$ & $-0.003 \pm .010$ \\
$\delta\tilde{g}^{ss}_L$ & $-0.0009\pm .0017$ & $-0.003\pm .015$ \\
$\delta\tilde{g}^{ss}_R$ & $-0.0052\pm .00095$ & $+0.002\pm .085$ \\
$\delta\tilde{g}^{cc}_L$ & $-0.0011\pm .0021$ & $+0.001\pm .018$ \\
$\delta\tilde{g}^{cc}_R$ & $+0.0028\pm .0047$ & $+0.009\pm .029$ \\
$\delta\tilde{g}^{bb}_L$ & $-0.0005\pm .0016$ & $-0.0015\pm .0094$ \\
$\delta\tilde{g}^{bb}_R$ & $+0.0019\pm .0083$ & $0.013\pm .054$ \\
 $\delta\tilde{g}^{\nu_e \nu_e}_L$ & $-0.0048\pm .0052$ & --- \\
$\delta\tilde{g}^{\nu_\mu \nu_\mu}_L$ & $-0.0021\pm .0027$ & 
	$+0.0023 \pm .0097$ \\
$\delta\tilde{g}^{\nu_\tau \nu_\tau}_L$ & $-0.0048\pm .0052$ &--- \\
$\delta\tilde{g}^{\nu_e \nu_e}_L + \delta\tilde{g}^{\nu_\tau \nu_\tau}_L$ &
	--- & $-0.004 \pm .033$ \\ 
$\delta\tilde{g}^{ee}_L$ & $-0.00029\pm .00043$ & $-0.0001\pm .0032$ \\ 
$\delta\tilde{g}^{ee}_R$ & $-0.00014\pm .00050$ & $+0.0001\pm .0030$ \\ 
$\delta\tilde{g}^{\mu\mu}_L$ & $+0.0040\pm .0051$ & $+0.005\pm .032$ \\ 
$\delta\tilde{g}^{\mu\mu}_R$ & $-0.0003\pm .0047$ & $+0.001\pm .028$ \\ 
$\delta\tilde{g}^{\tau\tau}_L$ & $-0.0021\pm .0032$ & $\; 0.000\pm .022$
\\
$\delta\tilde{g}^{\tau\tau}_R$ & $-0.0034\pm .0028$ & 
	$-0.0015\pm .019$ \\ \hline\hline
\end{tabular}
\caption{\small Neutral-Current Parameters.
More results of the fits of the new-physics parameters to the data. 
As before, blanks indicate where the
corresponding fit would be inappropriate, such as for when a parameter
always appears in a particular combination with others, and so 
cannot be individually fit. }
\label{cliff3}
\end{table}

\section{$g-2$ of the Muon}

Magnetic moments of elementary particles receive radiative contributions which 
can in principle be sensitive to new degrees of freedom and interactions. 
The combination of larger mass and relatively long lifetime of the muon
allows measurements of its anomalous magnetic moment, $a_{\mu} \equiv {1 \over 
2} (g-2)$, which are sensitive to 
large energy scales and very high order radiative corrections.  The current 
experimental value of $a_{\mu}^{\rm exp} = 116~592~300(840) \times 10^{-11}$ 
\cite{muonold} is in good agreement with the theoretical calculation of 
$a_{\mu}^{\rm th} = 116~591~877(176) \times 10^{-11}$ \cite{muontheory}, where 
the numbers in parentheses are the uncertainties.  Agreement at this level 
includes QED  corrections to ${\cal O}(\alpha^5)$ and hadronic vacuum 
polarization to ${\cal O}(\alpha^3)$.  The Brookhaven E821 muon $g-2$ 
experiment is expected to reduce the experimental error in $a_{\mu}^{\rm exp}$
to below $\pm 40 \times 10^{-11}$ \cite{muonbrook}. 
At this level of precision, electroweak radiative corrections are important
and new physics at the weak scale can be probed. 

In order to exploit the experimental precision of the Brookhaven experiment
as a probe for new physics it is necessary to understand the SM contributions. 
These are usually given as $a_{\mu} = a_{\mu}^{\rm QED} + a_{\mu}^{\rm had} + 
a_{\mu}^{EW}$.  The QED contributions have been calculated to 
${\cal O}(\alpha^5)$, including the $\tau$ vacuum polarization 
contributions, and give 
$a_{\mu}^{\rm QED} = 116~584~708(5) \times 10^{-11}$ \cite{muonqed}.
The hadronic contributions are of two types.  The first corresponds to 
effects which represent the contribution of running $\alpha$ from low to high 
scales.  These can not be calculated from first principles, but can be related 
to  $R(s) = \sigma(e^+e^- \rightarrow hadrons)/\sigma(e^+e^- \rightarrow \mu^+ 
\mu^-)$ by means of a dispersion relation. 
A recent evaluation at ${\cal O}(\alpha^2)$ of the available data gives 
$a_{\mu}^{\rm had-vac-pol} = 7024(153) \times 10^{-11}$ \cite{muonhadron}.
The ${\cal O}(\alpha^3)$ corrections to the hadronic vacuum polarization have 
been calculated to be $a_{\mu}^{\rm had-vac-pol^{\prime}} = -90(5) \times 
10^{-11}$ \cite{muontheory}.  The other type of hadronic contributions are 
from light by light hadron amplitudes.  Unfortunately these can not be related 
to other experimental observables but they can be estimated within
some theoretical model.  Recent estimates in a $1/N_c$ expansion of a 
Nambu-Jona-Lasinio model give 
$a_{\mu}^{{\rm had}-\gamma-\gamma} = 8(9) \times 10^{-11}$ \cite{muonll}.
The electroweak contributions which arise at one-loop from integrating out the
$W$ and $Z$ bosons are $a_{\mu}^{\rm EW-1-loop} = 195 \times 10^{-11}$,
which is roughly 5 times the expected experimental precision of the Brookhaven 
experiment.  Because of the large number of diagrams, the two-loop 
electroweak contributions are not insignificant.  A complete calculation of the 
two-loop fermionic and partial bosonic contributions gives
$a_{\mu}^{\rm EW-2-loop} = -43(3) \times 10^{-11}$ \cite{muontloop}.
Of all the SM contributions, the hadronic vacuum polarization is by far the 
most uncertain.  In order to test the one-loop electroweak corrections,
the uncertainty in $R(s)$ at hadronic energies must be reduced by roughly a 
factor of four.  Ongoing experiments at VEPP-2M together with future experiments
at DA$\Phi$NE and BEPC will hopefully reach this level.  

Beyond the electroweak corrections, new degrees of freedom or interactions at 
the weak scale can in principle give important contributions to $a_{\mu}$.
The anomalous magnetic moment operator 
\beq
\label{muonop}
{\cal L} = - a_{\mu} {1 \over 4 m_{\mu}} 
  \bar{\psi} \sigma_{\nu \rho} \psi F^{\nu \rho}
\eeq
is chirality violating, and must therefore vanish with the muon mass.  This
operator is therefore effectively dimension-six, being suppressed by two
powers of the scale $M$ characterizing the new physics, $ a_{\mu}^{\rm new} 
\propto m_{\mu}^2 / M^2$.  The magnitude of $a_{\mu}^{\rm new}$ is sensitive 
to the specific form of the new physics. 
In strongly coupled theories such as composite or technicolor
models, contributions to $a_{\mu}^{\rm new}$ can arise which are
suppressed only by the scale $M$, \ie, $a_{\mu}^{\rm new} \sim e m_{\mu}^2/M^2$ 
\cite{muoncomp}.  In this case the expected precision of the Brookhaven 
experiment would be sensitive at the $2 \sigma$ level
to physics up to a scale $M \sim 2$ TeV. 
The electroweak loop corrections are also sensitive to possible composite
structures of the gauge bosons or gauge couplings.  For example, the Brookhaven 
experiment will be sensitive to the anomalous moment of the $W^{\pm}$ boson,
$e (\kappa -1) F^{\mu \nu} W_{\mu}^+ W_{\nu}^-$, at the $2 \sigma$ level of 
$| \kappa -1 | > .07$ for $M \sim$ 1 TeV \cite{muoncomp}.

For weakly coupled extensions of the SM the contributions 
to $a_{\mu}^{\rm new}$ arise from radiative corrections and are
suppressed by a perturbative loop factor of
$a_{\mu}^{\rm new} \sim e (\alpha/4 \pi) m_{\mu}^2 / M^2$.
As an example, supersymmetric extensions of the SM can give contributions that 
are generally the same order as the electroweak contributions \cite{muonsusy}.
Imposing universality and radiative electroweak symmetry
breaking, one finds that chargino-sneutrino loops typically
dominate the supersymmetric contribution \cite{muonnath}.  Large values of 
$\tan \beta$ enhance the coupling of the Higgsino component of the 
chargino to the muon, and therefore maximize these types of contributions. 
The sign of $a_{\mu}^{\rm SUSY}$ turns out to be correlated with that of 
the $\mu$ parameter in this region of parameter space. 
Assuming $a_{\mu}^{\rm exp} = a_{\mu}^{\rm SM} + a_{\mu}^{\rm SUSY}$,
and that the Brookhaven experiment would be consistent
with $a_{\mu}^{\rm SM}$ at the $2 \sigma$ level, it would then force
the universal scalar mass and gluino mass (assuming gaugino universality)
to be $\lsim 500-600$ GeV for $\tan \beta = 30$ \cite{muonnath}. 
The precise bound will depend on how well $a_{\mu}^{\rm had}$ can be
determined from $R(s)$. 

Models in which the muon mass has a sizeable (or sole) 
component arising from radiative corrections
can give interesting contributions to $a_{\mu}^{\rm new}$.  In these types 
of models the muon (and perhaps other light SM fermions) is protected at 
tree level from obtaining a mass by some approximate chiral symmetry. 
The chiral symmetry is not respected radiatively, and the 
muon obtains a mass for example at one-loop, $m_{\mu} \sim (\lambda^2 / 16 
\pi^2) m_{\chi}$, where $\lambda$ is a Yukawa coupling, and
$m_{\chi}$ is a parameter characterizing the breaking of the chiral symmetry 
(such as the mass of another fermion).   A contribution to $a_{\mu}^{\rm new}$ 
is generated at the same order as $m_{\mu}$, giving $a_{\mu}^{\rm new} \sim e 
m_{\mu}^2/M^2$.  Notice that this is {\it not} suppressed by a loop factor
even though it arises perturbatively.   This is because {\it both} the mass 
and anomalous magnetic moment arise perturbatively. 
If the muon mass is (largely) generated radiatively at or just above the
weak scale, ``sizeable'' deviations of $a_{\mu}$ from the SM model prediction 
can therefore result.  As an example, in supersymmetric theories with flavor 
violation in the slepton sector, the muon can receive a radiative
contribution from the tau, $\delta m_{\mu} \sim (\sin \theta_{\mu \tau})^2 
(\alpha/4 \pi)  m_{\tau}$ where $\sin \theta_{\mu \tau}$ is a slepton mixing 
angle.  The supersymmetric contributions to $a_{\mu}$ could then be a factor 
$\sim (\sin \theta_{\mu \tau})^2 (m_{\tau}/m_{\mu})$
larger than the supersymmetric contributions mentioned above. 

\section{Rare Processes in the Quark Sector}

We next investigate the indirect effects of new physics in processes
which are rare or forbidden in the SM.  In this Section we turn our attention 
to the quark sector, examining each quark flavor separately. 

\subsection{Kaons}

Numerous processes involving Kaons 
occur through CP violation  or flavor-changing neutral currents.
As these two effects are small in the SM,
Kaons provide a fruitful testing ground for virtual effects
of new physics.   For this reason, rare Kaon processes have played
a strong and historical role in constraining new interactions.  For
example, the strongest bound (albeit assumption dependent) on the mass of 
a right-handed $W$ boson in the LRM is derived from its contribution
to $K^0-\bar K^0$ mixing\cite{sonibb}, the requirement of near
degeneracy of squark masses results from a super-GIM mechanism imposed in
the $K$ sector\cite{ellis}, and $K^0-\bar K^0$ mixing and $K_L\to\mu e$ have
provided severe constraints on technicolor model building\cite{lrand}
forcing the introduction of a Techni-GIM mechanism.
The observation of FCNC in the mass difference of neutral Kaons
as well as  the first observation of indirect CP 
violation (parameter $\epsilon_K$) provide tight constraints on several other 
models of new physics.  Once these are taken into account and combined
with the constraints from $B-\bar{B}$ mixing (and $b\rightarrow s\gamma$)
the range of  predictions for  K decays are severely restricted.
Nevertheless several Kaon processes remain sensitive to 
FCNC generated by new interactions at tree or loop level 
as well as to new mechanisms of CP violation.

Considering the difficulties encountered with
model independent analyses of new physics in Kaon processes due to the 
large number of parameters involved, only
typical models will be discussed here. These include  
the MSSM, the LRM, 3HDM,  models with light leptoquarks or family symmetries,
as well as non-minimal SUSY models; all of which are described in Section 2.
For the discussion, we will concentrate on the processes with the least 
theoretical uncertainties and will specifically   
emphasize the models that could lead to enhancements of
the SM prediction as experiments are expected, at best, to
reach the SM level (as demonstrated in Table \ref{gb1}).

\begin{table}
\begin{center}
\begin{tabular}{|c|c|c|c|}\hline\hline
Observable &Standard &Present Limit &Expected (Exp.)\\
           &Model    & (Exp.)       &Sensitivity \\ \hline
&&$(23\pm 7)\times 10^{-4}$ (NA31) &$10^{-4}$ (CERN-NA31)  \\
$\epsilon'/\epsilon$&$(1-15)\times 10^{-4}$ &$(7.4\pm 5.9)\times 10^{-4}$
(E731) & $10^{-4}$ (FNAL-E731) \\
&&&$10^{-4}$ (DA$\Phi$NE)  \\
$\kmue$&0&$3.3\times10^{-11}$ (AGS-791) &$10^{-12}$ (BNL-E871)  \\
$\kpimue$&0&$2.1\times10^{-10}$ (AGS) &$10^{-12}$ (BNL-E865)  \\
$\kpinunu$&$(.7-3)\times 10^{-10}$ &$5.2\times10^{-9}$ (AGS-787)&$ 10^{-10}$ 
(BNL-E787) \\
$\klpinu$&$.2-8\times 10^{-11}$ &$2.2\times10^{-4}$ (FNAL) 731&$10^{-8}$  
(FNAL-799)   \\
$\klpiee$&$.1-2\times 10^{-11}$ &$3.2\times10^{-7}$ (BNL E780)& 
$10^{-11}$ (FNAL-799II)  \\
$P_T(\kmu3)$ & $10^{-6}$ &$-3.1\pm5.3\times10^{-3}$ (BNL)& $5\times 10^{-4}$  
(KEK-E246) \\
\hline\hline
\end{tabular}
\end{center}
\caption{\small {Present limit and future prospects for Kaon
processes[103,104].}}
\label{gb1}
\end{table}

\subsubsection{Direct CP violation, $\epsilon'/\epsilon$}

The importance of this measurement to understand more about the mechanism of 
CP violation cannot be overemphasized, although,
due to conflicting experimental results, constraints on new physics from
$\epsilon'/\epsilon$ will not be taken into account here. 
The next round of experiments, which will reach a precision of
$10^{-4}$, might settle the issue of whether or not $\epsilon'/\epsilon\neq 0$. 
The SM predicts a non-zero value but allows for a wide range
(see Table~\ref{gb1}). Ultimately one wants to
establish whether CP violation is milliweak ($\Delta S=1$) as in the SM
and/or superweak ($\Delta S=2$). The latter occurs in
 multi-Higgs doublet models through scalar interactions, in SUSY 
models\cite{masieroeps}, or in the LRM to give a few examples\cite{winwolf}.
While beyond the standard models  often meet difficulties in
reproducing the value of $\epsilon$, this can be circumvented by
allowing  for the standard source of CP violation
in addition to the new superweak interaction.
A summary of potential effects of new physics in   a chosen
set of Kaon processes are given in Table \ref{gb2}.

\subsubsection{Lepton-number violation} 

The most severe upper limits on rare K decays have been obtained
for $\kmue$ and $\kpimue$. These decays feature a good sensitivity to
new physics as they are strictly forbidden in the SM unless neutrinos 
have masses.   Furthermore, the present limits on neutrino masses and mixings 
imply a decay rate which is orders of magnitude below the sensitivity of 
planned experiments\cite{langacker}.  The high sensitivity to new interactions
is best illustrated by a generic model
independent bound on the scale of new physics that generates $\kmue$,
($\Lambda > 108(420)$~TeV) for purely left-handed (scalar) operators. The 
process $\kpimue$ while being less sensitive,
probes a different set of operators; 
axial-vector or pseudoscalar operators as opposed to the
vector or scalar case\cite{valenciak}.

These limits can be translated into mass  bounds on the new particles that
could generate these decays at tree-level. In 
general, new particles that can induce tree-level flavor
changes, in particular vector bosons,
will also contribute to $\Delta M_K$ (as well as to $\kpinunu$).
Unless some symmetry allows one to avoid the latter constraints, 
it is unlikely that a signal would be observed in $\kmue$.
However, several models possess this type of symmetry, for example 
models with family symmetries.
Leptoquark models naturally avoid the constraint as they do not
contribute to $\Delta M_K$\cite{ritchie} at tree-level.
The same arguments apply also to cases where the lepton number violating
decays are induced at the loop level. For example in the MSSM, the
flavor change arises through mixing among the quark and lepton
superpartners. These mixings are constrained both by
$\Delta M_K$ and $\mu\rightarrow e \gamma$, so that the
rate for $\kmue$ is expected to be at most $10^{-14}$\cite{ritchie}.

\begin{table}
\begin{center}
\begin{tabular}{|c|c|c|c|c|c|c|}\hline\hline
& 3HDM& MSSM& LR& LQ&Hor.&Comments\\\hline 
$\epsilon'/\epsilon$&**  & ---&**&--- &*&{\small contributions from}\\
                    &    &    &  &    & &{\small non-minimal SUSY} \\
$\kmue$&---&---&---&**&**&\\
$\kpinunu$& *&--- &--- &** &** &{\small large in SUSY}\\
          &  &    &    &   &   &{\small with ${\not R}$ parity}\\
$P_T(\kmu3)$&** &--- &--- &** &---& \\
$P_L(\kmumu)$&--- &--- &** &**&--- & \\
\hline\hline
\end{tabular}
\end{center}
\caption{\label{gb2}\small
{Effect of new physics in Kaon processes. A double(single) star 
indicates possibly strong (mild) enhancement over the SM
while a dash stands for no measurable effect.}}
\end{table}

\subsubsection{Rare decays}

The standard model level for the theoretically clean decay, $\kpinunu$  
should be reached in the next decade.   This transition is theoretically
clean as it is short-distance dominated\cite{kpnnld}, the relevant hadronic 
operator is extracted from $K^+\to\pi^0 e^+\nu$, and the next-to-leading order 
QCD corrections are fully known\cite{kpnnqcd}.  An enhancement
over the SM rate  would clearly signal
new physics although such enhancements are not
expected in most minimal extensions of the SM 
once the constraints from $B-\bar{B}$ and $\epsilon_K$
%from $\Delta M,\epsilon_K$ and $b\rightarrow s\gamma$
are taken into account\cite{bigi}.
These processes are to a large extent governed by the same
parameters, limiting the impact of new physics in this case.
A possible exception concerns the  MSSM with SUSY particles in the 100 GeV
range where there can be some enhancement\cite{hagelin}. 
There remains the possibility of large enhancements in 
SUSY models  with broken R-parity, 
models with family symmetry producing a new type of neutrino,
as well as certain leptoquark models\cite{bigi}. Typically 
these models are more weakly constrained overall and could also lead
to non standard signals in other rare processes (for example
B or D decays).  The 3HDM also can lead to a moderate enhancement (by a 
factor 3) of the standard rate for this decay\cite{threehdm}. This is to be 
contrasted with the 2HDM where the existing  constraints 
preclude any significant effect in future kaon decays measurements\cite{bhp}.

The process $\kmumu$ shares several features of the preceding one as
far as sensitivity to new physics is concerned. However, the bounds obtained 
are not as reliable due to large and uncertain long distance contributions.
One interesting  aspect of this process is the
sensitivity to other sources of CP violation in the measurement
of the longitudinal polarization of the muon, which is
expected to be $P_L\approx 2\times 10^{-3}$ in the SM.
Both SUSY models with leptoquarks and the LRM with neutrino mixing
can produce large polarizations, while non-supersymmetric
models with leptoquarks predict a small polarization\cite{gengkmumu}. 

Both processes $\klpinu,\klpiee$ are sensitive to new sources
of CP violation\cite{dibg}, the former being basically free of long distance 
effect but representing a true challenge for experimentalists (to wit the 
present limit on this decay) while the second has larger theoretical 
uncertainties. Many models predict rates higher than the standard one and the
hope to observe those decays rests on the presence of new physics.
Leptoquarks for example can possibly lead to large
enhancements\cite{hagelin}.

\subsubsection{Transverse muon polarization in $\kmu3$}

The T-violating, transverse polarization of the muon ($P_T$)
in $\kmu3$ provides sensitive tests of models with new sources of CP
violation.  Indeed, in the SM, or in any model with only vector and/or
axial-vector interactions, $P_T$ is expected to be very small;
such is the case in the MSSM and the LRM.  On the
other hand, in multiple Higgs models where the CP or T violation arises in
the scalar sector from a phase in the charged Higgs mixing matrix a
polarization can be induced.  It has the form 
$P_T\propto Im(\alpha_1\beta_1^*)(m_K/m_{H^+})^2(v_2/v_3)^2$
where $\alpha_1$ and $\beta_1$ are the Yukawa coupling of 
the charged Higgs to quarks (corresponding to $X_1$ and $Y_1$, respectively,
in Eq. 2) and $v_2/v_3$ is the ratio of vev's.  At present the 
rate for $B(B\rightarrow X\tau\nu_\tau)$ 
which depends on the same parameters gives the strongest
constraint on the 3HDM (see Fig.~\ref{pt}).
The planned order of magnitude improvement on the search for this polarization
effect (see Table \ref{gb1}) should probe a large region of the
remaining parameter space as illustrated in Fig. \ref{pt} from Ref.~\cite{kuno}.
The Imaginary part of the Yukawa coupling in some
otherwise unconstrained leptoquark models
can also induced large polarization through tree-level leptoquark
exchanges\cite{gbgengkmu3}.

%%%
\vspace*{-0.5cm}
\nn
\begin{figure}[htbp]
%\centerline{
%\psfig{figure=kaon.ps,height=8cm,width=8cm,angle=0}}
\vspace*{8cm}
\caption{\small Constraints on the parameters of the
3HDM for $M_{H^+}=2 M_Z$ from (a)the neutron electric dipole moment;
(b) $b\rightarrow s\gamma$; (c) $P_T(\kmu3)$;
(d) $b\rightarrow X\tau\nu$. The dotted line shows the
expected sensitivity in $P_T$ at KEK-E246.}
\label{pt}
\end{figure}
\vspace*{0.4mm}

\subsubsection{Discussion} 

From inspection of Table \ref{gb2} one concludes that rare K decays are
not very sensitive to indirect effects from the MSSM, although they can probe
more general SUSY models.  In general  leptoquark models can potentially 
give strong signals in all rare decays and polarization measurements. 
This is simply a reflection of the fact
that these models are poorly constrained at present.
Other models could have measurable effects in only a few processes.
As it is expected to be too small, the effect of an anomalous $WW\gamma$ 
coupling  is not included.  Indeed, the best limit obtained on the
C and P violating coupling $g_5$ from the decays $\kmumu$ and $\kpinunu$
is of order 1  while we  expect $g_5<10^{-2}$ \cite{eeg5}.

\subsection{Charm-Quark Sector}

While investigations of the $K$ and $B$ systems have and will continue to
play a central role in our quest to understand flavor physics, in-depth
examinations of the charm-quark sector have yet to be performed, leaving a
gap in our knowledge.  Since charm is the only heavy charged $+2/3$ quark
presently accessible to experiment in copious amounts, it provides the sole
window of opportunity to examine flavor physics in this sector.  In addition,
charm allows a complimentary probe of SM physics (and beyond) to that
attainable from the down-quark sector.

Due to the effectiveness of the GIM mechanism, short distance SM contributions
to rare charm processes are very small.  Most reactions are thus dominated by
long range effects which are difficult to reliably calculate.  However, for
some interactions, there exists a window for the potential observation of new
physics.  In fact, it is precisely because the SM flavor changing neutral 
current rates are so small that charm provides an untapped
opportunity to discover new effects and offers a detailed test of the SM in
the up-quark sector.  Here, we will examine leptonic decays, rare decays,
\dmix\ mixing, and CP violation in the decays of charmed mesons.

\subsubsection{Leptonic Decays of Charmed Mesons}

The SM transition rate for the purely leptonic decay of a pseudoscalar charm
meson is 
\begin{equation}
\Gamma(D^+_{(q)}\to\ell^+\nu_\ell)={G^2_F\over 8\pi}
f_{D_{(q)}}|V_{cq}|^2m_{D_{(q)}}m^2_\ell\left( 1-{m^2_\ell\over m^2_{D_{(q)}}}
\right)^2 \,,
\label{leptdk}
\end{equation}
with $q=d, s$ and $f_{D_(q)}$ is the weak decay constant defined as usual by
\begin{equation}
\langle 0|\bar q\gamma_\mu\gamma_5 c| D_{(q)}({\bf p})\rangle = if_{D_{(q)}}
p_\mu \,,
\end{equation}
where $f_\pi=131\mev$ in this normalization.  The resulting branching
fractions are small due to the helicity suppression and are listed in 
Table \ref{charmfd} using the central values of the CKM parameters given 
in Ref.~\cite{pdg}.
\begin{table}
\centering
\begin{tabular}{|c|c|c|} \hline\hline
Meson & $\mu^+\nu_\mu$ & $\tau^+\nu_\tau$   \\ \hline
$D^+$   & $3.52\times 10^{-4}$ & $9.34\times 10^{-4}$ \\
$D^+_s$ & $4.21\times 10^{-3}$ & $4.11\times 10^{-2}$ \\ \hline\hline
\end{tabular}
\caption{\small SM branching fractions 
for the leptonic decay modes, assuming $f_D=200$ MeV and $f_{D_s}=230$ MeV.}
\label{charmfd}
\end{table}
Assuming that the CKM matrix elements are well-known, the leptonic decays can 
provide important information on the value of the 
pseudoscalar decay constants.  Precise measurements
of these constants are essential for the study of \dmix\ mixing, CP
violation, and non-leptonic decays.  

Non-SM contributions may affect these purely leptonic decays.  Signatures for 
new physics include the measurement of non-SM values for the absolute
branching ratios, or the observation of a deviation from the SM prediction
for the ratio
\begin{equation}
{B(D^+_{(s)}\to \mu^+\nu_\mu)\over B(D^+_{(s)}\to\tau^+\nu_\tau)}
={ {m^2_\mu\left( 1- m^2_\mu/ m^2_{D_{(s)}}\right)^2 }\over {m^2_\tau
\left( 1- m^2_\tau/ m^2_{D_{(s)}}\right)^2 }} \,.
\label{charmrat}
\end{equation}
This ratio is sensitive to violations of $\mu-\tau$ universality.
As a specific  example, we consider the case where the SM Higgs sector is
enlarged by an additional Higgs doublet.  These models generate
important contributions\cite{btaunu} to the decay $B\to\tau\nu_\tau$ and
it is instructive to examine their effects in the charm sector. 
In 2HDM Models I and II, we obtain 
\begin{equation}
B(D^+\to\ell^+\nu_\ell)  =  B_{\rm SM}\left( 1+ {m_D^2\over m^2_{H^\pm}}
\right)^2 \,, 
\end{equation}
where in Model II the $D_s^+$ decay receives an additional modification
\begin{equation}
B(D^+_s\to\ell^+\nu_\ell)  =  B_{\rm SM}\left[ 1+ {m_{D_s}^2\over m^2_{H^\pm}}
\left(1- \tan^2\beta{m_s\over m_c}\right)\right]^2 \,. 
\end{equation}
In this case, we see that the effect of the $H^\pm$ exchange is independent
of the leptonic final state and the above prediction for the ratio in Eq.
\ref{charmrat} is unchanged.  

\subsubsection{Rare and Forbidden Decays of Charm Mesons }

FCNC decays of the $D$ meson include the processes $D^0\to\ell^+\ell^-, 
\gamma\gamma$, and $D\to X_u+\gamma, X_u+\nu\bar\nu, X_u+\ell^+\ell^- $, with
$\ell=e, \mu$.  They proceed via electromagnetic or weak penguin
diagrams as well as receiving contributions from box diagrams in some
cases.  The calculation of the SM short distance rates for these processes is 
straightforward and the transition amplitudes and standard loop integrals,
which are categorized in Ref.\cite{inlim} for rare $K$ decays, are easily
converted to the $D$ system.  
The values of the resulting inclusive short distance branching fractions, 
before QCD corrections are applied, are 
shown in Table \ref{charmbr}, along with the current experimental 
bounds\cite{pdg,raredk}.  The leading order QCD corrections have recently
been calculated\cite{radcharm} for the radiative decay and are found to
greatly enhance the inclusive branching fraction giving
$B(D\to X_u\gamma)=(4-8)\times 10^{-12}$.  In all decay modes, the corresponding
exclusive rates are typically an order of magnitude less than the inclusive
case.  We note that the
transition $D^0\to\ell^+\ell^-,$ is helicity suppressed and hence has the 
smallest branching fraction.  The range given for this branching fraction,
$(1-20)\times 10^{-19}$, indicates the effect of varying the parameters
in the ranges $f_D=0.15-0.25\gev$ and $m_s=0.15-0.40\gev$.
\begin{table}
\centering
\begin{tabular}{|l|c|c|c|} \hline\hline
Decay Mode & Experimental Limit & $B_{S.D.}$ & $B_{L.D.}$ \\ \hline
$D^0\to\mu^+\mu^-$ & $<3.3\times 10^{-6}$ & $(1-20)\times 10^{-19}$ &
$<3\times 10^{-15}$ \\
$D^0\to e^+e^-$ & $<1.3\times 10^{-5}$ & $(2.3-4.7)\times 10^{-24}$ & \\
$D^0\to\mu^\pm e^\mp$ & $<1.9\times 10^{-5}$ & $0$ & $0$ \\ \hline
$D^0\to\gamma\gamma$ & --- & $10^{-16}$ & $<3\times 10^{-9}$ \\ \hline
$D\to X_u+\gamma$ & & $1.4\times 10^{-17}$ & \\
$D^0\to\rho^0\gamma$ & $<1.4\times 10^{-4}$ & & $<2\times 10^{-5}$ \\
$D^0\to\phi^0\gamma$ & $<2.0\times 10^{-4}$ & & $<10^{-4}$ \\
$D^+\to\rho^+\gamma$ & --- & & $<2\times 10^{-4}$ \\ \hline
$D\to X_u+\ell^+\ell^-$ & & $4\times 10^{-9}$ & \\
$D^0\to\pi^0ee/\mu\mu$ & $<4.5/54\times 10^{-5}$ & & \\
$D^0\to\bar K^0 ee/\mu\mu$ & $<1.1/6.7\times 10^{-4}$ & & 
$<2\times 10^{-15}$ \\
$D^0\to\rho^0 ee/\mu\mu$ & $<1.0/4.9\times 10^{-4}$ & & \\
$D^+\to\pi^+ee/\mu\mu$ & $<6.6/1.8\times 10^{-5}$ & few$\times 10^{-10}$ &
$<10^{-8}$ \\
$D^+\to K^+ee/\mu\mu$ & $<480/8.5\times 10^{-5}$ & & $<10^{-15}$ \\ 
$D^+\to\rho^+\mu\mu$ & $<5.8\times 10^{-4}$ & & \\
\hline
$D^0\to X_u+\nu\bar\nu$ & & $2.0\times 10^{-15}$ & \\
$D^0\to\pi^0\nu\bar\nu$ & --- & $4.9\times 10^{-16}$ & $<6\times 10^{-16}$ \\
$D^0\to\bar K^0\nu\bar\nu$ & --- & & $<10^{-12}$ \\
$D^+\to X_u+\nu\bar\nu$ & --- & $4.5\times 10^{-15}$ & \\
$D^+\to\pi^+\nu\bar\nu$ & --- & $3.9\times 10^{-16}$ & $<8\times 10^{-16}$ \\
$D^+\to K^+\nu\bar\nu$ & --- & & $<10^{-14}$ \\ \hline\hline
\end{tabular}
\caption{\small Standard Model predictions for the branching fractions due to 
short and long distance contributions for various rare $D$ meson decays. Also
shown are the current experimental limits[5,119].}
\label{charmbr}
\end{table}

The calculation of the long distance branching fractions are plagued with
the usual hadronic uncertainties and the estimates listed in the table convey
an upper limit on the size of these effects rather than an actual value.
These estimates have been computed by considering various intermediate
particle states (\eg, $\pi, K, \bar K, \eta, \eta', \pi\pi,\ {\rm or}\ K\bar K$)
and inserting the known rates for the decay of the intermediate particles
into the final state of interest.  In all cases we see that the long
distance contributions overwhelm those from SM short distance physics, and hence
would hide potential contributions from new physics.  This is shown explicitly
in Fig. \ref{cugamma}, where the branching fraction $B(D\to X_u\gamma)$ is
given in the four generation model as a function of the relevant fourth
generation CKM mixing factor.  We see that a sizable enhancement of the three
generation rate is possible, however, the short distance rate is still
overpowered by the long range effects.

\begin{figure}[htbp]
%\centerline{
%\psfig{figure=slac6821_fig1.ps,height=9cm,width=12cm,angle=90}}
\vspace*{8.5cm}
\caption{\small Branching fraction for $D\to X_u\gamma$ in the four generation 
SM as a function of the appropriate CKM mixing factor, with the solid, dashed,
dotted, dash-dotted curve corresponding to fourth generation quark masses
$M_{b'}=100, 200, 300, 400$ GeV, respectively.}
\label{cugamma}
\end{figure}

Lepton flavor violating decays, \eg, $D^0\to\mu^\pm e^\mp$ and
$D\to X+\mu^\pm e^\mp$, are strictly forbidden in the SM with massless
neutrinos.  In a model with massive non-degenerate neutrinos and 
non-vanishing neutrino mixings, such as in four generation models,
$D^0\to\mu^\pm e^\mp$ would be mediated by box diagrams with the massive
neutrinos being exchanged internally.  LEP data restricts\cite{neutr} heavy
neutrino mixing with $e$ and $\mu$ to be $|U_{Ne}U^*_{N\mu}|^2<7\times
10^{-6}$ for a neutrino with mass $m_N>45\gev$.  Consistency with
this bound constrains the branching fraction to be $B(D^0\to\mu^\pm 
e^\mp)<6\times 10^{-22}$.  This same results 
also holds for a heavy singlet neutrino
which is not accompanied by a charged lepton.  The observation
of this decay would be a clear signal for the existence of
a different class of models with new physics.
For example, leptoquarks can mediate $D^0\to \mu^\pm e^\mp$ by tree-level
exchange, although their contributions are suppressed by angular momentum
conservation.  From the present experimental bound on this process (as given
in Table \ref{charmbr}), Davidson \etal\cite{sacha} derive the constraint on the
leptoquark mass $m_{LQ}$ and coupling parameters (as defined in Section 2),
\begin{equation}
\sqrt{F_{eu}F_{\mu c}} < 4\times 10^{-3} {\alpha\over 4\pi}
\left[{m_{LQ}\over 100\gev}\right]^2 \,.
\end{equation}

\subsubsection{$D^0-\bar D^0$ Mixing and New Physics}  

Currently, the best bound\cite{pdg} on \dmix\ mixing is from fixed target 
experiment, with $x_D\equiv\Delta m_D/\Gamma<0.083$ (where $\Delta m_D=m_2-m_1$
is the mass difference), yielding $\dm<1.3\times 10^{-13}$ GeV.  However, the 
data analysis in this case\cite{browpak} was based on the assumption 
that there is no interference between the mixing signal and the dominant 
background of doubly Cabbibo suppressed decays.  It has been recently
noted\cite{browpak} 
that while this assumption may be valid in the SM (since the 
expected size of mixing is small), it does not necessarily apply in models
with new physics where \dmix\ mixing is potentially large.

The short distance SM contributions to \dm\ proceed through a $W$ box diagram 
with internal $d,s,b$-quarks.  In this case the external momentum, which is
of order $m_c$, is communicated to the light quarks in the loop and
can not be neglected.  The effective Hamiltonian is
\begin{equation}
{\cal H}^{\Delta c=2}_{eff} = {G_F\alpha\over 8\sqrt 2\pi x_w}\left[
|V_{cs}V^*_{us}|^2 \left(I_1^s {\cal O}-m_c^2I_2^s {\cal O'}\right)+
|V_{cb}V^*_{ub}|^2\left( I_3^b {\cal O}-m_c^2I_4^b {\cal O'}\right) \right] \,,
\end{equation}
where the $I_j^{q}$ represent integrals\cite{datta} that are functions of
$m_{q}^2/M_W^2$ and $m_{q}^2/m_c^2$, and ${\cal O}=[\bar u\gamma_\mu
(1-\gamma_5)c]^2$ is the usual mixing operator while ${\cal O'}=[
\bar u(1+\gamma_5)c]^2$ arises in the case of non-vanishing external
momentum.  The numerical value of the short distance contribution is 
$\dm\sim 5\times 10^{-18}$ GeV (taking $f_D=200$ MeV).  The long distance
contributions have been computed via two different techniques: (i) the 
intermediate particle dispersive approach
(using current data on the intermediate states) yields\cite{bghp}
$\dm\lsim 10^{-16}$ GeV, and (ii) heavy quark
effective theory which results\cite{hqet} in $\dm\sim 10^{-17}$ GeV.  
Clearly, the SM predictions lie far below the
present experimental sensitivity!

\vspace{3mm}
\noindent $\bullet$ {\bf Fourth Generation Model}
\vspace{2mm}

One reason the SM short distance 
expectations for \dmix\ mixing are so small is that there
are no heavy particles participating in the box diagram to enhance the rate.  
Hence the first extension to the SM that we consider is the
addition\cite{four} of a heavy $Q=-1/3$ quark. 
We can now neglect the external momentum and \dm\ is given
by the usual expression\cite{inlim},
\begin{equation}
\dm={G_F^2M_W^2m_D\over 6\pi^2}f_D^2B_D|V_{cb'}V_{ub'}^*|^2F(m^2_{b'}/M_W^2) \,.
\end{equation}
The value of \dm\ is displayed in this model in Fig.\ \ref{newdmix}(a) 
as a function of the 
overall CKM mixing factor for various values of the heavy quark mass.  We see 
that \dm\ approaches the current experimental range for large values of the
mixing factor.  

\vspace{3mm}
\noindent $\bullet$ {\bf Multi-Higgs-Doublet Model}
\vspace{2mm}

Next we examine two-Higgs-doublet models discussed above which
avoid tree-level FCNC by introducing a global symmetry.  
The expression for \dm\ in these models can be found in Ref.\ 18.
From the Lagrangian in Eq. (2) it is clear that Model I will only modify the SM
result for very small values of $\tan\beta$, and this region is already
excluded\cite{bhp,bsgcleo} from existing data on 
$b\to s\gamma$ and $B_d^0-\overline B_d^0$ 
mixing.  However, enhancements can occur in Model II for large values of
$\tan\beta$, as demonstrated in Fig.\ \ref{newdmix}(b).

\vspace{3mm}
\noindent $\bullet$ {\bf Supersymmetry}
\vspace{2mm}

Virtual exchange of squarks and gluinos in a SUSY-box diagram can have a
strong contribution to \dmix\ mixing.  In this case mixing can be induced 
by flavor changing radiatively generated mass insertions.  
These are thought to be small in the MSSM, but can be large in non-minimal
models\cite{yossi}.  The resulting $\Delta C=2$ Hamiltonian is
\begin{eqnarray}
{\cal H}^{\Delta c=2}_{SUSY} & = & {\alpha_s^2\over 216\tilde m_0}\left[
{\delta \tilde m^2_{u_Lc_L}\over \tilde m^2_0} G\left( {m_{\tilde g}^2\over
\tilde m_0^2}\right) (\bar u_L\gamma_\mu c_L)(\bar u_L\gamma^\mu c_L)
\right. \nonumber\\
& & \qquad \left. +(RR)^2+(LL)(RR)+(LR)^2+(LR)(RL)\right]\,,
\end{eqnarray}
where $\delta\tilde m_{u_Lc_L}$ characterizes the mass insertion, the G(x)'s
are known functions\cite{masieroeps}, and the remaining parameters are defined 
in Section 2.  The experimental bound on \dm\ can be translated into 
constraints on the various terms in the above Hamiltonian
as shown in Table \ref{dmixsusy}.

\begin{table}
\centering
\begin{tabular}{|c|c|}\hline\hline
(AB) & Upper Bound \\ \hline
(LL)$^2$, (RR)$^2$ & $(0.2)^2$ \\
(LL)(RR) & $(3.6\times 10^{-2})^2$ \\
(LR)$^2$, (RL)$^2$ & $(5.0\times 10^{-2})^2$ \\
(LR)(RL) & $(0.1)^2$ \\ \hline\hline
\end{tabular}
\caption{\small Bounds on $\delta\tilde m^2_{u_Ac_B}/\tilde m_0^2$ from \dm,
assuming a limit of \dm$<10^{-13}$ GeV.}
\label{dmixsusy}
\end{table}

\vspace{3mm}
\noindent $\bullet$ {\bf Flavor Changing Neutral Higgs Model}
\vspace{2mm}

We now consider the case of extended Higgs sectors without natural flavor 
conservation.  In this case, the lightest neutral higgs  
$h^0$ can now contribute to \dm\ through tree-level exchange 
as well as mediating \dmix\ mixing by $h^0$ and t-quark virtual
exchange in a box diagram.  These latter contributions only compete with those 
from the tree-level process for large values of $\Delta_{ij}$ (where 
$\Delta_{ij}$ is defined in Section 2).  In Fig. \ref{newdmix}(c-d)
we show the value of \dm\ in this model from these two types of contributions.

\nn
\begin{figure}[htbp]
%\centerline{
%\psfig{figure=slac6821_fig2a.ps,height=7.cm,width=8cm,angle=-90}
%\hspace*{-5mm}
%\psfig{figure=slac6821_fig2b.ps,height=7.cm,width=8cm,angle=90}}
%\vspace*{-0.75cm}
%\centerline{
%\psfig{figure=slac6821_fig2c.ps,height=7.cm,width=8cm,angle=-90}
%\hspace*{-5mm}
%\psfig{figure=slac6821_fig2d.ps,height=7.cm,width=8cm,angle=-90}}
%\vspace*{-0.75cm}
%\centerline{
%\psfig{figure=slac6821_fig2e.ps,height=7.cm,width=8cm,angle=90}
%\hspace*{-5mm}
%\psfig{figure=alrmdmix.ps,height=7.cm,width=8cm,angle=-90}}
\vspace*{16cm}
\caption{\small \dm\ in (a) the four generation SM with the same labeling as in
Fig. 11, (b) in two-Higgs-doublet model 
II as a function of $\tan\beta$ with, from top to bottom, the solid, dashed,
dotted, dash-dotted, solid curve representing $m_{H^\pm}=50, 100, 250, 500,
1000$ GeV.  The solid horizontal line corresponds to the present experimental
limit.  (c) Tree-level and (d) box diagrams contributions to \dm\
in the flavor changing Higgs model described in the text as a function of
the mixing factor for $m_h=50, 100, 250, 500, 1000$ GeV corresponding to the
solid, dashed, dotted, dash-dotted, and solid curves from top to bottom.
(e) Constraints in the leptoquark coupling-mass plane from \dm. (f) Values of
\dm\ in the Alternate Left-Right Symmetric Model as a function of the ratio of
masses of the exotic fermion $h_R$ to the right-handed $W$.  The curves 
represent the generational mass ratios for $h_R$ of 0.1, 0.3, 0.5, 0.7, and 
0.9, from top to bottom.} 
\label{newdmix}
\end{figure}

\vspace{3mm}
\noindent $\bullet$ {\bf Leptoquark Models}
\vspace{2mm}

Leptoquarks participate in \dm\ via virtual exchange inside a box 
diagram\cite{sacha}, together with a charged lepton or neutrino.  
Assuming that there is no leptoquark-GIM mechanism, and taking both exchanged
leptons to be the same type, we obtain the restriction
\begin{equation}
{F_{\ell c}F_{\ell u}\over m^2_{LQ}} < 
{196 \dm\over (4\alpha f_D)^2m_D} \,,
\end{equation}
where $F_{\ell q}$ is defined in Section 2.
The resulting constraints in the leptoquark coupling-mass plane are presented
in Fig. \ref{newdmix}(e), assuming 
that a limit of $\dm<10^{-13}$ GeV could be obtained from experiment.

\vspace{3mm}
\noindent $\bullet$ {\bf Alternate Left-Right Symmetric Model}
\vspace{2mm}

As discussed in Section 2, in this model the right-handed $W$ boson
couples the right-handed up-quarks to the exotic charged $-1/3$ $h_R$ fermion
present in the {\bf 27} representation of $E_6$.  The $W_R$ and $h_R$ can then
participate in the box diagram for \dmix\ mixing\cite{ema}, and can lead to 
large enhancements as shown in Fig. \ref{newdmix}f.

\subsubsection{CP Violation}

CP violation in the $Q=2/3$ quark sector is complimentary to that of the
$K$ and $B$ systems, but has yet to be explored.  In the SM, the CKM phase is 
responsible for generating CP violation, and in the charm system the resulting 
rates are small.  However, new
sources of CP violating phases could greatly enhance the rates thus 
rendering CP violation in the charm system a sensitive probe for physics 
beyond the SM.  CP violation requires the interference of at least two
amplitudes with non-vanishing phases.  This can occur indirectly via \dmix\
mixing, or directly via asymmetries induced in the decay amplitude, or
kinematically in final state distributions.

\vspace{3mm}
\noindent $\bullet$ {\bf Indirect CP Violation}
\vspace{2mm}

Indirect CP violation corresponds to
the interference of a $D^0$ decaying to a final state $f$ at time $t$,
with a $D^0$ which mixes into a $\bar D^0$ and then decays to $f$
at time $t$.  This process is theoretically clean as the hadronic
uncertainties cancel in the asymmetry.  However, since \dm\ is extremely
small in the SM the induced CP violation is negligible.
If new physics were to enhance \dmix\ mixing, as seen to occur in the previous
section for some models, then this mechanism could yield sizeable 
CP violating effects.  This interaction between mixing and CP violation in
the $D$ meson system has recently received much attention in the
literature\cite{browpak,bghp}.

\vspace{3mm}
\noindent $\bullet$ {\bf Direct CP Violation}
\vspace{2mm}

In order for direct CP violation to occur, the
decay amplitudes must have two separate weak phases and two 
different strong phases.  This can be easily seen as follows.
Let us assume that the decay amplitude to final state $f$ has the form
\begin{equation}
A_f=A_1 e^{i\delta_1}+A_2 e^{i\delta_2} \,,
\end{equation}
with $A_{1,2}$ being the two amplitudes after the strong phases $\delta_{1,2}$
have been factored out.  For the CP conjugate amplitude, the weak phases
are conjugated, but the strong phases are not.  The CP asymmetry is then
given by
\begin{equation}
{|A_f|^2-|\bar A_{\bar f}|^2 \over |A_f|^2+|\bar A_{\bar f}|^2} =
{2\Im (A_1^*A_2)\sin(\delta_1-\delta_2)\over |A_1|^2+|A_2|^2+2\Re (A_1^*A_2)
\cos(\delta_1-\delta_2)} \,,
\end{equation}
which clearly vanishes if $A_{1,2}$ contain the same weak phase and if
$\delta_1=\delta_2$.  Before estimating the typical size of this asymmetry
in the SM, we first note that 
in contrast to $B$ decays, the branching fractions for
the relevant modes, \ie, $\pi^+\pi^-, K^+K^-$, etc., are rather sizeable 
in the charm system, and for once, the large effects of final state
interactions are welcomed!  The size of the CP asymmetry in the SM is
estimated\cite{bucella} to be at most a few $\times 10^{-3}$.  The present
experimental sensitivity for various modes is in the vicinity of $10\%$
\cite{cpcharmexp}.

An interesting example of the potential size of CP violating effects
from new physics is that of left-right symmetric models\cite{orsay}.
In this case reasonably large values for CP asymmetries can be obtained
for the Cabbibo allowed decay modes.  This occurs due to the existence of
an additional amplitude from the $W_R$ exchange, which carries a
different weak phase from that of the $W_L$ mediated decay.  The estimated
values of the CP asymmetries in these models is of order a few $\times 10^{-2}$.
CP asymmetries at the percent level are expected\cite{susyd} in some 
non-minimal SUSY models for the decays $D^0\to K_S^0\pi^0,K_S^0\phi$.

\subsection{Bottom-Quark Sector}

A large amount of data on the $B$-meson system has been and will continue to 
be acquired during the next decade at LEP, CESR, the Tevatron, HERA, the SLAC 
and KEK B-factories, as well as the LHC\cite{browd,babar}, and promises to yield
exciting new tests of the SM.  FCNC processes in 
the $B$-sector are not as suppressed as in the other meson systems and can 
occur at reasonable rates in the SM.  This is due to a sizable loop-level
contribution from the top-quark, which results from the combination of the 
large top mass (giving a big GIM splitting) and the diagonal nature of the 
CKM matrix.  Long distance effects are expected to play less of role due
to the heavy $B$ mass, and hence rare processes are essentially short distance 
dominated.  Many classes of new models can also give significant 
and testable contributions to rare $B$ transitions.  The benchmark process
for this type of new physics search is the inclusive decay $B\to X_s\gamma$ 
(and the related exclusive process $B\to K^*\gamma$) which has been recently
observed by CLEO\cite{bsgcleo}.  It has since provided strong restrictions on 
the parameters of several theories beyond the SM\cite{jlh}  (which will be
reviewed below).  This constitutes the first direct observation 
of a penguin mediated process (!) and demonstrates the fertile ground ahead
for the detailed exploration of the SM in rare $B$ transitions.
 
\subsubsection{Leptonic Decays}

The SM transition rate for the purely leptonic decays $B\to\ell\nu_\ell$ is the
same as that given for the charm system in Eq. \ref{leptdk}, with appropriate 
substitutions. It is helicity suppressed and yields tiny branching 
fractions in the SM as shown in Table \ref{blept}.  These SM predictions are
somewhat imprecise due to the uncertainty in $f_B$ and $V_{ub}$, and
hence can vary over the range (where $B_{SM}$ is the result listed in the Table)
\begin{equation}
B_{SM}\left( {f_B\over 180\mev}\right)^2\left( {V_{ub}\over 0.0035}\right)^2\,.
\end{equation}
We see from the Table that the $90\%$ C.L.
experimental bounds are roughly two orders of magnitude above the SM
predictions for the cases of $B\to\mu\nu_\mu,\tau\nu_\tau$\cite{bleptlim}.  The
B-Factories presently under construction at SLAC and KEK should be able
to observe $B\to\tau\nu_\tau$ (and eventually the $\mu\nu_\mu$ mode as huge 
amounts of luminosity are accumulated over several years).  This would provide 
a classic measurement of the decay constant $f_B$ (assuming $V_{ub}$ is known
from other sources), but only if no new physics 
contributes to the decay.  For example, in models
with an enlarged Higgs sector, tree-level charged Higgs exchange can also 
mediate this transition.  In the 2HDM of Type II the branching fraction is 
modified by
\begin{equation}
B(B\to\ell\nu_\ell)=B_{SM}\left( 1-\tan^2\beta{m_B^2\over\mch^2}\right)^2 \,.
\end{equation}
Taking the SM and ALEPH bound on $B\to\tau\nu_\tau$ given in Table \ref{blept}
then implies $\tan\beta/\mch<0.47$ GeV$^{-1}$.  This constraint varies in the
range $\tan\beta/\mch<(0.38-0.68)$ GeV$^{-1}$ 
as one takes $f_B=180\pm 40$ MeV and
$|V_{ub}|=0.002-0.005$.  Once this decay is detected, tests for this type of 
scalar exchange can be performed by measuring the helicity of the 
final state $\tau$.  The measured branching fraction from LEP for the decay 
$B\to X\tau\nu$ yields\cite{btaunu,bleptlim} a similar constraint of
$\tan\beta/\mch<0.52$ GeV$^{-1}$, which is independent of the uncertainties
discussed above.  

\begin{table}
\centering
\begin{tabular}{|c|c|c|} \hline\hline
Mode & SM Prediction & Experimental Bound \\ \hline
$e\nu_e$ & $6.9\times 10^{-12}$ & $< 1.5\times 10^{-5}$ (CLEO) \\
$\mu^+\nu_\mu$ & $2.9\times 10^{-7}$ & $<2.1\times 10^{-5}$ (CLEO) \\
$\tau^+\nu_\tau$ & $6.6\times 10^{-5}$ & $<2.2\times 10^{-3}$ (CLEO) \\
 & & $<1.8\times 10^{-3}$ (ALEPH) \\ \hline\hline
\end{tabular}
\caption{\small SM branching fractions for the $B_d$ leptonic decay modes, 
assuming $f_B=180$ MeV and taking the central values of the relevant CKM
matrix elements[5].  The results of experimental searches[135] are also shown.}
\label{blept}
\end{table}

\subsubsection{Radiative Decays}

As discussed above, radiative $B$ decays have become one of the best testing 
grounds of the SM.  The CLEO Collaboration has reported\cite{bsgcleo} the 
observation of the inclusive decay $B\to X_s\gamma$ with a branching fraction of
$(2.32\pm 0.57\pm 0.35)\times 10^{-4}$, as well as an updated measurement for
the related exclusive process $B(B\to K^*\gamma)=(4.3^{+1.1}_{-1.0}\pm 0.6)
\times 10^{-5}$.  This yields a value of $0.19\pm 0.07\pm 0.04$ for the ratio 
of exclusive to inclusive rates.  On the theoretical
side, the reliability of the calculation of the quark-level process \bsg\
is improving\cite{qcd} as agreement on the leading-logarithmic QCD corrections 
has been reached and calculations at the next-to-leading logarithmic
order are underway.  These new results have inspired a large number of
investigations of this decay in various classes of models\cite{jlh}.

In the SM, the quark-level transition \bsg\ is mediated by $W$-boson and
t-quark exchange in an electromagnetic penguin diagram.  To obtain
the branching fraction, the inclusive rate is scaled to that of the
semi-leptonic decay $b\to X\ell\nu$.  This procedure removes uncertainties
from the overall factor of $m_b^5$, 
and reduces the ambiguities involved with the imprecisely
determined CKM factors.  The result is then
rescaled by the experimental value for the semi-leptonic branching fraction.
The calculation of $\Gamma(\bsg)$ employs the
renormalization group evolution\cite{qcd} for the 
coefficients of the $b\to s$ transition operators in the effective Hamiltonian
at the leading logarithmic level.  The participating operators consist of
the current-current operators $O_{1,2}$, the QCD penguin operators $O_{3-6}$,
and the electro- and chromo-magnetic operators $O_{7,8}$.
The Wilson coefficients are evaluated perturbatively
at the $W$ scale, where the matching conditions are imposed, and evolved
down to the renormalization scale $\mu$, usually taken to be $\sim m_b$.  
This procedure yields 
\begin{equation}
B(\bsg)={6\alpha\over\pi g(z)}\left| 
{V_{tb}V^*_{ts}\over V_{cb}}\right|^2
|c_7^{eff}(\mu)|^2 B(B\to X\ell\nu) =2.97^{+0.77}_{-0.59}\times 10^{-4} 
\end{equation}
for a top-quark mass of 180 GeV, with $g(z)$ being the phase space 
corrections for
the semi-leptonic decay.  The central value corresponds to
$\mu=m_b$, while the upper and lower errors represent the deviation due to
assuming $\mu=m_b/2$ and $\mu=2m_b$, respectively.  We see that (i) this result
compares favorably to the recent CLEO measurement and (ii) the freedom of
choice in the value of the renormalization scale introduces an uncertainty
of order $25\%$.  Clearly, this uncertainty must be taken into account
when determining constraints on new physics.  Comparison with the experimental
result gives the bound $|V_{ts}/V_{cb}|=0.91\pm 0.12(exp)\pm 0.13(th)$
in the SM\cite{skwar}.

We note here that it has been pointed out by numerous authors\cite{ldbdg}
that long distance contributions to $B\to X_d\gamma$ may be significant and
hence these decays may not yield a good determination of the CKM element 
$|V_{td}$.  However, separate measurements of charged and neutral $B$
decays into $\rho\gamma$ and $\omega\gamma$ may be useful in sorting out
the magnitude of the long distance contributions.

Before discussing explicit models of new physics, we first investigate the
constraints placed directly on the Wilson coefficients of the magnetic moment
operators from the CLEO measurement of \bsg.  
Writing the coefficients at the matching scale in the form 
$c_i(M_W)=c_i(M_W)_{SM}+c_i(M_W)_{new}$, where $c_i(M_W)_{new}$ represents
the contributions from new interactions, we see that the CLEO measurement
limits the possible values of $c_i(M_W)_{new}$ for $i=7,8$.  These bounds
are depicted in Fig. \ref{bsgamma}(a) for $m_t=175$ GeV, 
where the allowed regions lie inside the diagonal
bands.  We note that the two bands occur due to the overall sign ambiguity in
the determination of the coefficients.  The
horizontal lines correspond to potential limits on $B(b\to sg)<
(3-30)\times B(b\to sg)_{SM}$.
We see that such a constraint on $b\to sg$ is needed to further restrict the
values of the Wilson coefficients at the matching scale.

\vspace{3mm}
\noindent $\bullet$ {\bf Fourth Generation}
\vspace{2mm}

In the case of four families there is an additional contribution to \bsg\
from the virtual exchange of the fourth generation up-quark $t'$\cite{bsgfour}.
The Wilson coefficients of the dipole operators are then modified by
\begin{equation}
c_{7,8}(M_W)=c_{7,8}^{SM}(m_t^2/M_W^2)+{V_{t'b}V^*_{t's}\over V_{tb}V^*_{ts}}
c_{7,8}^{SM}(m_t'^2/M_W^2) \,.
\end{equation}
$V_{ij}$ represents the 4x4 CKM matrix which now contains nine parameters;
six angles and three phases.  The values of the elements of the 4x4 CKM matrix
are much less restricted than their 3 generation counterparts, as one can
no longer apply the 3-generation unitarity constraints\cite{pdg}. Hence,
even the overall CKM factor in the \bsg\ branching ratio, 
$|V_{tb}V^*_{ts}/V_{cb}|$, can take on different values.  Fig. \ref{bsgamma}(b)
displays the resulting branching fraction as a function of $m_{t'}$ for
$m_t=180$ GeV; here the vertical lines represent the range of possible values
as the CKM elements are varied.  These ranges were determined by
generating $10^8$ sets of the nine parameters in the 4x4 CKM matrix and
demanding consistency with (i) 4 generation unitarity and the extraction 
of the CKM elements from charged current measurements, (ii) the value of the
ratio $|V_{ub}/V_{cb}|$, (iii) $\epsilon$, and (iv) $B^0-\bar B^0$ mixing.
We see that there is little or no sensitivity to the $t'$-quark mass, and
that the CLEO measurement places additional constraints on the
4x4 CKM matrix.  In fact, we find that consistency with CLEO demands
$0.20\leq |V_{tb}V_{ts}|\leq 1.5\times 10^{-2}$ and 
$0.23\leq |V_{t'b}V_{t's}|\leq 1.1\times 10^{-3}$.

\vspace{3mm}
\noindent $\bullet$ {\bf Two-Higgs-Doublet Models}
\vspace{2mm}

In 2HDM the \ch\ contributes to \bsg\ via virtual 
exchange together with the top-quark.  At the $W$ scale the
coefficients of the dipole operators take the form (in Model II described above)
\begin{equation}
c_i(M_W)=c_i^{SM}(\mts/\mws)+A_{1_i}^{\ch}(\mts/\mchs)
+{1\over \tan^2\beta}A_{2_i}^{\ch}(\mts/\mchs) \,,
\end{equation}
where $i=7,8$.  The analytic
form of the functions $A_{1_i}, A_{2_i}$ can be found in Ref. \cite{bhp,bsgch}.
In Model II large enhancements appear for small values of \tb, but
more importantly, we see that $B(\bsg)$ is always larger than that of the SM,
independent of the value of \tb\ due to the presence of
the $A_{1_i}^{\ch}$ term.  In this case, the CLEO upper bound 
excludes\cite{bsgcleo,me} the region to the left and beneath the curves shown 
in Fig. \ref{bsgamma}(c) for $m_t=180\pm 12$ GeV.  

\vspace{3mm}
\noindent $\bullet$ {\bf Supersymmetry}
\vspace{2mm}

There are several new classes of contributions to \bsg\ in Supersymmetry.
The large \ch\ contributions from Model II discussed above are 
present, however, the limits
obtained in supersymmetric theories also depend on the size of the other
super-particle contributions and are generally much more 
complex.  In particular, it has been shown\cite{bert,okada} that large
contributions can arise from stop-squark and chargino exchange (due to the
possibly large stop-squark mass splitting), as well as from the gluino and
down-type squark loops (due to left-right mixing in the sbottom sector).
The additional neutralino-down-squark contributions are expected to be small.
Some regions of the parameter space can thus cancel the \ch\ contributions 
resulting in predictions for the branching fraction at (or even below) the 
SM value, while other regions always enhance the amplitude.  In minimal
supergravity models with radiative breaking, the sign of the sparticle
loop contributions is found to be correlated with the sign of the higgsino mass
parameter $\mu$\cite{okada,leszek}.  This is demonstrated in 
Fig. \ref{bsgamma}(d) from Goto and Okada\cite{okada}, where the points in
this figure represent a scan of the remaining parameter space.
We see that taking $\mu<0$ $(>0)$ enhances (suppresses) the
branching fraction from the predictions in the 2HDM of Type II.  
We also note here that \bsg\ has been found to constrain dark matter candidates
in Supersymmetric models\cite{drees}.

\vspace*{-0.5cm}
\nn
\begin{figure}[htbp]
%\centerline{
%\psfig{figure=slac6822_fig2a.ps,height=8.cm,width=8cm,angle=90}
%\hspace*{-5mm}
%\psfig{figure=bsg4.ps,height=8.cm,width=8cm,angle=-90}}
%\vspace*{-0.75cm}
%\centerline{
%\psfig{figure=bsgch.ps,height=8.cm,width=8cm,angle=-90}
%\hspace*{-5mm}
%\psfig{figure=br.tan05.eps,height=7.8cm,width=10.5cm,angle=0}}
\vspace*{15cm}
\caption{\small (a) Bounds on the contributions from new physics to $c_{7,8}$.
The region allowed by CLEO corresponds to the area inside the diagonal bands.
The horizontal lines represent potential measurements of $R\equiv 
B(b\to sg)/B(b\to sg)_{SM}<30,20,10,5,3$ corresponding to the set of solid,
dotted, dash-dotted, dashed, and dotted lines, respectively.  The point
`S' represents the SM. (b) The range of values for $B(\bsg)$ in the 4 
generation SM as a function of $m_{t'}$.  (c) Limits from \bsg\
in the charged Higgs mass - \tb\ plane.  The excluded region is that to the
left and below the curves.  The three curves correspond to the values
$m_t=192, 180, 168$ GeV from top to bottom.
(d) $B(\bsg)$ as a function of the charged Higgs mass with $m_t=175$ GeV
and $\tan\beta=5$ from Ref. 143.  The solid curve corresponds to the
2HDM Model II value, while the dashed-dot curve represents the SM.  Each
dot corresponds to a sample point of the SUSY parameter space.}
\label{bsgamma}
\end{figure}

\vspace{3mm}
\noindent $\bullet$ {\bf Anomalous Trilinear Gauge Couplings}
\vspace{2mm}

The trilinear gauge coupling of the photon to $W^+W^-$ can also be tested
in radiative $B$ decays.  \bsg\ naturally avoids the problem of 
introducing cutoffs to regulate the divergent loop integrals due to the 
cancellations provided by the GIM mechanism,
and hence cutoff independent bounds on anomalous couplings can be obtained.
In this decay only the coefficient
of the magnetic dipole operator, $O_7$, is modified
by the presence of the additional terms in Eq. \ref{anomw} and can 
be written as
\begin{equation}
c_7(M_W) = c_7^{SM}(\mts/\mws) +\Delta\kappa_\gamma A_1(\mts/\mws)
+\lambda_\gamma A_2(\mts/\mws) \,.
\end{equation}
The explicit form of the functions $A_{1,2}$ can be found in 
Ref. \cite{tgrtwo}.  As both of these parameters are varied, either large 
enhancements or suppressions over the SM prediction for the \bsg\ branching
fraction can be obtained.  When one demands consistency with both the upper 
and lower CLEO bounds, a large region of the 
$\Delta\kappa_\gamma-\lambda_\gamma$ parameter plane is excluded; this
is displayed in Fig. \ref{bsgamma2}(a) from Ref. \cite{bsgcleo} 
for $m_t=174$ GeV.
Here, the allowed region is given by the cross-hatched area, where
the white strip down the middle is excluded by the
lower bound and the outer white areas are ruled out by the upper limit on
$B(\bsg)$.  The ellipse represents the region allowed by D0\cite{dzero}.
Note that the SM point in the $\Delta\kappa_\gamma-
\lambda_\gamma$ plane (labeled by the dot) lies in the center of one of the
allowed regions.
We see that the collider constraints are complementary to those from \bsg.

\vspace*{-0.5cm}
\nn
\begin{figure}[htbp]
%\centerline{
%\psfig{figure=slac6822_fig2d.ps,height=8.cm,width=8cm,angle=0}
%\hspace*{-5mm}
%\psfig{figure=anomtop.ps,height=8.cm,width=10cm,angle=-90}}
\vspace*{8cm}
\caption{\small (a) Constraints on anomalous $WW\gamma$ couplings.  The 
shaded area is that allowed by CLEO and the interior of the ellipse is the
region allowed by D0.  The dot represents the SM values. (b) Bounds on
anomalous top-quark photon couplings from \bsg.  The solid and dashed curves
correspond to the cases described in the text.
In each case, the allowed regions lie inside the semi-circles.}
\label{bsgamma2}
\end{figure}

\vspace{3mm}
\noindent $\bullet$ {\bf Anomalous Top-Quark Couplings}
\vspace{2mm}

If the top-quark has anomalous couplings to on-shell photons or gluons,
the rate for \bsg\ would be modified.  The effect of an anomalous magnetic
and/or electric dipole moment in the Lagrangian of Eq. 6 on
the Wilson coefficients is
\begin{equation}
c_{7,8}(M_W)=c^{SM}_{7,8}(m_t^2/M_W^2)+\kappa_{\gamma,g}F_{1_{7,8}}(m_t^2/M_W^2)
+\tilde\kappa_{\gamma,g}F_{2_{7,8}}(m_t^2/M_W^2) \,.
\end{equation}
The functions $F_{1,2}$ can be found in Ref. \cite{meanomtop}.  The effects
of anomalous chromo-dipole moments arise from operator mixing.  When the
resulting branching fraction and the CLEO data are combined, the constraints
in Fig. \ref{bsgamma2}(b) are obtained for $m_t=180$ GeV.  In this
figure, the allowed region is given by the area inside the solid (dashed) 
semi-circle when $\kappa_g,\tilde\kappa_g=0 
(=\kappa_\gamma,\tilde\kappa_\gamma)$.
These bounds are considerably weaker than those obtainable from direct 
top-quark production at colliders\cite{frey}.

\subsubsection{Other Rare Decays}

Other FCNC decays of $B$ mesons include $B^0_{d,s}\to\ell^+\ell^-, 
\gamma\gamma$, $B\to X_{s,d}+\ell^+\ell^-, X_{s,d}\nu\bar\nu$, with
$\ell=e\mu\tau$.  In the SM they are mediated by appropriate combinations of 
electromagnetic and weak penguins as well as box diagrams, and generally have 
larger rates, as discussed above, due to the heavy top-quark and the diagonal 
nature of the CKM matrix.  The SM predictions\cite{aligreub} and current 
experimental situation\cite{pdg,browd,bdkexp} for these decays are summarized 
in Table \ref{bdkbr}, taking $m_t=180$ GeV.  
The purely leptonic decays, $B^0\to\ell^+\ell^-$ can be 
enhanced by contributions from new physics at both the loop-level, for example 
in Extended Technicolor models\cite{sund} or by
virtual \ch\ exchange\cite{bmumu} in 2HDM, and at tree-level, \eg, with 
leptoquark exchange\cite{sacha}.  However, as can be
seen from the Table, the experimental probes of these purely leptonic decays 
are orders of magnitude above the expected rates, and hence only potentially
large tree-level contributions can currently be tested.  Indeed, the most
stringent constraints on tree-level leptoquark contributions in $B$ decays
are obtained from the exclusive reaction $B\to Ke\mu$\cite{sacha}.  However,
in this case there exist large uncertainties associated with the
hadronic matrix elements, yielding some sloppiness in the resulting bounds.

\begin{table}
\centering
\begin{tabular}{|l|c|c|} \hline\hline
Decay Mode & Experimental Limit & $B_{SM}$ \\ \hline
$B^0_d\to e^+e^-$    & $<5.9\times 10^{-6}$ (CLEO) & $2.6\times 10^{-15}$ \\
$B^0_d\to\mu^+\mu^-$ & $<1.6\times 10^{-6}$ (CDF) & $1.1\times 10^{-10}$ \\
$B^0_d\to \tau^+\tau^-$    &   ---                 & $2.1\times 10^{-8}$ \\
$B^0_s\to e^+e^-$    &            --- & $5.3\times 10^{-14}$ \\
$B^0_s\to\mu^+\mu^-$ & $<8.4\times 10^{-6}$ (CDF)  & $2.4\times 10^{-9}$\\
$B^0_s\to \tau^+\tau^-$    & ---               & $5.1\times 10^{-7}$ \\ \hline
$B^0\to e^\pm\mu^\mp$ & $<5.9\times 10^{-6}$ (CLEO)& 0 \\
$B^0\to e^\pm\tau^\mp$& $<5.3\times 10^{-4}$ (CLEO)& 0 \\
$B^0\to \mu^\pm\tau^\mp$ & $<8.3\times 10^{-4}$ (CLEO)& 0 \\ \hline
$B^0_d\to\gamma\gamma$  & $<3.8\times 10{-5}$ (L3) & $1.0\times 10^{-8}$ \\ 
$B^0_s\to\gamma\gamma$  & $<1.1\times 10^{-4}$ (L3)&$3\times 10^{-7}$ \\ \hline
$B\to X_s+\gamma$ & $(2.32\pm 0.57\pm 0.35)\times 10^{-4}$ (CLEO) & 
$(2.97^{+0.77}_{-0.59})\times 10^{-4}$ \\
$B\to K^{*}\gamma$    & $(4.3^{+1.1}_{-1.0}\pm 0.6)\times 10^{-5}$ (CLEO) & 
$(4.0\pm 2.0)\times 10^{-5}$\\ 
$B^+\to\rho^+\gamma$ & $<0.34\times B(B\to K^{*}\gamma)$ (CLEO) & 
$(1.9\pm 1.6)\times 10^{-6}$\\
$B^0\to\rho^0(\omega)\gamma$ & $<0.34\times B(B\to K^{*}\gamma)$ (CLEO) & 
$(0.85\pm 0.65)\times 10^{-6}$\\ \hline
$B\to X_s+e^+e^-$ & --- & $7.0\times 10^{-6}$ \\
$B\to X_s+\mu^+\mu^-$ & $<5.0\times 10^{-5}$ (UA1)  & $6.2\times 10^{-6}$ \\
$B\to X_s+\tau^+\tau^-$ &  ---               & $3.2\times 10^{-7}$ \\
$B^0\to K^0ee/\mu\mu$ & $<1.5/2.6\times 10^{-4}$ (CLEO) & $(5.0\pm 3.0)/(3.0\pm
1.8)\times 10^{-7}$\\ 
$B^-\to K^-ee/\mu\mu$ & $<1.2/0.9\times 10^{-5}$ (CLEO) & $(5.0\pm 3.0)/(3.0\pm
1.8)\times 10^{-7}$ \\ 
$\bar B^0\to\bar K^{*0}ee/\mu\mu$& $<1.6/2.5\times 10^{-5}$ (CLEO/CDF) & 
$(2.0\pm 1.0)/(1.25\pm 0.62)\times 10^{-6}$ \\
$\bar B^-\to\bar K^{*-}ee/\mu\mu$& $<6.3/11\times 10^{-4}$ (CLEO) & 
$(2.0\pm 1.0)/(1.25\pm 0.62)\times 10^{-6}$ \\
$B^+\to K^+e^\pm\mu^\mp$ & $<1.2\times 10^{-5}$ (CLEO) & $0$ \\ 
$\bar B^0\to \bar K^{*0}e^\pm\mu^\mp$& $<2.7\times 10^{-5}$ (CLEO)&$0$\\ \hline
$B\to X_s+\nu\bar\nu$ & $<3.9\times 10^{-4} \dagger$ 
& $5.0\times 10^{-5}$  \\ \hline
\end{tabular}
\caption{\small Standard Model predictions[149] for the branching fractions 
for various rare $B$ meson decays with $f_{B_d}=180$ MeV. Also
shown are the current experimental limits[5,150]. $\dagger$ This is an inferred
bound[151] from limits on $B\to\tau\nu_\tau$.}
\label{bdkbr}
\end{table}

The transition $b\to s\ell^+\ell^-$ merits further attention as it offers 
an excellent opportunity to search for new physics.  For example, it has
been found\cite{grin} that Extended Technicolor models with a GIM mechanism
already violates (!) the experimental upper bound on $B\to X_s\mu\mu$, 
but more traditional ETC models yield a rate which is close to the SM 
prediction.  The decay proceeds via electromagnetic and $Z$ 
penguin as well as by $W$ box diagrams and hence can probe different coupling
structures than the pure electromagnetic process $b\to s\gamma$.  The matrix
element can be written as
\begin{equation}
{\cal M}={\sqrt 2 G_F\alpha\over\pi}V_{tb}V^*_{ts}\left[ c_9^{eff}\bar s_L
\gamma_\mu b_L\bar\ell\gamma^\mu\ell +c_{10}\bar s_L\gamma_\mu b_L\bar\ell
\gamma^\mu\gamma_5\ell - 2c_7m_b\bar s_Li\sigma_{\mu\nu}{q^\nu\over q^2}
b_R\bar\ell\gamma^\mu\ell\right] \,,
\end{equation}
where $q^2$ is the momentum transferred to the lepton pair.  Here we take
the sign convention of Ali \etal\cite{ali} for the Wilson coefficients.  
These short distance contributions are theoretically well-known as the NLO
QCD corrections have recently been computed\cite{burasbsll} for the 
coefficient $c_9$, and $c_{10}$ does not receive any large contributions from
the renormalization evolution.
This reaction also receives long distance contributions
from the processes $B\rightarrow K^{(*)}\psi^{(')}$ followed by $\psi^{(')}
\rightarrow\ell^+\ell^-$ and from $c\bar c$ continuum intermediate states.
The short distance contributions  lead to the inclusive branching 
fractions given in the Table;  we see that these modes will likely 
be observed during the next few years!  The best technique of separating the 
long and short distance contributions, as well as observing any deviations 
from the SM predictions, is to measure the various kinematic distributions 
associated with the final state lepton pair, such as the lepton pair invariant 
mass distribution\cite{bsll}, the lepton pair forward-backward 
asymmetry\cite{ali}, 
and the tau polarization asymmetry\cite{mebsll} in the case $\ell=\tau$.
These distributions are presented in Fig. \ref{bsll}, with and without the 
resonance contributions.  Note that both asymmetries are large for this value
of the top-quark mass.  As an example of how new physics can 
affect this process, we display in Fig. \ref{bsll}(d) the tau polarization 
asymmetry for various changes of sign of the
contributing Wilson coefficients.
Measurement of all three kinematic distributions
would allow for the determination of the sign and magnitude of the Wilson 
coefficients for the contributing electroweak loop
operators and thus provide a completely model independent analysis.
We present a $95\%$ C.L. Monte Carlo fit\cite{mebsll} to these coefficients in 
Fig. \ref{bsllfit}, assuming the SM is realized in nature, and taking an 
integrated luminosity of $5\times 10^8\, B\bar B$ pairs.  (This clearly 
requires the high statistics samples which will be available at future 
B-factories.)  This procedure demonstrates that the coefficients 
$c_{7,9,10}(\mu)$ can be measured to an accuracy 
of roughly $7.5\%, 15\%$, and $5\%$, respectively, which would yield a very
stringent test of the SM.  

\vspace*{-0.5cm}
\nn
\begin{figure}[htbp]
%\centerline{
%\psfig{figure=slac6820_fig1a.ps,height=8.cm,width=8cm,angle=-90}
%\hspace*{-5mm}
%\psfig{figure=bsllasym.ps,height=8.cm,width=8cm,angle=-90}}
%\vspace*{-0.75cm}
%\centerline{
%\psfig{figure=slac6820_fig1b.ps,height=8.cm,width=8cm,angle=-90}
%\hspace*{-5mm}
%\psfig{figure=slac6820_fig2a.ps,height=8.cm,width=8cm,angle=-90}}
\vspace*{15cm}
\caption{\small (a) Differential branching fraction, (b) lepton pair
forward backward asymmetry, and (c) tau polarization
asymmetry as a function of $\hat s$ for $\ell=\tau$ (solid and dashed curves)
and $\ell=e$ (dotted and dash-dotted curves), with and without the long
distance contributions. (d) Tau polarization asymmetry with changes in sign
of the Wilson coefficients at the electroweak scale, corresponding to
$c_{10}, c_9, c_{9,10}, SM$, and $c_{7,8}$ from bottom to top. }
\label{bsll}
\end{figure}

\nn
\begin{figure}[htbp]
%\centerline{
%\psfig{figure=slac6820_fig3a.ps,height=8.cm,width=8.cm,angle=-90}
%\hspace*{-5mm}
%\psfig{figure=slac6820_fig3b.ps,height=8.cm,width=8.cm,angle=-90}}
\vspace*{8cm}
\caption{$95\%$ C.L. projected contour in the (a) $c_9-c_{10}$ and (b)
$c_7-c_{10}$ plane.  `S' labels the SM prediction and the diamond represents
the best fit values. Here we use the sign convention of Ali[155] for the
coefficients.}
\label{bsllfit}
\end{figure}

Presently, there have been no direct searches for the `invisible' decay,
$B\to X_s\nu\bar\nu$, however, bounds on this process may be inferred from 
searching for events with large missing energy in $B$ decays, such as 
$B\to\tau\nu_\tau$.  The limit obtained in this manner\cite{yuval} is 
quoted in Table \ref{bdkbr}.  This transition proceeds via $Z$ penguin and
$W$ box diagrams in the SM, with the rate being roughly one order of
magnitude lower than the inferred bound.  Various classes of new interactions 
can contribute substantially to this decay and they have been categorized in
Ref. \cite{yuval}; these include models with leptoquarks, Supersymmetry
with R-parity violating couplings, Topcolor models, and horizontal gauge
symmetries.  Defining a most general form of a four-fermion interaction
responsible for this decay as ${\cal L}=C_L\bar s_L\gamma_\mu b_L\bar\nu_L
\gamma^\mu\nu_L + C_R\bar s_R\gamma_\mu b_R\bar\nu_L\gamma^\mu\nu_L$, gives
\begin{equation}
B(B\to X_s\nu\bar\nu)={C_L^2+C_R^2\over |V_{cb}|^2g(z)} B(B\to X\ell\nu)\,,
\end{equation}
from which these authors have found the model independent bound
\begin{equation}
C_L^2+C_R^2<3.0\times 10^{-6} \left[ {B(B\to X_S\nu\bar\nu)\over
3.9\times 10^{-4} }\right] \,.
\end{equation}
In some models, the restrictions obtained from this process either surpass or
are competitive with those from $B\to X_s\ell^+\ell^-$.

\subsubsection{$B^0-\bar B^0$ Mixing}

The quark level process which is dominantly responsible for $B^0-\bar B^0$ 
mixing in the SM is that of top-quark exchange in a $W$ box diagram.
The mass difference for $B_d$ meson mixing is then given by
\begin{equation}
\Delta M_d={G_F^2M_W^2m_B\over 6\pi^2}f^2_{B_d}B_{B_d}\eta_{B_d}
|V_{tb}V^*_{td}|^2F(m_t^2/M_W^2) \,,
\label{bmixeq}
\end{equation}
with $\eta_{B_d}$ being the QCD correction factor which is calculated to
NLO\cite{buras2}, and F(x) being the usual Inami-Lim function\cite{inlim}.  An
equivalent expression for $B_s$ mixing is obtained with $d\to s$.  This
yields the SM values of $\Delta M_d=(3.0^{+9.0}_{-2.7})\times 10^{-13}$ GeV  
and $\Delta M_s=(7.4^{+8.6}_{-4.3})\times 10^{-12}$ GeV, where the ranges
correspond to taking $m_t^{phys}=180\pm 12$ GeV, $|V_{td}|=0.009\pm 0.005$
and $|V_{ts}|=0.040\pm 0.006$ as given in Ref. \cite{pdg}, and 
$f_{B_d}\sqrt{B_{B_d}}=180\pm 40$ MeV, $f_{B_s}\sqrt{B_{B_s}}=200\pm 40$ MeV
as suggested by lattice gauge theory\cite{sonilat}.  This agrees well with
the experimental bounds\cite{saulan} of $\Delta M_d=(3.01\pm 0.13)\times
10^{-13}$ GeV and $\Delta M_s>4.0\times 10^{-12}$ GeV.  This situation is
summarized in Fig. \ref{bmixfig}.  

The ratio of hadronic matrix elements, 
$f_{B_d}\sqrt{B_{B_d}}/f_{B_s}\sqrt{B_{B_s}}$, is more accurately
calculable in lattice gauge theory\cite{sonilat}, hence a measurement of
$\Delta M_d/\Delta M_s$ would be an important determination of the value of the 
CKM ratio $|V_{td}/V_{ts}|$ in the SM.  Remarkably, this remains true in
many scenarios beyond the SM.  In this class of models, the virtual exchange of
new particles alters the Inami-Lim function in Eq. \ref{bmixeq} above, but
not the factors in front of the function.  The effects of the new physics
then cancels in the ratio.  Models of this type include, 2HDM and
Supersymmetry in the super-CKM basis.
Notable exceptions to this feature can be found in models which (i) change the
structure of the CKM matrix, such as the addition of a fourth generation, or
extra singlet quarks, and in Left-Right Symmetric models, (ii) have couplings
proportional to fermion masses, such as flavor changing Higgs models, or
(iii) have generational dependent couplings, \eg, leptoquarks or SUSY with
R-parity violation.

\vspace*{-0.5cm}
\nn
\begin{figure}[htbp]
%\centerline{
%\psfig{figure=bmix.ps,height=9cm,width=12cm,angle=-90}}
\vspace*{8cm}
\caption{\small The SM expectation for the $\Delta M_d - \Delta M_s$ plane,
where the predicted region lies inside the solid curves.  The experimental
bounds lie in between the solid horizontal lines and to the right of the
solid vertical line.}
\label{bmixfig}
\end{figure}

It is difficult to use $\Delta M_d$ alone to restrict new physics due to
the enormous errors on the theoretical predictions for this quantity from the
imprecisely determined CKM factors and unmeasured $B$ hadronic matrix elements.
(This is unfortunate as $\Delta M_d$ is so precisely measured!)
In most cases, the restrictions obtained from \bsg\ surpass those
from $B^0-\bar B^0$ mixing.  As a demonstration of this point, we note the 
results in Ref. \cite{okadadm} where $\Delta M_d$ is calculated in 2HDM of 
Type II and in minimal Supergravity models.  These models contribute to
$B^0-\bar B^0$ mixing via \ch-top-quark, chargino-stop-squark, and 
gluino-down-squark virtual exchange in box diagrams (the neutralino 
contributions are found to be small).  These authors find that 
although substantial enhancements are possible (up to a factor of $50\%$
over the SM), $\Delta M_d$ remains well within 
the overall theoretical errors.  Another example is given in 
Fig. \ref{bsgamma}(b), where \bsg\ is shown to greatly restrict the parameter 
space of the 4 generation SM, even after constraints from $B^0-\bar B^0$
mixing were applied.

\subsubsection{CP Violation in $B$ Decays}

CP violation in the $B$ system will be examined\cite{babar} during the next
decade at dedicated B-Factories.  CP violation arises in the SM from the
existence of the phase in the 3 generation CKM matrix as first postulated by
Kobayashi and Maskawa\cite{km}.  The
relation $V_{tb}V^*_{td}+V_{cb}V^*_{cd}+V_{ub}V^*_{ud}=0$, which is required
by unitarity, can be depicted as a triangle in the complex plane as shown
in Fig. \ref{triangle}, where
the area of the triangle represents the amount of CP violation.  
It can be shown that the apex of the triangle is located at the 
point $(\rho,\eta)$ in the complex plane, where $\rho$ and $\eta$ are 
parameters describing the CKM matrix in the Wolfenstein notation\cite{wolf}.
The present status of these parameters is summarized in Fig. \ref{cpvio}(a),
where the shaded area is that allowed in the SM.
This region is determined by measurements
of the quantities (i) $|V_{ub}|$ and $|V_{cb}|$, (ii) $\epsilon_K$, and (iii) 
the rate for $B^0_d-\bar B^0_d$ mixing,  
together with theoretical estimates for the parameters which relate these
measurements to the underlying theory, such as $B_K,\ f_B,$ and $B_B$.
The value of $\overline{m_t}(m_t)$ is taken to be consistent with
the physical range $180\pm 12$ GeV.  This yields the
allowed ranges for the angles of the triangle: $-0.89\leq\sin 2\alpha\leq 1.00,\
0.18\leq\sin 2\beta\leq 0.81$, and $-1.00\leq\sin 2\gamma\leq 1.00$.

\vspace*{-0.5cm}
\nn
\begin{figure}[htbp]
%\centerline{
%\psfig{figure=triangle.ps,height=6cm,width=12cm,angle=0}}
\vspace*{5.5cm}
\caption{\small The rescaled Unitarity triangle.}
\label{triangle}
\end{figure}

\vspace*{-0.5cm}
\nn
\begin{figure}[htbp]
%\centerline{
%\psfig{figure=slac6822_fig4a.ps,height=8cm,width=8cm,angle=90}
%\hspace*{-5mm}
%\psfig{figure=slac6822_fig4b.ps,height=8cm,width=8cm,angle=90}}
\vspace*{8cm}
\caption{\small Constraints in the (a) SM and (b) two-Higgs-doublet Model II in
the $\rho-\eta$ plane from $|V_{ub}|/|V_{cb}|$ (dotted circles), $B_d^0-
\bar B_d^0$ mixing (dashed circles) and $\epsilon$ (solid hyperbolas).
The shaded area corresponds to that allowed for the
apex of the Unitarity triangle.}
\label{cpvio}
\end{figure}

It is important to remember that this picture can be dramatically altered if 
new physics is present, even if there are no new sources of CP violation.
Figure \ref{cpvio}(b) displays the constraints in the $\rho-\eta$ plane
in the two-Higgs-doublet Model II.  In this case the presence of the
extra Higgs doublet is felt by the virtual exchange of the \ch\ boson in
the box diagram which mediates $B^0_d-\bar B^0_d$ mixing and governs the
value of $\epsilon_K$.  For this $\rho-\eta$ region, the allowed ranges of
the angles of the unitarity triangle become $-1.00\leq\sin 2\alpha\leq 1.00,\
0.12\leq\sin 2\beta\leq 0.81$, and $-1.00\leq\sin 2\gamma\leq 1.00$.  In fact,
this opens up a new allowed region in the $\sin2\alpha-\sin2\beta$ plane,
as shown in Fig. \ref{yuvalfig} from Ref. \cite{grossnir}.  Similar effects
have also been pointed out in Supersymmetric models\cite{oshimo}.
We see that the SM predictions for CP violation are thus modified.  Clearly,
caution must be exercised when relating the results of future CP violation
experiments to the $\rho-\eta$ plane.

The B-Factories presently under construction should be able to discern
whether new physics contributes to CP violation.
Signals for new sources of CP violation include, (i) non-closure of the
3 generation unitarity triangle, (ii) new contributions to $B^0-\bar B^0$ mixing
which yield a non-vanishing phase for this process, 
(iii) non-vanishing CP asymmetries for the
channels $B^0_d\to \phi\pi^0, K_S^0K_S^0$, (iv) inconsistency of separate
measurements of the angles of the unitarity triangle, 
and (v) a deviation of CP rates from SM predictions. 
Models which contain additional CP phases include, non-minimal Supersymmetry,
Multi-Higgs Doublets, Left-Right Symmetric Models, and the Superweak
Model.  A concise review of the effects of these models on CP violating
observables is given by Nir\cite{ynir}.  We present here, as an example,
the case of Multi-Higgs models with three or more Higgs doublets.  In
this scenario $B^0-\bar B^0$ mixing receives additional contributions from
the $H^\pm_{1,2}$ exchange which depend on the phase in the charged scalar
mixing matrix (this phase is discussed in Eq. 2).  Interference between
these contributions and the SM yield an overall non-zero phase in $\Delta M_d$.
Denoting this phase as $\theta_H$ the unitarity angles measured by CP
asymmetries in $B$ decays are thus shifted by
\begin{equation}
a_{CP}(B\to\psi K_S)=-\sin(2\beta+\theta_H),\, 
a_{CP}(B\to\pi\pi)=\sin(2\alpha+\theta_H) \,.
\end{equation}
The magnitude of this effect depends on the size of $\theta_H$, which has
recently\cite{grossnir} been constrained by \bsg.  Another interesting
example is provided in models with an extra iso-singlet down quark; in this
scenario, it has been found\cite{silverman} that  measurements of the 
unitarity angles $\alpha$ and $\beta$ alone
are not enough to distinguish and bound the new contributions, and that
observation of both the third angle $\gamma$ and $B_s$ mixing are also needed.
In summary, the large data sample which will become available will provide a 
series of unique consistency tests of the quark
sector and will challenge the SM in a new and quantitatively precise manner.

\vspace*{-0.5cm}
\nn
\begin{figure}[htbp]
%\centerline{
%\psfig{figure=yossi.ps,height=6cm,width=14cm,angle=0}}
\vspace*{5cm}
\caption{\small The allowed region in the $\sin 2\alpha - \sin 2\beta$ plane
in the SM (solid) and in 2HMD (dot-dashed). From [165].}
\label{yuvalfig}
\end{figure}

\subsection{Top-Quark}

Loop induced flavor changing top-quark decays are small in the SM, as in the 
charm-quark system, due to the effectiveness of the GIM mechanism and the 
small masses of the $Q=-1/3$ quarks.  However, these transitions are
anticipated to be theoretically clean as long distance effects
are expected to be negligible.  The SM rates for $t\to c\gamma, cZ, cg$ are
given by $4.9\times 10^{-13},\, 1.4\times 10^{-13},\, 4.4\times10^{-11}$,
respectively, for $m_t=180$ GeV\cite{ehs}.  The branching fraction for
$t\to ch$ as a function of the Higgs mass is represented by the solid curve 
in Fig. \ref{tch}(a-b).  We see that this rate is also tiny, being in the
$10^{-13}$ range over the entire kinematically allowed region for the Higgs 
mass.  Loop contributions from new physics have been
studied in 2HDM\cite{ehs,jack} and in SUSY\cite{susytop}, and generally can 
enhance these transition rates by 3-4 orders of magnitude for some regions
of the parameter space.  The effects of virtual \ch\ exchange in 2HDM of
Type II on the reactions $t\to cV$, $V=\gamma,Z,g$, are displayed in 
Fig. \ref{tch}(c) for $m_t=180$ GeV.  We see that, indeed, enhancements are 
present for large values of $\tan\beta$.  We also examine the decays
$t\to ch, cH$ in Model II, where $h$ and $H$ respectively represent the 
lightest and heaviest physical neutral scalars present in 2HDM.  The resulting 
rates are depicted in Fig. \ref{tch}(a-b) 
for the demonstrative case of $\mch=600$ GeV and $\tan\beta=2 (30)$,
corresponding to the dashed (solid) curves.  Here we have made use of the
SUSY Higgs mass relationships in order to reduce the number of free
parameters.  We note that the effects of super-partner virtual exchange should
also be included (with, of course, a corresponding increase in the number
of parameters!).  We have also studied these modes in Model I, and found
similar rate increases for regions of the parameter space.
Even if new physics were to produce such enhancements, the resulting
branching fractions would still lie below the observable level in future
experiments at an upgraded Tevatron, the LHC, or the NLC.  

On the other hand, if these FCNC decays were to be detected, they would provide
an indisputable signal for new physics.  Hence a model independent approach 
in probing anomalous FCNC top-quark couplings has recently 
been taken by a number of authors\cite{fcnctop}.  By parameterizing the
general $tcV$ vertex in a manner similar to that presented in Eq. 6, and
performing a Monte Carlo study of the signal rate versus potential backgrounds,
Han \etal\cite{fcnctop} have found that such anomalous couplings can 
be probed down to the level of $\kappa_{\gamma,Z}\equiv
\sqrt{g_L^2+g_R^2}|_{\gamma,Z}\simeq0.1(0.01)$ at the 
Tevatron (LHC).  This corresponds to values of the branching fractions for 
$t\to cZ,c\gamma$ at the level of few$\times 10^{-3}$ for the Tevatron bounds
and $10^{-4}$ for the LHC.  CDF has,
in fact, already performed a search for these FCNC decays from their present
top sample, and has placed the bounds\cite{lecompte} $B(t\to c\gamma+u\gamma)
<2.9\%$ and $B(t\to cZ+uZ)<90\%$ at $95\%$ and $90\%$ C.L., respectively.

\vspace*{-0.5cm}
\nn
\begin{figure}[htbp]
%\centerline{
%\psfig{figure=tch.ps,height=8cm,width=8cm,angle=-90}
%\hspace*{-5mm}
%\psfig{figure=tchh.ps,height=8cm,width=8cm,angle=-90}}
%\vspace*{-0.75cm}
%\centerline{
%\psfig{figure=tcvch.ps,height=8cm,width=10cm,angle=-90}}
\vspace*{17cm}
\caption{\small Branching fractions for (a) $t\to ch$ (b) $t\to cH$ as a
function of the neutral Higgs mass in 2HDM of Type II.  The SM rate is
represented by the solid curve.  (c) $B(t\to cV)$ where $V=g,\gamma,Z$ as a
function of $\tan\beta$ in Model II.  In all cases the top-quark mass is
taken to be 180 GeV.}
\label{tch}
\end{figure}

Potential non-SM tree-level decays of the top-quark could feasibly
occur at measurable rates in future colliders.  
Examples of these possible transitions are: (i) the decay of top into a
charged Higgs, $t\to bH^+$ in multi-Higgs models\cite{bp}, (ii) the tree-level 
flavor-changing decay $t\to ch$, which can occur, if kinematically accessible, 
in multi-higgs models without natural flavor conservation\cite{fcnch,hou}, 
(iii) $t\to \tilde t\tilde\chi^0$ which can take place in 
Supersymmetry if the stop-squark is sufficiently light\cite{xerxes}
(this possibility is related to the large value of the  top Yukawa coupling,
and is thus special to the top system), and (iv) $t\to\tilde\ell^+ d$ in 
SUSY models with R-parity violation\cite{lblguys}.  
For favorable values of the parameters, each of these modes could be
competitive with the SM decay $t\to bW^+$.  The observation of the top-quark
by CDF and D0, which relies heavily on the expected signal from SM top 
decay\cite{TMASS}, can thus restrict the values of the branching fractions 
for these potential new modes.  The possible constraints that could be obtained
on the models which would allow the decays (i) $t\to bH^+$ and (ii) $t\to ch$ 
to occur, if these collaborations were to make 
the statement that the observed $t\bar t$ production
rate is $50-90\%$ of that expected in the SM are given in Fig. \ref{toptree}.
We have examined the case of the decay into a \ch\ in Model II, taking
$m_t=180$ GeV, and find that the potentially excluded regions lie below the 
curves.  Clearly, large regions of the parameter space have the potential to
be ruled out.  In the case of $t\to ch$ decay, we have parameterized the
tree-level $tch$ coupling as $(\sqrt 2 G_F)^{1/2}m_t(\alpha-\beta\gamma_5)$
and displayed the restrictions in the $k\equiv \sqrt{\alpha^2+\beta^2} - m_h$
plane.  The region above the curves would be excluded.

CP violation in top-quark production and decay is expected to be very small in 
the SM\cite{ehstwo}, however, numerous models with new interactions, such as
multi-Higgs models and Supersymmetry, can give
rise to CP violation in the top system at interesting levels.  Since the
top-quark decays before it has time to hadronize, it provides a particularly
good laboratory for the study of such effects.  Searches for CP violating
effects can be carried out by studying CP-odd spin-momentum correlations
in the top-quark decay products.  $e^+e^-$ colliders, with polarized beams, 
are especially suited to carry out such investigations.  Numerous studies of
CP symmetry tests can be found in Ref. \cite{frey,cptop}.

\vspace*{-0.5cm}
\nn
\begin{figure}[htbp]
%\centerline{
%\psfig{figure=tbh.ps,height=8cm,width=8cm,angle=-90}
%\hspace*{-5mm}
%\psfig{figure=tchtree.ps,height=8cm,width=8cm,angle=-90}}
\vspace*{8cm}
\caption{\small Constraints placed on the non-standard decays
(a) $t\to bH^+$ and (b) $t\to ch$ from demanding that the observed event rate
for top-quark pair production is at least $50,60,70,80,90\%$ of that 
expected in the SM, corresponding to the dashed-dot, solid, dotted, dashed,
and solid curves.  $m_t=180$ GeV is assumed.}
\label{toptree}
\end{figure}

\section{Electric Dipole Moments}

Experiments sensitive to the electric dipole moments (EDMs) of atoms
\cite{Cs,Tl,Xe,Hg}, molecules \cite{molecule}, and the neutron \cite{neutron},
provide by far the most sensitive tests of low
energy flavor conserving CP violation \cite{bounds1,bounds2,barrrev}.
The current experimental bounds given in 
Table \ref{edmexp} represent an extraordinary  level of precision.
New techniques, such as atomic traps, may allow improvements
of up to two orders of magnitude by the turn of the century.  
Although no EDM has yet been observed, as discussed
below, the current bounds already place stringent limits on CP violating 
extensions of the SM.

The SM possesses two possible sources of CP violation:
the phase in the CKM quark mixing matrix, and the QCD
vacuum angle, $\bar{\theta}_{QCD}$.  The CKM phase contributes to EDMs only 
at three loops, and requires mixings through all three generations.   
As such, it is highly suppressed, and gives contributions
well below current experimental sensitivity.
In contrast, the QCD vacuum angle contribution is not 
suppressed, and represents a potential background to 
any non-standard model contributions.
However, as discussed below, positive measurements in the
atomic or molecular systems could distinguish $\bar{\theta}_{QCD}$
from non-standard model CP violation. 

\begin{table}%[h]
\begin{center}
\begin{tabular}{|c|c|c|}
\hline\hline
Particle & EDM ($e$~cm) & Reference\\ \hline
$^{133}$Cs  & $ ~~< 7 \times 10^{-24}$ & \cite{Cs}  \\
$^{205}$Tl  & $ ~~< 2 \times 10^{-24}$ & \cite{Tl}  \\
$^{129}$Xe  & $ ~~< 1 \times 10^{-26}$ & \cite{Xe}  \\
$^{199}$Hg  & $ ~~< 9 \times 10^{-28}$ & \cite{Hg}  \\
$^{205}$TlF & $ ~~< 5 \times 10^{-23}$ & \cite{molecule} \\
 neutron    & $ ~~< 8 \times 10^{-26}$ & \cite{neutron} \\
\hline\hline
\end{tabular}
\end{center}
\caption{Experimental bounds on electric dipole moments.}
\label{edmexp}
\end{table}

A systematic determination of the limits placed on CP violating extensions
of the SM from experimental bounds requires evaluating the effective CP 
violating interactions at the weak, nuclear, and atomic scales. 
The results of such an analysis indicate that, in general,
atoms with an unpaired electron ($^{133}$Cs and $^{205}$Tl)
are most sensitive to weak sector CP violation, while
atoms and molecules with paired electrons ($^{129}$Xe, $^{199}$Hg, and
$^{205}$TlF) are sensitive to strong sector CP violation \cite{bounds2}. 
At the atomic scale, the dominant weak sector CP violation is from 
the electron EDM.  In the strong sector, for nuclear spin $j = {1 \over 2}$,
the most important interaction is the electric dipole moment
of the nucleus, which gives rise to a local electromagnetic 
interaction between the electrons and nucleus, generally referred
to as the Schiff moment. 
For nuclear spin $j \geq 1$ the most important strong sector interaction
is the nuclear magnetic quadrapole moment.
At the nuclear scale both these interactions arise predominantly
from the CP-odd pion-nucleon coupling, and at the 
microscopic scale from the light quark chromo-electric
dipole moment (CDM) \cite{bounds2}.
The disparate sensitivity of atoms and molecules with and without
net electronic spin allows CP violation in the weak and strong sectors
to be distinguished.  To illustrate this, the relative EDMs which arise
from an electron EDM and CP-odd pion nucleon coupling
(assumed for simplicity to be isoscalar) are given in Table \ref{wsedm}.
For simplicity the strong sector contributions 
are normalized to that of $^{199}$Hg,
for which the best experimental bound is available. 
Notice that all the atoms, independent of electronic spin, 
are roughly equally sensitive to CP violation which arises in the strong sector.
The exception is $^{129}$Xe for which the atomic matrix elements
are suppressed because of the closed electron shell. 
The slightly increased sensitivity for $^{133}$Cs 
with nuclear spin $j = {7 \over 2}$ is due to the nuclear
magnetic quadrapole moment, which does not exist in the 
other atoms with nuclear spins $j = {1 \over 2}$.
The highly increased sensitivity of the molecule $^{205}$TlF
relative to the atoms is due to the small energy splitting
for rotational levels of opposite parity.  This is offset, however, by 
increased sensitivity to experimental systematics. 

\begin{table}%[h]
\begin{center}
\begin{tabular}{|c|c|c|}
\hline\hline
Particle & Weak  & Strong \\ \hline
$^{133}$Cs  &  ~~120 & ~~5 \\
$^{205}$Tl  &  ~~$-$600 & ~~0.5 \\
$^{129}$Xe  &  ~~$-$0.0008 & ~~0.09 \\
$^{199}$Hg  &  ~0.012 & 1\\
$^{205}$TlF &  ~~80  & 2000 \\
 neutron    &  ~~ -- & 150 \\
\hline\hline
\end{tabular}
\end{center}
\caption{Relative sensitivity of EDMs to CP violation arising
in the weak sector (electron EDM) and strong sector
(CP-odd pion-nucleon coupling).}
\label{wsedm}
\end{table}

Extensions of the SM at or just above the weak
scale often contain additional CP violation. 
This generally gives contributions to EDMs which are not
accidentally suppressed, as is the CKM contribution. 
As an example, supersymmetric theories generally 
possess a large number of phases in the most general soft supersymmetry
breaking terms.  The magnitude of these phases depends on the CP properties
of the supersymmetry breaking sector. 
It has recently been emphasized that even the CKM phase can 
induce non-zero supersymmetric phases at the electroweak scale
from renormalization group running between the 
Planck and GUT scales \cite{GUTphases}.
The existence of CP violating supersymmetric phases is therefore
an important test of supersymmetric GUT theories \cite{GUTphases,brsEDM}.
Both electron and quark EDMs, and quark
CDMs arise from one-loop diagrams with internal 
superparticles \cite{bounds1}.  Since electroweak gauginos are usually much 
lighter than gluinos, they typically dominate the one-loop diagrams.  
For a lightest superpartner mass of 100 GeV  (and assuming 
universality of the soft masses at a high scale)
the experimental bound from $^{205}$Tl on the electron EDM
limits the phases to be $<0.01$ \cite{bounds2,brsEDM}.
For the same set of parameters the bound from $^{199}$Hg on the light quark 
CDM also coincidentally limits the phases to be $<0.01$ \cite{bounds2}.
A similar constraint is obtained from the bound on the neutron 
EDM \cite{bounds1,bounds2,brsEDM}.  In assessing the reliability of these 
limits it is important to keep in mind that for both $^{199}$Hg and the neutron 
there are large uncertainties introduced in the 
evaluation of the hadronic matrix elements of quark CDMs and 
EDMs (in the case of the neutron).  In contrast, the bounds on the electron EDM 
from $^{133}$Cs and $^{205}$Tl are
comparatively free from any atomic or nuclear uncertainties. 

The smallness of the CP violating phases in supersymmetric
models amounts to a mild naturalness problem.  Various suppression mechanisms 
for the phases have been suggested, including CP conservation at 
the GUT scale \cite{smallkm}, heavy superpartners \cite{heavy},
very light neutralinos \cite{smallmu}, and
dynamical relaxation of the phases \cite{dynrelax}.

As another example of a CP violating extension of the SM, 
multi-Higgs theories can possess phases in the 
Higgs potential which gives rise to EDMs at low energy.
In this case since the Higgs-fermion coupling is proportional
to the fermion mass, the dominant diagrams are two-loop,
with an internal top quark or $W$ boson \cite{higgstloops,higgschromo}.
Since here both the electron and quark EDMs and quark CDMs
arise only at two loops, the bounds on the CP violating Higgs
phases are not nearly as stringent as for supersymmetry.
For a single light Higgs boson of mass 100 GeV,
the current bounds on the electron EDM from $^{133}$Cs,
and on the light quark CDM from $^{199}$Hg and the neutron
do not yet place limits on multi-Higgs theories. 
However, improvements in any of the bounds would begin to constrain
the Higgs sector phases. 

Finally, we consider the role of the CP violating QCD vacuum angle,
$\bar{\theta}_{QCD}$.  Its contribution to strong sector CP violation is not 
suppressed by any large mass scale, making the EDM experiments 
very sensitive probes of non-zero $\bar{\theta}_{QCD}$.
The current bounds from $^{199}$Hg and the neutron restrict
the QCD vacuum angle to be $<$ few $\times 10^{-10}$,
leading to the well known strong CP problem.  With the atomic and molecular 
systems it is actually possible to distinguish $\bar{\theta}_{QCD}$ from 
non-standard model CP violation.  Since $\bar{\theta}_{QCD}$ arises in the 
strong sector, it gives EDMs in the ratios indicated in the Strong column
of Table \ref{wsedm}.  The main uncertainties in these ratios is from nuclear 
structure rather than hadronic matrix elements.
Any deviation from this pattern would imply some component
of non-standard model CP violation. 
In fact, given the current (extraordinary) bound on $^{199}$Hg,
a positive measurement in $^{133}$Cs above the level 
of $5 \times 10^{-27}$ $e$ cm, or in $^{205}$Tl above the
level of $5 \times 10^{-28}$ $e$ cm would be a clear signal
for (CP violating) physics beyond the standard model. This would be an 
important result of positive measurements in these atomic systems. 

\section{Lepton Number Violation}

Individual lepton number is conserved in the SM with massless neutrinos.
However, in many extensions of the SM, lepton flavor is violated by new 
physics which lies just above the electroweak scale\cite{lfci}.
The purely leptonic rare processes for which the best experimental
bounds are available are given in Table \ref{leptonflavorv}[198-203].
In the case of $\mu \rightarrow e$ conversion, we define
B $\equiv \sigma(\mu~{\rm Ti} \rightarrow e~{\rm Ti}) / 
  \sigma (\mu~{\rm Ti} \rightarrow capture)$.
Considerable improvements in some of the bounds are possible. 
At LAMPF the level B($\mu \rightarrow e \gamma$) $\sim 6 \times 10^{-13}$
could be accessible \cite{muegprop}, while at PSI B($\mu~{\rm Ti} \rightarrow 
e~{\rm Ti}$) $\sim 3 \times 10^{-14}$ could be reached \cite{mueprop}.  As 
discussed below these levels of precision would begin to probe much of the 
parameter space of supersymmetric GUT theories. 

\begin{table}%[h]
\begin{center}
\begin{tabular}{|c|c|c|}
\hline\hline
  & Branching Ratio & Reference\\ \hline
$\mu \rightarrow e \gamma$ & ~~$<5 \times 10^{-11}$ & LAMPF-Crystal 
Box\cite{muephoton} \\
$\mu \rightarrow eee$  & ~~$<1 \times 10^{-12}$ & SINDRUM\cite{mueee} \\
$\mu~{\rm Ti} \rightarrow e~{\rm Ti}$  & $ ~~<4 \times 10^{-12}$ 
     & SINDRUM\cite{mueconv}  \\
$\tau \rightarrow \mu \gamma$ & $ ~~<5 \times 10^{-6}$ & 
CLEO\cite{taumuphoton}\\
$Z\to e\mu$ & $~~<1.7\times 10^{-6}$ & OPAL \cite{opal}\\
$Z\to e\tau$ & $~~<7.3\times 10^{-6}$ & L3\cite{lthree}\\
$Z\to \mu\tau$ & $~~<1.0\times 10^{-5}$ & L3\cite{lthree}\\
\hline\hline
\end{tabular}
\end{center}
\caption{Experimental bounds on purely leptonic flavor violation.}
\label{leptonflavorv}
\end{table}

All the effective operators which contribute to the processes in Table 
\ref{leptonflavorv} are effectively dimension-six and are suppressed by two 
powers of the scale characterizing lepton flavor violation. 
In most models the rate for $l \rightarrow l^{\prime} \gamma$
is proportional to $m_l^2$.  The bound from $\mu \rightarrow e \gamma$ 
therefore typically gives more stringent limits on the microscopic physics 
than that from  $\tau \rightarrow \mu \gamma$.  In strongly coupled composite 
models in which lepton flavor is not a good symmetry, the bound from $\mu 
\rightarrow e \gamma$ gives a lower limit on the composite scale of roughly 
100 TeV.

Lepton flavor is not necessarily a symmetry of supersymmetric theories. 
The soft supersymmetry breaking terms which give the sleptons a
large mass need not conserve flavor, and can lead to interesting levels 
for the processes listed in Table \ref{leptonflavorv}.  The dominant diagrams 
arise at one-loop and involve mixing of internal sleptons. 
For superpartner masses of roughly 100 GeV, the bound from 
$\mu \rightarrow e \gamma$ limits the selectron-smuon mixing
angle to be $\sin \theta_{\tilde e \tilde\mu} < 3 \times 10^{-3}$.
It has been emphasized recently that non-vanishing mixing
between the sleptons is a generic feature of supersymmetric
GUT theories which are unified below the Planck or compactification
scale (and if supersymmetry breaking is transmitted by gravitational
strength interactions) \cite{GUTflavor}.  Above the GUT scale, quarks and 
leptons are unified within GUT multiplets. 
Renormalization group running between the Planck or compactification
and GUT scales then mixes sleptons through CKM mixings. 
The large top-quark Yukawa coupling enhances this effect. 
For slepton masses in the few $\times 100$ GeV range, the branching ratio for 
$\mu \rightarrow e \gamma$ is within the range $10^{-13}-10^{-11}$
over much of the SUSY parameter space\cite{GUTflavor}.  This is at or directly
below the present experimental bound, and hence
the future improvements \cite{muegprop,mueprop}
therefore represent an important test of supersymmetric GUT theories. 

Since the $\tau$ is the 
heaviest and least well-studied lepton, one might expect on rather general 
grounds that it is most likely to experience LFCI. In addition, due to the 
rather strong constraints arising from the decays $\mu \to e\gamma$, $\mu \to 
3e$, and $\mu-e$ conversion in atoms we might expect LFCI to be highly 
suppressed within the first two generations.  LFCI involving the $\tau$ may 
appear in several ways. First, it is possible that the $\tau$ may have 
sizeable radiative decay modes, \ie, $\tau \to e\gamma,\mu \gamma$, or a 
significant decay to three lepton final states, 
\ie, $\tau \to 3e,~3\mu,~ee\mu,~\mu\mu e$. Secondly, the $\tau$ may have LFCI 
with the SM $Z$ so that the width for $Z\to \tau+e,\mu$ is of reasonable 
size. Within any given model it is likely that the rates for all of these 
processes are related so that, \eg,  if LFCI of the $\tau$ appear in $Z$ 
decays, extensive detailed searches for LFCI in $\tau$ decays are warranted. 
As an example of this scenario, we consider a slightly more general version 
of the model of Eilam and Rizzo{\cite{lfci}} wherein the existence of exotic 
fermions which mix with the ordinary SM leptons induces an off-diagonal 
coupling of the $Z$ to leptons. In such a model, the diagonal couplings of the 
leptons to the $Z$ differ little from the SM case
\begin{equation}
{\cal L}_{SM} = {g\over {2c_w}}\bar l \gamma_\mu(v-a\gamma_5) lZ^\mu  \,,
\end{equation}
where as usual $v=-1/2+2\sin^2 \theta_w$ and $a=-1/2$. The off-diagonal 
couplings can be written in a similar form
\begin{equation}
{\cal L}_{LFCI} = {g\over {2c_w}}\bar \tau \gamma_\mu(v_l'-a_l'\gamma_5) 
lZ^\mu  +h.c. \,,
\end{equation}
where $l\neq \tau$ and with $v_l',a_l'$ depending on the details of the 
model. In this notation, the branching fraction for the $\tau$ 
flavor-violating decay can be written as 
\begin{equation}
B(Z\to \tau^-l^++\tau^+l^-)=2B_l {{(v_l')^2+(a_l')^2}\over {v^2+a^2}} \,,
\end{equation}
with $B_l=0.034$ being the conventional SM leptonic branching fraction. 
Presently, the strongest $95\%$ CL bound on such decays are given in Table
\ref{leptonflavorv}.  These limits imply 
$(v_\mu')^2+(a_\mu')^2 \leq 3.7 \times 10^{-5}$ and 
$(v_e')^2+(a_e')^2 \leq 2.7 \times 10^{-5}$.

Of course, the existence of these $Zl\tau$ couplings directly induces the 
decay modes of $\tau$ into three leptons.  For the processes 
$\tau \to ee\mu$ and $\tau \to \mu\mu e$ we obtain the branching fractions 
\begin{equation}
B=(v^2+a^2)[(v_{e,\mu}')^2+(a_{e,\mu}')^2]B_\tau      \,,
\end{equation}
where $B_\tau$ is the usual $\tau$ leptonic branching fraction. Using the L3 
limits immediately implies that $B(\tau \to ee\mu) \leq 1.7\times 10^{-6}$
and $B(\tau \to \mu\mu e) \leq 1.2\times 10^{-6}$. The branching fractions for 
the $\tau \to 3e,~3\mu$ modes are a bit more complex due to the existence of 
identical particles in the final state; we obtain
\begin{equation}
B={1\over {2}}\left[3(v^2+a^2)[(v_{e,\mu}')^2+(a_{e,\mu}')^2]+(2va)
(2v_{e,\mu}'a_{e\mu}')\right]B_\tau      \,.
\end{equation}
If the second term can be neglected, we then obtain bounds which are $50\%$ 
larger than those stated above. A short calculation shows that the influence of 
the second term relative to the first is at most $5-6\%$ and can occur with 
either sign. 

\section{Double Beta Decay}

No-neutrino double beta decay is a low-energy process which tests
mass scales beyond the reach of present accelerators.  Thus, it is another
promising way to search for physics beyond the SM.  For
it to occur would require not only that lepton number not be
conserved but also that there be at least one additional piece of
new physics.  The latter might be the existence of electron neutrino
mass, of a heavy Majorana neutrino which mixes with the electron
neutrino, of right-handed currents, of a Goldstone boson (such as
the Majoron), or of supersymmetric particles violating $R$ parity.
Limits set on the lifetime for no-neutrino double beta decay
therefore give corresponding restrictions on all these areas of
possible new physics.  Since this second-order weak process with
potentially large phase space is so sensitive, these limits are
generally better than can be set in any other existing experimental
process.  Should a positive effect ever be observed, however, most
of these different potential sources of double beta decay are in
principle indistinguishable experimentally from this process alone.

Since nuclei with even numbers of protons ($Z$) and neutrons
($A-Z$) are generally more stable than their odd-odd neighbors, 
such a nucleus can only decay via the emission of two electrons and two 
neutrinos to the next even-even nucleus, 
$(A,Z) \rightarrow (A,Z+2)+2e^-+2\bar\nu_e$, if it is energetically possible. 
Such very long-lived decays $(\sim 10^{20}$ years) have now
been observed in several nuclei and serve as a test of nuclear
matrix element calculations.

Of interest to particle physics are two potential lepton-number violating
no-neutrino decays.  One is $(A,Z)\rightarrow (A,Z +
2) + 2e^-+\chi^0$, by which limits have been set on neutrino-Majoron
coupling, $\langle g_{\nu\chi}\rangle < 10^{-4}$.  The other
is the more generally useful neutrinoless double beta decay,
$(A,Z)\rightarrow (A,Z+2)+2e^-$, which will be designated as
$\beta\beta_{0\nu}$.  Even if neutrinos are Majorana particles,
these decays would be highly suppressed because of the helicity
reversal required in the neutrino exchange, however, the vastly increased
phase space compared to the two-neutrino case makes these decays
very sensitive probes of non-standard physics.  Experimentally, the
search for $\beta\beta_{0\nu}$ is sensitive because one is looking
for a spike in the summed electron energy spectrum.  $\beta\beta_{0\nu}$ is
worth pursuing because it is an excellent method of experimentally
determining whether neutrinos are Majorana or Dirac
particles, and probably the only way one can ever hope to measure
neutrino masses (as opposed to mass differences) in this very small
mass range. 

\subsection{Experimental Situation}

The current best limit on $\beta\beta_{0\nu}$ is from experiments
using kilograms of enriched $^{76}$Ge.  
The Heidelberg-Moscow group\cite{bal}, which has the best bound at present,
has reported a $90\%$ C.L. he half-life limit of 
$5.6\times 10^{24}$ years corresponding to an effective neutrino mass,
$\langle m_\nu\rangle \approx |\sum_i\eta_iU^2_{ei}m_{\nu_i}|< 0.65$ eV, a 
value rather dependent on the nuclear matrix elements.  The
effective neutrino mass is a sum over those Majorana neutrino mass
states, $m_{\nu_i}$, to which the electron neutrino couples via
mixing matrix elements $U_{ei}$, and $\eta_i=\pm1$ denotes the CP
phase of the i-th neutrino.

While such bounds can provide interesting limits on new physics,
no one has yet claimed a positive result.  However, it is enough to intrigue 
experimentalists and to raise the question as to how much better one can do 
in limiting neutrino mass and other nonstandard physics by $\beta\beta_{0\nu}$.
This will be characterized by $\langle m_\nu\rangle$, but clearly
these limits have many applications.  The enriched Ge experiments,
which now employ $6-8$ kg, have the potential to use about twice
that amount of material.  They could reach an order of magnitude
better lifetime limit, or go from $\sim 1$ eV to $\sim 0.3$ eV in
$\langle m_\nu\rangle$.  A very large experiment, NEMO III, is being
designed which uses sources separate from the detector, and a
calorimeter with tracking, enabling it to reach a 
$\langle m_\nu\rangle$ limit $\sim
0.1$ eV.  Because the resolution is poor (27\%\ at the $^{100}$Mo
endpoint of 3030 keV), this device would have difficulty
establishing a positive effect, but it does have the potential of improving
the present bounds.

Two other isotopes would be competitive and indeed hold promise of
going to still smaller values of $\langle m_\nu\rangle$, but they are only at
the test phase.  One is $^{136}$Xe, which has been used successfully
in a gas Time Projection Chamber, but liquid Xe is probably
required, along with the use of scintillation and/or ionization, to get 
sufficient mass.  Scaling up to ton quantities is possible.  The other isotope
is $^{150}$Nd, which has $\sim$ 70 times the sensitivity of
$^{76}$Ge.  The best form of a detector here, is one in which the source and
detector are the same.  This use of $^{150}$Nd has not been
possible until recently with the advent of cryogenic bolometers. 
Indications are that one could have 1 keV resolution (three
times better than Ge ionization detectors) at the endpoint of 3467
keV, which is higher in energy than all natural $\beta$ and
$\gamma$ radioactivity, thus reducing background problems.  The UCSB,
CfPA, Stanford, Baksan group is testing this approach and will
probably add a $\sim 0.1$ kg\, $^{150}$Nd detector to their dark
matter experiment.

\subsection{Neutrinoless Double Beta Decay 
 and Physics Beyond The Standard Model}

The bounds obtained by the Heidelberg-Moscow collaboration\cite{bal}
imply strong upper limits on the strength of 
lepton number violating interactions. Since
the standard electroweak model conserves $B-L$ quantum number,
the above upper limits would
provide important information on new physics scenarios beyond the
standard model that involve lepton number violation. There exist several
well motivated scenarios of new physics that fall into this category.  The
models that we will discuss here are:
(i) the left-right symmetric models of weak interaction with
the see-saw mechanism for neutrino masses\cite{MS,langsank},
(ii) the MSSM where without the 
additional assumption of R-parity conservation
one has both lepton and baryon number violating terms\cite{rpar}, and
(iii) composite models for leptons.

To see how one extracts constraints on new physics from the observed
lower limit on the lifetime for $\beta\beta_{0\nu}$ process, let us
parameterize the amplitude for this process as
\be{mnu}
A_{\beta\beta}\simeq {G^2_F}\left(\mu^2_{\beta\beta}\right) \,,
\ee
\noindent where we have hidden all nuclear physics effects in the
effective mass $\mu_{\beta\beta}$. The width for the
decay can then be written as 
\be{width}
\Gamma_{\beta\beta}\simeq {{Q^5|A_{\beta\beta}|^2}\over{60\pi^3}} \,.
\ee
Here, $Q$ is the available energy for the two electrons.
 To get a feeling for the upper limit on $\mu_{\beta\beta}$, note that
the present $^{76}$Ge experiment\cite{bal} implies that 
$\Gamma_{\beta\beta}\leq 3.477\times 10^{-57}$ GeV; using for a rough
estimate $Q\simeq 2.3$ MeV, we find that $\mu_{\beta\beta}\leq 10^{-5}$ GeV.
We will now estimate the parameter $\mu_{\beta\beta}$ for various extensions
of the SM and translate this upper limit into constraints on the
parameters of the model.

Let us first consider the classic contribution of the Majorana neutrino mass
to this process\cite{mnu}. In this case, one gets $\mu^{2}_{\beta\beta}
=\langle m_{\nu}\rangle p_F f_N$ 
( $p_F$ is the Fermi momentum in the nucleus $\approx 50 $ MeV,
and $\langle m_{\nu}\rangle$ is as defined above).  Barring nuclear
uncertainties\cite{klapdor} ( hidden in the factor $f_N$ ), 
detailed calculations lead to the upper bound $\langle m_{\nu} \rangle\leq
.65$ eV.

\vspace{3mm}
\noindent{\bf $\bullet$ Left-right symmetric model}
\vspace{2mm}

There are four new contributions to the $\beta\beta_{0\nu}$ process in the
LRM in addition to the neutrino mass diagram just discussed:
 (i) the first one arises from the exchange of heavy Majorana
right handed neutrinos ( $N_R$) and the right handed $W$ bosons 
(see Fig. \ref{rabi1});
(ii) that arising from a $W_L-W_R$ exchange and therefore necessarily
involving the mixing between the light and heavy neutrinos; (iii) exchange
of a doubly charged Higgs boson as shown in Fig. \ref{rabi2}; 
(iv) vector scalar exchange\cite{babu} involving a singly 
charged Higgs boson and the
left-handed $W_L$ boson . The present upper bound on the 
$\beta\beta_{0\nu}$ amplitude then leads to restrictions on the 
parameters of the model involved in these various graphs. We summarize the
constraints obtained on these separate contributions below, where we
assume that the strength of the right-handed interactions and the
right-handed CKM matrix are equal to their left-handed counterparts
(\ie, $g_R=g_L$ and $V^R_{ij}=V^L_{ij}$).

%%%
\vspace*{-0.5cm}
\nn
\begin{figure}[htbp]
%\centerline{
%\psfig{figure=rabifig1.ps,height=6cm,width=6cm,angle=0}}
\vspace*{7cm}
\caption{\small Heavy right handed neutrino contribution to 
$\beta\beta_{0\nu}$ in the LRM.}
\label{rabi1}
\end{figure}
\vspace*{0.4mm}

%%%
\vspace*{-0.5cm}
\nn
\begin{figure}[htbp]
%\centerline{
%\psfig{figure=rabifig2.ps,height=6cm,width=6cm,angle=0}}
\vspace*{6cm}
\caption{\small Contribution of the doubly charged Higgs boson in the
LRM.}
\label{rabi2}
\end{figure}
\vspace*{0.4mm}

\vspace{3mm}
\noindent{\underline{\it Heavy $N_R$ exchange }}
\vspace{2mm}

The effective mass parameter in this case can be written as (where $\zeta$
denotes the amount of heavy-light neutrino mixing)
\be{nuR}
\mu^2_{\beta\beta}\simeq (p^3_{eff})\left(
{M^4_{W_L}\over{M^4_{W_R}}}+\zeta^2\right){{1}\over{m_N}} \,.
\ee
Here $p_{eff}$ is an effective momentum chosen to have a value of 50 MeV.
The present limits on the neutrinoless double beta decay lifetime then
impose a correlated constraint on the parameters $M_{W_R}$ and 
$m_N$\cite{moha1}.

\vspace{3mm}
\noindent{\underline{\it Light-heavy neutrino mixing contribution:}}
\vspace{2mm}

In this case, one finds $\mu^2_{\beta\beta}\simeq \zeta\left(M^2_{W_L}/
M^2_{W_R}\right)(p^2_{eff})$. This leads to the
correlated constraint on $\zeta$ and $M_{W_R}$ shown in Fig. \ref{rabi4}.
If we combine the theoretical constraints of vacuum stability then, 
the present $^{76}$Ge data provides a lower limit on the masses of
the right handed neutrino ($N_e$) and the $W_R$ of 1 TeV, which is
a rather stringent constraint. The limits on $\zeta$ on the other hand
are not more restrictive than what would be expected from the structure
of the theory. We have of course assumed that the leptonic mixing
angles are small, so that there is no cancellation between the parameters.

%%%
\vspace*{-0.5cm}
\nn
\begin{figure}[htbp]
%\centerline{
%\psfig{figure=rabifig4.ps,height=8cm,width=10cm,angle=0}}
\vspace*{8cm}
\caption{\small Bounds on the light and 
heavy neutrino mixing parameter
in the LRM from $^{76}Ge$ data.}
\label{rabi4}
\end{figure}
\vspace*{0.4mm}

\vspace{3mm}
\noindent{\underline{\it Higgs contributions:}}
\vspace{2mm}

The two types of Higgs induced contributions to 
$\beta\beta_{0\nu}$ decay indicated above lead to the following
expressions for $\mu_{\beta\beta}$. The one arising 
from the coupling of the doubly charged Higgs boson to electrons leads to
\be{w5w}
\mu^2_{{\beta\beta}_H}= ({{f_{11}}\over{M^2_{\Delta}}})
2^{7/4}G^{3/2}_F\left({{M_{W_L}}\over{M_{W_R}}}\right)^3 \,.
\ee
The present $^{76}$Ge data implies that (assuming $M_{W_R}\geq 1$ TeV ) 
$M_{\Delta^{++}}\geq \sqrt{f_{11}}~~ 80$ GeV. The second contribution
arises from the mixing amongst the charged Higgs fields after 
the full gauge symmetry is broken
down to $U(1)_{em}$. Denoting this mixing term by the angle $\theta$,
one can get in this case,
\begin{equation}
\mu^2_{{\beta\beta}_H'}\simeq
{{h_u f_{11} sin 2\theta p^2_{eff}}\over {4\sqrt{2}G_F M^2_{H^{+}}}}~,
\end{equation}
where we have assumed
that the $H^+$ is the lighter of the two Higgs fields, and $f_{11}$ and
$h_u$ are Yukawa couplings associated with the triplet and bi-doublet Higgs
fields, respectively.  One then 
finds\cite{babu} $h_u f_{11}{\rm sin}2\theta 
\leq 6\times 10^{-9}(M_{H^+}/ 100\gev)^2$, which is quite
a stringent constraint on the parameters of the theory. To appreciate
this somewhat more, we point out that one expects $h_u\approx m_u/ M_W
\approx 5 \times 10^{-5}$ in which case, we get an upper limit for the coupling
of the Higgs triplets to leptons $f_{11}{\rm sin}2\theta
\leq 10^{-4}$ (for $m_{H^+} = 100 ~GeV$).  
Taking a reasonable choice of $\theta \sim M_{W_L}/M_{W_R}
\sim 10^{-1}$ would correspond to a limit $f_{11} \le 10^{-3}$.  
Bounds on this parameter from an analysis\cite{swartz}
of Bhabha scattering are only or order $\sim 0.2$ 
for the same value of the Higgs mass.

\vspace{3mm}
\noindent{\bf $\bullet$ MSSM with R-parity violation}
\vspace{2mm}

The next class of theories we will consider is the case of supersymmetry
with R-parity violation\cite{rpar}. The R-violating 
part of the potential is given in Section 2.
In order to maintain proton stability we set $\lambda^{''}=0$ in the
superpotential of Eq. (3).
The first contribution to $\beta\beta_{0\nu}$ decay is
dominantly mediated by heavy gluino exchange\cite{moha2} as shown in
Fig. \ref{rabi5}.  Detailed evaluation of the nuclear 
matrix element for this class of models\cite{hirsch} has led to the
following bound on the R-violating parameter,
\be{rp}
\lambda^{\prime}_{111}\leq 3.9\times 10^{-4}\left({{m_{\tilde{q}}}\over{100
GeV}}\right)^2\left({{m_{\tilde{g}}}\over{100 GeV}}\right)^{1/2} \,.
\ee
The second class of contribution is presented in Fig. \ref{rabi6}.
This leads to a contribution to $\mu^2_{\beta\beta}$ given by
\begin{equation}
\mu^2_{\beta\beta}~\simeq~\left({{(\lambda^{\prime}_{113}}
\lambda^{\prime}_{131})\over{2\sqrt{2}G_F M^2_{\tilde{b}}}}
\right)\left({{ m_b}\over{M^2_{\tilde{b}^c}}}\right)\left(\mu\tan\beta
+A_bm_0\right)(p^3_{eff}) \,.
\end{equation}
Here $A_b,m_0$ are supersymmetry breaking parameters, and $\mu$ is the
supersymmetric mass of the Higgs bosons, as discussed in Section 2.  
For the choice of all squark masses, as well as $\mu$ and the SUSY
breaking mass parameters,
being of order of 100 GeV, $A_b=1$, tan$\beta=1$,the following bound
on R-violating couplings is obtained,
\be{w9}
\lambda^{\prime}_{113}\lambda^{\prime}_{131}\leq 3\times 10^{-8} \,.
\ee
If the exchanged scalar particles in Fig. \ref{rabi6} are the
$\tilde{s}-\tilde{s}^c$ pair, one obtains the constraint 
\be{w10}
\lambda_{121}'\lambda_{112} 
\leq 1 \times 10^{-6} \,.
\ee

%%%
\vspace*{-0.5cm}
\nn
\begin{figure}[htbp]
%\centerline{
%\psfig{figure=rabifig5.ps,height=6cm,width=8cm,angle=0}}
\vspace*{6cm}
\caption{\small Gluino mediated contribution in MSSM with R-parity
violation.}
\label{rabi5}
\end{figure}
\vspace*{0.4mm}

%%%
\vspace*{-0.5cm}
\nn
\begin{figure}[htbp]
%\centerline{
%\psfig{figure=rabifig6.ps,height=6cm,width=8cm,angle=0}}
\vspace*{6cm}
\caption{\small Vector-scalar contribution in MSSM with R-parity violation.}
\label{rabi6}
\end{figure}
\vspace*{0.4mm}

\vspace{3mm}
\noindent{\bf $\bullet$ Limits on the scale of lepton compositeness}
\vspace{2mm}

If the quarks and leptons are composite particles, it is natural to
expect excited leptons which will interact with the electron via some
effective interaction involving the $W_L$ boson. If the excited neutrino
is a majorana particle, then there will be contributions to $\beta\beta_{0\nu}$
decay mediated by the excited neutrinos ($\nu^*$). The effective interaction
responsible for this is obtained from the primordial interaction
\be{nunu}
H_{eff}^{\nu^*}= g{{\lambda^{(\nu^*)}_W}\over{2 m_{\nu^*}}}\overline{e}
\sigma^{\mu\nu}(\eta^*_L(1-\gamma_5)+\eta^*_R(1+\gamma_5))\nu^*W_{\mu\nu}
+~h.c. 
\ee
Here L and R denote the  left and right chirality states. This has been
studied in detail in two recent papers\cite{sriva} and leads to the bound 
(taking $|\eta^*_L|^2+|\eta^*_R|^2=1$ with $\eta^*_L\eta^*_R=0$, and
assuming that the excited neutrino is a Majorana particle)
\begin{equation}
 m_{\nu^*}\geq 5.9\times 10^{4}\tev
\end{equation}
for $\lambda^{(\nu^*)}_W\geq 1$.  Here, the compositeness scale has been set
to be the mass of the excited neutrino.  However, this yields a rather 
stringent bound on the compositeness scale!

\vspace{3mm}
\noindent{\bf $\bullet$ Models with heavy sterile neutrinos}
\vspace{2mm}

The see-saw mechanism for understanding small neutrino masses always
requires the introduction of heavy neutral sterile fermions. If
there is a single sterile neutrino within the standard gauge model,
then constraints from the see-saw mechanism suppress the heavy sterile
contribution to the $\beta\beta_{0\nu}$ process.
Since the heavy sterile sector is largely unknown, a possibility to
consider is to have two heavy sterile leptons which participate in a
$3\times 3$ see-saw with the light neutrino to make $m_{\nu}$ small.
The analog of the mixing parameter $\zeta$ is then not constrained to be
small\cite{bamert} by the see-saw considerations and also a larger
range of masses for the heavy sterile particles are then admissible. 
Such models are however subject to a variety of cosmological and
astrophysical constraints. These constraints have been analyzed in
detail in \cite{bamert} and it is found that there is a large range
of the parameter space for the sterile particles which can be probed by
the ongoing neutrinoless double beta experiments.

\section{Summary}

In conclusion, we have shown that a large number of processes are influenced
by the virtual effects of new physics in higher order interactions and thus 
have tremendous power in probing physics beyond the SM.  This attack
on the search for new physics is important as it can probe higher energy scales
and it provides a complementary search reach to direct production at colliders.
The drawbacks are, however, that (i) simultaneous virtual 
contributions of many new particles have the potential to cancel each other's 
effects, (ii) not all models have large indirect effects, and (iii) an
enormous amount of very precise data must be gathered.
Considering our lack of knowledge on what lies beyond the Standard
Model, we stress the importance of searching for effects of
new physics via every possible means and cross-checking any positive
signal in the largest number of processes available.  However, 
these indirect probes should not be recommended as providing a complete
substitute for continuing our search new physics at ever-higher energy 
colliders.

\vspace{6mm}
\noindent {\bf Acknowledgements}
\vspace{4mm}

We would like to thank G. Burdman, S. Chu, M. Cooper, D. Demille, N. Fortson,
E. Golowich, A. Grant, L. Hall, D. Iteinzen, S. Lamoreaux, W. Marciano, 
S. Pakvasa, J. Rosner, and E. Thorndike for useful discussions.
%
%%%%%%%%%%%%%%%%%%--- References
%%%%%%%%%%%%%%%%%%%%%%%%%%%%%%%%%%%%%%%%%%%%%%%%%%%%%%%
\def\MPL #1 #2 #3 {Mod. Phys. Lett. {\bf#1},\ #2 (#3)}
\def\IJMP #1 #2 #3 {Int. J. Mod. Phys. {\bf#1},\ #2 (#3)}
\def\NPB #1 #2 #3 {Nucl. Phys. {\bf#1},\ #2 (#3)}
\def\PLB #1 #2 #3 {Phys. Lett. {\bf#1},\ #2 (#3)}
\def\PR #1 #2 #3 {Phys. Rep. {\bf#1},\ #2 (#3)}
\def\PRD #1 #2 #3 {Phys. Rev. {\bf#1},\ #2 (#3)}
\def\PRL #1 #2 #3 {Phys. Rev. Lett. {\bf#1},\ #2 (#3)}
\def\RMP #1 #2 #3 {Rev. Mod. Phys. {\bf#1},\ #2 (#3)}
\def\ZPC #1 #2 #3 {Z. Phys. {\bf#1},\ #2 (#3)}
\def\PLA #1 #2 #3{Phys. Lett. {\bf#1},\ #2 (#3)}
\def\PTP #1 #2 #3{Prog. Thoer. Phys. {\bf#1},\ #2 (#3)}
\def\NCA #1 #2 #3 {Nuovo Cim.\ {\bf#1},\ #2 (#3)}

\end{document}